\documentclass{aa} 

\usepackage{graphicx}
\usepackage{txfonts}
\usepackage{xcolor}
\usepackage[]{graphicx}
\usepackage{subfig}
\usepackage{longtable}
\usepackage{lscape}
\usepackage[breaklinks=true]{hyperref} 

\graphicspath{
    {.} 
    {paper_figures/}
}

\newcommand{\lya}{\ensuremath{\mathrm{Ly}\alpha}}
\newcommand{\lae}{\ensuremath{\mathrm{Ly}\alpha} emitter}
\newcommand{\laes}{\ensuremath{\mathrm{Ly}\alpha} emitters}

\newcommand{\ergslum}{\ensuremath{\mathrm{erg\,s^{-1}}}}
\newcommand{\ergsline}{\ensuremath{\mathrm{erg\,s^{-1}\,cm^{-2}}}}
\newcommand{\ergsurfb}{\ensuremath{\mathrm{erg\,s^{-1}\,cm^{-2}\,arcsec^{-2}}}}

\newcommand{\invmpc}{\ensuremath{\mathrm{cMpc^{-3}}}}
\newcommand{\msun}{\ifmmode M_{\odot} \else M$_{\odot}$\fi}
\newcommand{\msunyr}{\ensuremath{\mathrm{M_{\odot} yr^{-1}}}}

\newcommand{\kms}{\ensuremath{\mathrm{km\,s^{-1}}}}
\newcommand{\degree}{\ensuremath{^\circ}}

\newcommand{\erglinesurf}[2]{\ensuremath{\mathrm{#1 \times 10^{#2}\,erg\,s^{-1}\,cm^{-2}\,arcsec^{-2}}}}
\newcommand{\erglsurf}[1]{\ensuremath{\mathrm{10^{#1}\,erg\,s^{-1}\,cm^{-2}\,arcsec^{-2}}}}

\newcommand{\hi}{\ifmmode {\textrm{H\textsc{i}}} \else H\textsc{i} \fi}

\newcommand{\mosaic}{\textsf{MOSAIC}}
\newcommand{\udft}{\textsf{UDF-10}}
\newcommand{\mxdf}{\textsf{MXDF}}
\newcommand{\origin}{\textsf{ORIGIN}}
\newcommand{\odhin}{\textsf{ODHIN}}
\newcommand{\galics}{\textsf{GALICS}}

\begin{document} 

\title{The MUSE Extremely Deep Field: the Cosmic Web in Emission at High Redshift\thanks{Based on observations made with ESO telescopes at the La Silla Paranal Observatory under the large program 1101.A-0127}
}

\author{
Roland Bacon\inst{1} \and David Mary\inst{2} \and Thibault Garel\inst{3,1} \and Jeremy Blaizot\inst{1} \and Michael Maseda\inst{4} \and Joop Schaye\inst{4} \and Lutz Wisotzki\inst{5} \and Simon Conseil\inst{6} \and Jarle Brinchmann\inst{7,4} \and Floriane Leclercq\inst{3} \and Valentina Abril-Melgarejo\inst{9} \and Leindert Boogaard\inst{4} \and Nicolas Bouch\'e\inst{1} \and Thierry Contini\inst{8} \and Anna Feltre\inst{13,1} \and Bruno Guiderdoni\inst{1} \and Christian Herenz\inst{14} \and Wolfram Kollatschny\inst{10} \and Haruka Kusakabe\inst{3} \and Jorryt Matthee\inst{11} \and L\'eo Michel-Dansac\inst{1} \and Themiya Nanayakkara\inst{12} \and Johan Richard\inst{1} \and Martin Roth\inst{5} \and Kasper B. Schmidt\inst{5} \and Matthias Steinmetz\inst{5} \and Laurence Tresse\inst{1} \and Tanya Urrutia\inst{5} \and Anne Verhamme\inst{3} \and Peter M. Weilbacher\inst{5} \and Johannes Zabl\inst{1} \and Sebastiaan L. Zoutendijk\inst{4}
}

\institute{
   Univ Lyon, Univ Lyon1, Ens de Lyon, CNRS, Centre de Recherche Astrophysique de Lyon UMR5574, F-69230, Saint-Genis-Laval, France
   \and Laboratoire Lagrange, CNRS, Universit\'e C\^ote d'Azur, Observatoire de la C\^ote d'Azur, CS 34229, 06304, Nice, France 
   \and Observatoire de Gen\`eve, Universit\'e de Gen\`eve, 51 Ch. des Maillettes, CH-1290 Versoix, Switzerland
   \and Leiden Observatory, Leiden University, P.O. Box 9513, 2300 RA Leiden, The Netherlands 
   \and Leibniz-Institut f{\"u}r Astrophysik Potsdam (AIP), An der Sternwarte 16, 14482 Potsdam, Germany
   \and Gemini Observatory/NSF’s NOIRLab, Casilla 603, La Serena, Chile
   \and Instituto de Astrof{\'\i}sica e Ci{\^e}ncias do Espaço, Universidade do Porto, CAUP, Rua das Estrelas, PT4150-762 Porto, Portugal  
   \and IRAP, Institut de Recherche en Astrophysique et Plan\'etologie, CNRS,  Universit\'e de Toulouse, 14, avenue Edouard Belin, F-31400 Toulouse, France 
    \and Aix Marseille Universit\'e, CNRS, LAM (Laboratoire d'Astrophysique de Marseille) UMR 7326, 13388, Marseille, France 
    \and Institut  f{\"u}r Astrophysik, Universit{\"a}t G{\"o}ttingen, Friedrich-Hund-Platz 1, D-37077 G{\"o}ttingen, Germany 
    \and ETH Zurich, Institute of Astronomy, Wolfgang-Pauli-Str. 27, CH-8093 Zurich, Switzerland 
    \and Centre for Astrophysics and Supercomputing, Swinburne University of Technology, Melbourne, VIC 3122, Australia
    \and INAF – Osservatorio di Astrofisica e Scienza dello Spazio di Bologna, Via P. Gobetti 93/3, 40129 Bologna, Italy
    \and  European Southern Observatory, Av. Alonso de C\'ordova 3107, 763 0355 Vitacura, Santiago, Chile
   }

\date{Submitted 11 November 2020, Accepted 10 February 2021}

\abstract {
   We report the discovery of diffuse extended \lya\ emission from redshift 3.1 to 4.5, tracing cosmic web filaments on scales of 2.5-4 comoving Mpc. These structures have been observed in overdensities of \lya\ emitters in the MUSE Extremely Deep Field, a 140 hour deep MUSE observation located in the Hubble Ultra Deep Field. Among the 22 overdense regions identified, 5 are likely to harbor very extended \lya\ emission at high significance with an average surface brightness of \erglinesurf{5}{-20}. Remarkably, 70\% of the total \lya\ luminosity from these filaments comes from beyond the circumgalactic medium of any identified \lae. 
Fluorescent \lya\ emission powered by the cosmic UV background can only account for less than 34\% of this emission at z$\approx$3 and for not more than 10\% at higher redshift.    
We find that the bulk of this diffuse emission can be reproduced by the unresolved \lya\ emission of a large population of ultra low luminosity \laes\ ($<10^{40}$ \ergslum), provided that the faint end of the \lya\ luminosity function is steep ($\alpha \lessapprox -1.8$),  it extends down to luminosities lower than $10^{38} - 10^{37}$ \ergslum\ and the clustering of these \laes\ is significant (filling factor $< 1/6$). 
If these \laes\ are powered by star formation, then this implies their luminosity function needs to extend down to star formation rates $\mathrm{< 10^{-4} M_\odot yr^{-1}}$. 
These observations provide  the first detection of the cosmic web in \lya\ emission in typical filamentary environments
and the first observational clue for the existence of a large population of ultra low luminosity \laes\ at high redshift. 
}


\maketitle
   
\section{Introduction}
\label{sect:intro}
The current paradigm of structure formation predicts that most of the gas in the intergalactic medium (IGM) is organized in a "cosmic web" composed of  non-uniform gaseous filaments  connecting galaxies on scales of megaparsecs (e.g. \citealt{White1987, Bond1996}).

These filaments are feeding the circumgalactic medium (CGM), the gaseous component that is responsible for regulating the gas exchange between galaxies and the surrounding IGM (e.g. \citealt{Tumlinson2017}).
This complex interplay between infall of gas from filaments through the CGM and the ejection of matter from the feedback processes is expected to play a key role in the regulation of galaxy growth through cosmic time.

The IGM has been explored over the past decades mainly using absorption line spectroscopy, which provides a powerful way to trace the neutral hydrogen observed in Lyman-alpha (\lya) absorption against bright background quasars \citep{Gunn1965, Meiksin2009}. 
However, it has not been possible to obtain a detailed picture of these filaments, as information is limited to one dimension along the line-of-sight to the background source. The low sky density of sufficiently bright background sources prevents the study of the cosmic web on scales smaller than a few megaparsecs.
Only very recently, sparse two-dimensional constraints on the IGM structure at a transverse spatial sampling of a few Mpc have started to become available (e.g. \citealt{Lee2018}).

Imaging the cosmic web in emission would provide the missing three-dimensional information. Filaments are predicted to emit the hydrogen \lya\ line by fluorescence induced by the ultraviolet background (UVB) radiation.
However, the low intensity of the UVB \citep{Haardt2012} translates into an expected surface brightness of a self-shielded filament of approximately \erglsurf{-20} at z=3 \citep{Gould1996}. This challenging level has meant that a direct detection of UVB induced fluorescent emission from IGM filaments has remained elusive \citep{Gallego2018}.

To overcome this obstacle, a solution is to observe selected regions where local ionizing sources such as bright quasars (QSO) or star forming galaxies, boost the \lya\ emission to detectable levels (e.g. \citealt{Cantalupo2005}). This technique has been successfully used to map the IGM at scales of a few 100 kpc around QSOs \citep{Cantalupo2014, Martin2015, Borisova2016, Fumagalli2016, Kikuta2019}.
To extend the mapping of the IGM, specific fields with multiple QSOs have been recently targeted using the MUSE instrument \citep{Bacon2010} at ESO/VLT (e.g. \citealt{Arrigoni2019, Lusso2019}).

However, the scale currently probed by all these observations is limited to a few hundred kiloparsecs, a scale larger than the CGM of the host galaxy but still too small to probe the filaments at the megaparsec scale relevant to the IGM. 
A notable exception are the observations of \cite{Umehata2019} who report the detection of a 1.3 Mpc long filament with a mean \lya\ surface brightness of \erglinesurf{3}{-19} in the SSA22 protocluster at $\rm z=3.1$ \citep{Steidel1998}. 

While these observations have revealed some filamentary structures of ionized gas in very massive structures, they are biased to specific environments. For example the SSA22 protocluster with its large overdensity of AGNs, sub-millimeter galaxies and \lya\ blobs \citep{Lehmer2009, Umehata2018, Umehata2019, Herenz2020}, is an extreme environment.  
Furthermore the environment of a QSO with anisotropic UV radiation and possible excess of tidal debris from past interactions \citep{Canalizo2001}, may not be representative of the generic IGM. 
  
Another approach was used by \cite{Daddi2020} who targeted a massive structure, corresponding to a dark matter halo of $\mathrm{\approx 4 \times 10^{13}M_\sun}$, embedded in a giant \lya\ nebula at $z$=2.9 with Keck/KCWI \citep{Martin2010}. Within 300 kpc they find three cold gas filaments connected to the central massive galaxy. Such observations give important constraints for models of the formation of galaxies in these massive structures located at the nodes of the cosmic web.   

However, such very massive structures are not representative of the filamentary environment which represent 60\% of the total gas mass of the Universe and where we expect most of the galaxy formation to occur (e.g. \citealt{Libeskind2017, Martizzi2019}). For example, the conclusions reached by \cite{Daddi2020} for the gravitational energy as the main physical mechanism responsible of the \lya\ emission, may not stand in two orders of magnitude less massive environments.

It is therefore highly desirable to obtain 3D information of the IGM in emission in more representative environments. Having multiple detections of the cosmic web structure and its evolution with redshift would also give fundamental constraints for the simulation and models of structure formation.

In the past years we have used a fraction of our MUSE guaranteed observing time to perform deep field observations in the Hubble Ultra Deep Field (HUDF, \citealt{Beckwith2006}) at 10 and 30 hours depth \citep{Bacon2017}. These observations have increased the amount of available spectroscopic information in the HUDF by more than an order of magnitude \citep{Inami2017} and enabled several breakthroughs in our understanding of the high redshift universe, notably the discovery of ubiquitous extended \lya\ emission from the CGM around individual galaxies at $\mathrm{z > 3}$ \citep{Wisotzki2016, Leclercq2017, Leclercq2020}.

We also performed stacking experiments to explore the extended \lya\ emission around galaxies at larger distance, extending the results for individual halos to a radius of 60 pkpc (physical kpc) or 240 ckpc (comoving kpc) at $\rm z=3$ \citep{Wisotzki2018}. Although such a distance is well beyond the predicted virial radii of the host dark matter haloes for the individual \laes\  of our sample, the stacking process erases all geometrical information and is not adapted to the morphological study of the filamentary cosmic web.

Given the successful experience with these MUSE spectroscopic deep fields, we have recently performed a new deep field to push forward in depth. The so-called MUSE Extremely Deep Field (hereafter \mxdf), is a single field located within the HUDF area (Fig.~\ref{fig:fields}). It reaches a maximum depth of 140 hours and benefits from improved spatial resolution thanks to the recent coupling of MUSE with the ESO Adaptive Optics Facility (AOF) as well as improved data reduction processes (Section~\ref{sect:data}).

In this paper we present new results regarding the cosmic web in emission at $\mathrm{z > 3}$ based on this new data set complemented by existing HUDF datacubes published in \cite{Bacon2017}.

Galaxy formation should take place preferentially in the densest part of the cosmic web filaments and thus by selecting moderately overdense regions of \laes, we should maximize our chance to detect diffuse \lya\ emission\footnote{
By diffuse emission we mean \lya\ emission that is spatially extended and not resolved in \laes\ at our limiting flux. See also Section~\ref{subsec:shape} for a practical definition.}
associated with these filaments. 

We have explored the datacube in redshift space to select overdense regions of \laes\ (Section~\ref{sect:overdensities}). For each overdense region, we search and study \lya\ diffuse emission (Section~\ref{sect:diffuse}).
We then discuss the implication of our discoveries in Sections~\ref{sect:analysis} and \ref{sect:lyasource}, and finally conclude in Section~\ref{sect:conclusion}.

We use the Planck 2018 cosmological model \citep{Planck.2018}
with $\mathrm{H_0 = 67.4 \, km\,s^{-1} Mpc^{-1}}$ and $\mathrm{\Omega_m = 0.315}$.
Unless specified, all distances are given in comoving scale (cMpc).

\section{Observations and data reduction}
\label{sect:data}
The details of the observations, data reduction and source catalogs will be given in a forthcoming paper (Bacon et al, in prep). We provide here a short summary of the process.

\subsection{Observations}
\begin{figure}[htbp]
\begin{center}
\includegraphics[width=0.7\columnwidth]{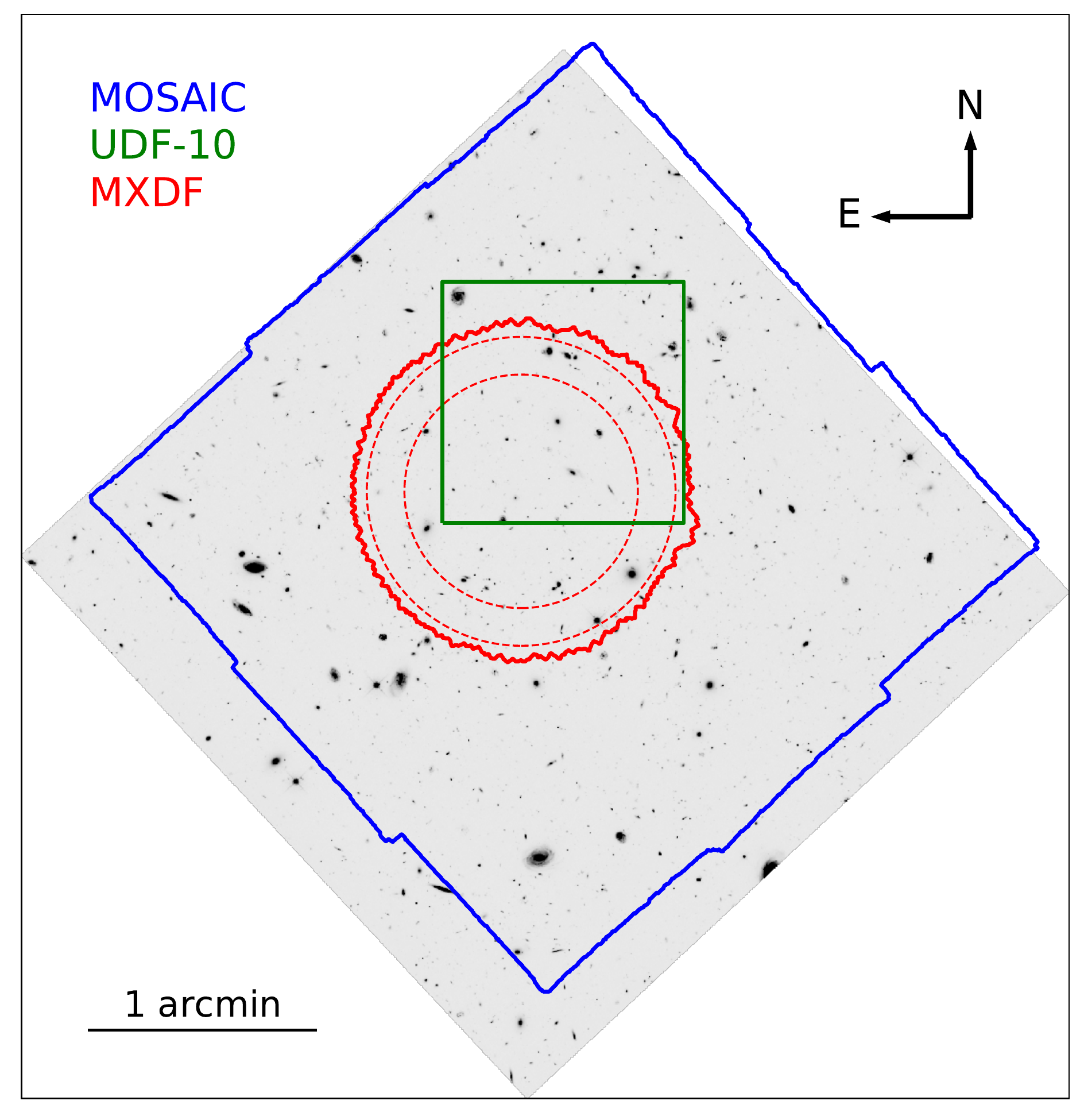}
\caption{Location of the 3 deep fields used in this paper: \mxdf\ (140 hours depth), \mosaic\ (10 hours depth) and \udft\ (30 hours depth) overlayed on the HST F775W UDF image. The two dotted red circles show the \mxdf\ 10 and 100 hours exposure time contours.}
\label{fig:fields}
\end{center}
\end{figure}

The observing campaign started in August 2018 and lasted until January 2019 for a total of 6 runs performed during the new moon periods. All observations were performed with the dedicated VLT GALACSI/AOF Ground-Layer Adaptive Optics system (\citealt{Kolb2017, Paufique2018}). A total of 155 hours of integration were obtained. After rejection of bad exposures the achieved final depth is 140 hours. To minimize systematics, the field of view was rotated by a few degrees between each observing block consisting of 4 $\times$ 25 min exposures.
Consequently, the final combined field of view is approximately circular (Fig.~\ref{fig:fields}) with a radius of 41\arcsec\ and  31\arcsec\ for respectively 10+ and 100+ hours depth. The field center celestial coordinates are 53{\degree}.16467, -27{\degree}.78537 (J2000 FK5).

\subsection{Data Reduction}
The data reduction is similar to the procedure developed in the first non-AO observations of the UDF \citep{Bacon2017}. It is based on the MUSE pipeline \citep{Weilbacher2020} and includes a number of new developments. 
A super flatfield is performed for each exposure by combining, without any recentering and derotation, a large number of observations obtained during the same run. This "superflat" is subtracted from each individual reduced datacube, all of which are then combined into the final datacube. Remaining sky subtraction residuals are removed with ZAP (version 2.0, \citealt{Soto2016}). 
The datacube propagated variance is rescaled to take the impact of noise covariance due to the interpolation process into account (see \citealt{Bacon2017}). In the region with more than 100 hours depth, the average 5$\sigma$ surface brightness detection limit at 7000 \AA\ is \erglinesurf{1.3}{-19}\ for an unresolved emission line summed over 3.75 \AA\ (3 spectral pixels) and 1 arcsec$^2$ (5$\times$5 spaxels) and the corresponding 5$\sigma$ point source limiting flux is  $2.3\times 10^{-19}$ \ergsline\ for the same emission line. 

Because of the absence of bright stars in the field, we use the \textsf{muse-psfrec} software \citep{Fusco2020} to estimate the spatial point-spread function (PSF) from the real-time Adaptive Optics telemetry information recorded by GALACSI. The PSFs of each observation are then combined to produce a measure of the final image quality. The PSF is modeled as a Moffat function \citep{Moffat1969} whose parameters (FWHM and $\beta$) change smoothly with wavelength. 
Thanks to the AO performance, an excellent image quality (Moffat FWHM) of 0.6\arcsec\ and $\rm \beta = 2.1$ at 4700 \AA\ to 0.4\arcsec\ and $\rm \beta = 1.8$ at 9300 \AA\ is achieved for this very deep exposure.   

\subsection{Source Detection and Classification}
\label{subsec:detection}
We perform two types of source detection and extraction: a blind source detection with \origin\ and 
source extraction and deblending using HST prior information with \odhin.

The \origin\ software \citep{Mary2020} has been developed to automatically detect faint line emitters in MUSE datacubes. It is optimized for the detection of compact sources with faint spatial-spectral emission signatures and provides an automated and reliable estimate of the purity (i.e. related to the proportion of false discoveries).
The software was run with a purity threshold value of 0.8 and resulted in the detection of 2137 emission lines grouped into 1002 sources.

The \odhin\ software \citep{Bacher2018} uses the higher spatial resolution provided by the HST images to perform deblending of sources in the MUSE data cube. The approach is similar
to \textsf{TDOSE} \citep{Schmidt2019} with the difference that it is non-parametric.

For the inspection we limit the \origin\ sources to the inner 16+ hours region (corresponding to an 80\arcsec\ diameter) to avoid an increase in the false detection rate at the edge of the field when the SNR decreases rapidly. This reduces the number of sources to inspect to 845. 
Similarly, only 389 HST sources from a total of 1387 within the \mxdf\ field have been selected for inspection, using a SNR continuum cut of 0.8 per spectral pixel.

Evaluation of the redshift solutions provided by the \textsf{Marz} cross-correlation software \citep{Hinton2016} is performed by 3 independent experts\footnote{Scientists already experienced in redshift measurement with MUSE or specifically trained for this activity.}. After reconciliation of the disagreement between experts, a final catalog of 733 sources with redshift, matching sources and confidence is produced.  

95\% of the \laes\ (hereafter LAEs) result from \origin\ detections. In this case, an optimal extraction is performed on the raw datacube using the \origin\ correlation pseudo narrow-band image as a weighting map. While this will produce a more precise flux estimation than continuum based extraction for the case of extended \lya\ emission, it will nevertheless miss part of the \lya\ extended halo flux\footnote{On average, 65\% of the \lya\ flux is in the extended halo \citep{Leclercq2017}.}.  

Flux and equivalent width measurement of the \lya\ line is performed with {\textsf{pyplatefit}}, a Python enhanced version of the {\textsf{PLATEFIT}} IDL software developed for the analysis of the SDSS survey \citep{Tremonti2004, Brinchmann.2004}. Among the improvements implemented in {\textsf{pyplatefit}} which are of interest here, one can mention the asymmetric Gaussian (simple and double) line fit and the bootstrap method of evaluating robust errors. 

The \lya\ redshift is based on the peak of the \lya\ line. Note that, because of the resonant scattering properties of the \lya\ photons in the interstellar medium, the \lya\ redshift is systematically different from the systemic redshift (e.g. \citealt{Shapley2003, McLinden2011, Rakic2011, Song2014}). Typical velocity offsets are $\approx$200 \kms\ for LAEs, with larger values ($\approx$500 \kms) for Lyman-break galaxies \citep{Shibuya2014, Muzahid2020}. As reported by \cite{Shibuya2014} and \cite{Muzahid2020}, there is a strong anti-correlation between the \lya\ velocity offset and the \lya\ equivalent width. Given the high equivalent widths of our LAE sample (see Fig.~\ref{fig:properties}) one can expect an average velocity offset $\lessapprox 200$ \kms.

For the vast majority of our LAE sample, there are no alternatives to \lya\ redshifts: non-resonant emission lines such as $\mathrm{CIII]_{1907,1909}}$ are outside the spectral window or too faint to derive a systemic redshift from and the continuum is too faint to measure any reliable absorption line/feature. 
A possibility is to use the empirical correlation found by \cite{Verhamme2018} between the FWHM of the \lya\ line and the velocity offset. 
For a double peaked \lya\ lines, the redshifts reported by the double asymmetric fits are measured from the averages of the two peak central wavelengths. As shown by \cite{Verhamme2018}, this value is a better approximation of the systemic redshift than the \lya\ peak location.

Note that very precise absolute redshifts are not essential for this paper given the width of the wavelength window used in the search (Section \ref{sect:overdensities}). No attempt is made to further correct the \lya\ redshift to the systemic value. What matters, however, is the scatter of the \lya\ redshifts around its median value at a given systemic redshift. Using the sample of 55 galaxies of \cite{Verhamme2018}, we derive a scatter of $\pm 95$ \kms\ (95\% probability percentile), which is smaller than the windows that we will use.

\subsection{Source catalog}
\label{subsec:catalog}

We combine the revised \udft\ (30 hours depth, 1x1 arcmin$^2$) and \mosaic\ (10 hours depth, 3x3 arcmin$^2$) catalogs (\cite{Inami2017}; Bacon et al in prep) with the new sources discovered in the \mxdf\ (Fig.~\ref{fig:fields}). Selecting all sources with $z > 2.9$ gives a total of 1258 LAEs in the 9 arcmin$^2$ field of view of the HUDF observed with MUSE. 
Note that a low fraction of sources (55 i.e. 4\%) are not strictly speaking \laes\ but are detected from their strong \lya\ absorption, often associated with  weak \lya\ emission. These galaxies are brighter in the continuum than the overall population of \laes\ and are more typical of Lyman-break galaxies.

While a few \laes\ have additional detected UV lines such as $\mathrm{CIII]_{1907,1909}}$ or $\mathrm{CIV_{1548,1550}}$, there is no source in the catalog without \lya\ detection at z $>$ 2.9.

In the deepest \mxdf\ region (depth $>$ 100 hours), the LAE density reaches 375 galaxies per arcmin$^{2}$. It is the densest collection of LAEs ever obtained in a single field.
To illustrate this point, one can perform a quantitative comparison with the large \lae\ narrow band survey performed around the SSA22 field by \cite{Yamada2012}. The authors identify 2161 candidate {\lae}s in the redshift range [3.062$-$3.125], leading to a LAE average surface density of $\mathrm{0.20 \, arcmin^{-2}}$. Using the same redshift range we obtain, thanks to the high sensitivity to low luminosity LAEs, 
 a LAE surface density of  $\mathrm{6.2}$ and $\mathrm{17.9 \, arcmin^{-2}}$ in, respectively, the \mosaic\ and 100+ hours depth \mxdf\ field of view.

\begin{figure}[htbp]
\begin{center}
\includegraphics[width=\columnwidth]{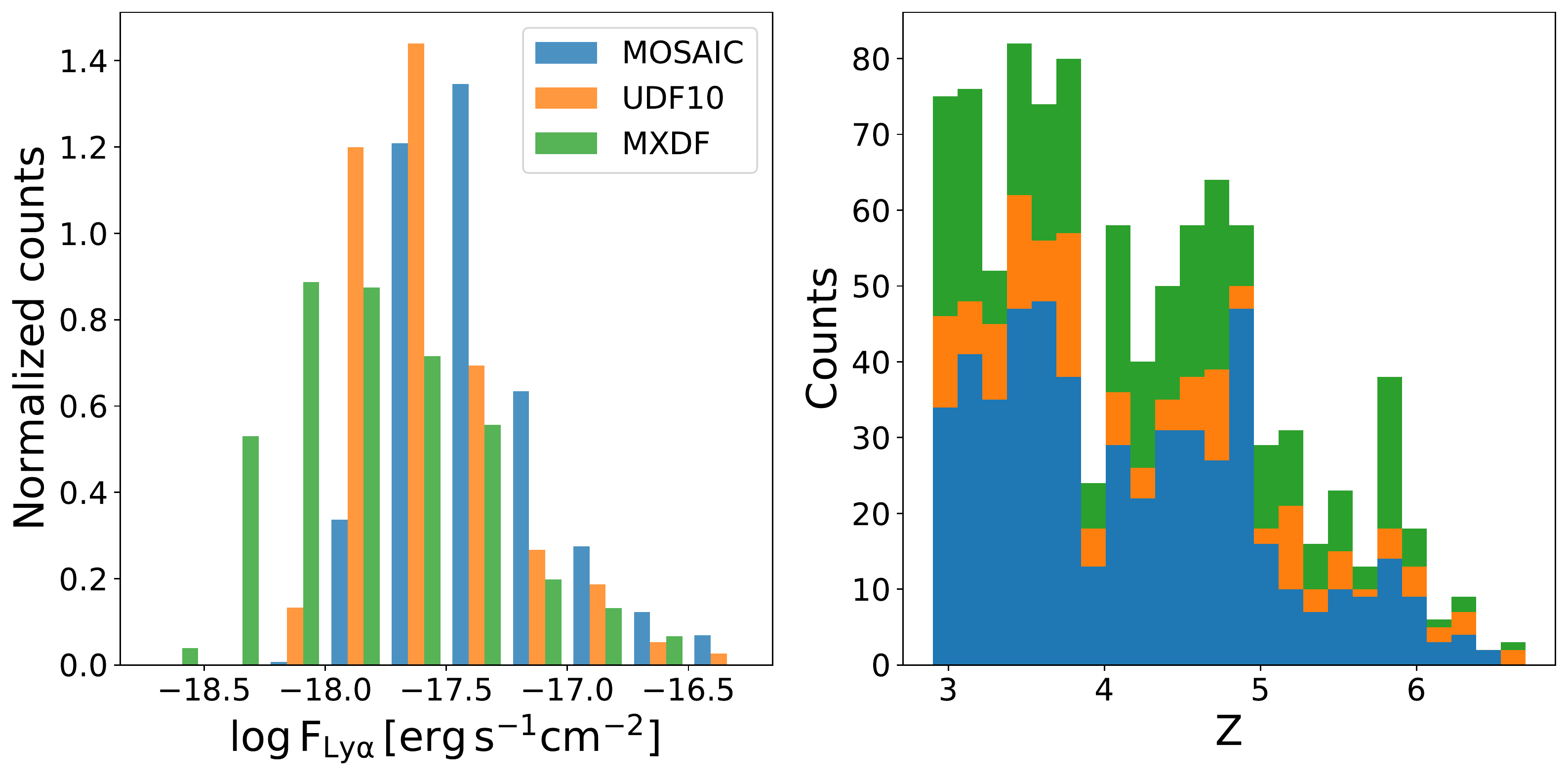}
\caption{\lya\ flux (left panel, bin width 0.25 \ergsline) and redshift (right panel, bin width 0.2) distribution of the 1258 \lya\ emitters.}
\label{fig:redshift}
\end{center}
\end{figure}

The catalog covers a wide redshift range from $z$=2.90 to 6.65 (Fig.~\ref{fig:redshift}), with a significant part at high redshift (346 LAEs at $z$>4.8). Redshift confidence has been attributed by experts, from 1 (low confidence) to 3 (very high confidence). 
There are 402, 513 and 343 \laes\ with confidence 1, 2 and 3, respectively.
The lower confidence LAEs are generally reliable detections (i.e., they are obtained with \origin\ set at a high purity level) but with lower SNR \lya\ line profiles where the \lya\ redshift solution is the most probable. In this paper we use all LAEs irrespective of their redshift confidence.

An important fraction (40\%, 504) of LAEs has no entry in the \cite{Rafelski2015} HUDF photometric catalog. For a small part (10\%, 50) it has not been possible to select a unique HST counterpart among the few possibilities.
In some other cases (6\%, 30) an HST counterpart can be seen in some of the HST bands that was missed by the automatic identification performed by \cite{Rafelski2015}. But for the vast majority of the 504 sources without a \cite{Rafelski2015} counterpart, namely 84\% (423 objects) no detectable HST counterpart is observed. A previous study of 103 similar sources in this field, anterior to MXDF observations, shows that these sources are the high equivalent width tail of the \lae\ population \citep{Maseda2018} and that they have on average a high ionizing photon production efficiency \citep{Maseda2020}. 

\section{Redshift overdensities of \lya\ emitters}
\label{sect:overdensities}

A visual inspection of the datacube shows that LAEs are not distributed uniformly but exhibit strong clustering in  redshift space.
Here we focus on the detection and properties of LAE redshift overdensities. The method used is described in the following section.


\subsection{Choice of coordinate system}

Given our small field of view, finding overdensities is nothing more than performing a binning of the redshift distribution followed by a peak detection. There are, however, some subtleties to take into account. One of them is the choice of coordinate system which is strategic given the very wide redshift range (2.9-6.7) probed by our observations.

A first possibility selected by \cite{Cohen1999} and \cite{Gilli2003} is to use bins in velocity space ($c \ln\left[1+z\right]$). A fixed bin in rest-frame velocity space has the advantage of sampling the \lya\ line, independent of the Hubble flow. The sampling must be large enough given that the \lya\ profile is generally broad and shifted in velocity\footnote{Using the expected $\pm 95$ \kms\ velocity shift scatter around the median value computed in section \ref{subsec:detection} and the  370 \kms\ rest-frame median FWHM measured line width, we derive a value of 420 \kms\ for the minimum spectral window size to capture the bulk of the \lya\ flux.}.

An alternative is to use a fixed physical scale which allows us to compare the sizes of physical objects independently of the redshift. In this coordinate system, a sampling of 2 Mpc corresponds to 600 \kms\ at $z$=3 and 1400 \kms\ at $z$=6. Given the small evolution in redshift of the \lya\ line profile \citep{Hayes2020}, such a large spread of the sampling with redshift will be sub-optimal for the diffuse emission signal extraction.

The last possibility is to use comoving coordinates. In this coordinate system, the cosmic web filaments keep their structural properties along the expansion of the Universe. The sampling evolution with redshift is also less important than for the physical scale: an 8 comoving Mpc sampling translates into 610 \kms\  at $z$=3 and 800 \kms\ at $z$=6. 
We have therefore selected this last option which is best-suited to the object of the study and is not far from optimal for signal detection.

\subsection{Detection of overdensities}
\label{sect:over_method}

Although the search for diffuse emission will be done in the \mxdf\ deep area, the search for overdensities is performed on the full LAE catalog , i.e. in the  entire \mosaic\ field of view. With an area ten times larger than in the deeper region of the \mxdf,  the $3 \times 3$ arcmin$^2$  \mosaic\ field of view improves the chance to detect  large-scale overdensities.
The search is done in two steps: we start by identifying peaks in the redshift distribution using a wide sampling. Group redshifts are then refined with a finer sampling and their over-density is estimated by comparing the number of LAEs in the group with the expected mean value.

\begin{figure*}[htbp]
\begin{center}
\includegraphics[width=0.9\paperwidth]{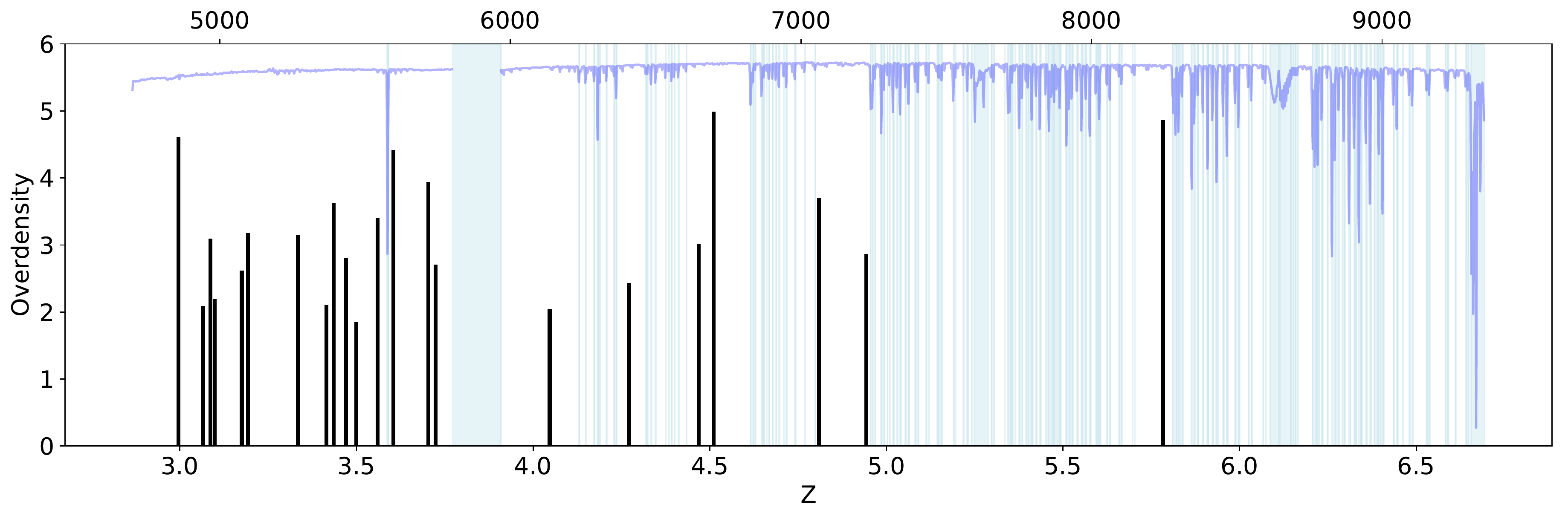}
\caption{LAE overdensities ($\delta$) found in the HUDF as function of redshift. The top axis displays the corresponding wavelengths in \AA. The noise spectrum is shown in blue on the reverse $y$-axis. Regions excluded from the overdensity search are displayed in pale blue. The 5800-5966 \AA\ masked region is due to the sodium notch filter used to filter out the light from the bright AO Laser guide star.}
\label{fig:overdensity}
\end{center}
\end{figure*}

We compute the histogram of the LAE population in comoving space, imposing a constant step of 8 cMpc over the whole redshift range. 
Given the expected faintness of the signal, we aggressively mask all redshift bins with sky lines. This removes 46\% of the bins, mostly at $z$>5 (Fig.~\ref{fig:overdensity}).

The background value ($B$) is estimated as the mean value of the number of LAEs in a 8 cMpc
bin after 3 sigma clipping. 
It should be noted that the possible redshift dependance of the background is ignored in this preliminary first step\footnote{This assumption will introduce some bias in the overdensity selection. This could be a problem for an exhaustive study of overdensities but our aim is restricted to find the most overdense regions. As shown later in this section, the final overdensity estimate takes into account the redshift evolution of the mean LAE density.}.
For each histogram peak, its signal to noise ratio (hereafter SNR) is estimated as the peak signal ($S$), i.e. the excess number of LAEs over the background and divided by the noise value $\mathrm{\sqrt{S+\sigma_B^2}}$, where $\sigma_B$ is the standard deviation of the background value. After some tests we select a SNR cut of 2.5 and a minimum number of LAEs of 7. This results in 24 peaks.

Naturally, the number and redshifts of overdensities is a function of the input parameters: i.e. the SNR cut, the minimum number of members and the window size. We check the sensitivity of the method by playing with these parameters. Increasing the SNR cut from 2 to 3 decreases the number of overdensities from 37 to 11. 
The number of overdensities is stable with the imposed minimum number of LAEs up to 10, and decreases rapidly from 24 to 15 for larger values of minimum number of members.
The selected redshift window size of 8 cMpc is optimal: the number of detections decreases from 24 to 21 and from 24 to 15 when the window size is changed to 6 and 10 cMpc, respectively.

We do not try to further optimize the detection parameters given that it is not
our goal to have an exhaustive list of overdensities, but to find the most overdense regions within a redshift window size suitable for the search for diffuse \lya\ emission (i.e. large enough with respect to the \lya\ line width but not too large to dilute the signal). We therefore proceed to use the optimal 8 cMpc window and the SNR cut of 2.5, which is a good compromise in the number of overdensities to explore.

In a second step we refine the group measurement for the selected peaks using a histogram with a finer grid of 50 kpc sampling. The 8 cMpc size of the window is kept fixed, but its center is slightly adjusted around the initial peak value to maximize the number of LAE members.
Groups overlapping in redshift are merged into a single fixed 8 cMpc size window. This reduces the number of overdensities by two.

The final group catalog, given in Table~\ref{tab:over}, is composed of 22 overdensities. Groups have on average 17 members with a minimum of 10 and a maximum of 26.

To estimate the overdensity, i.e. the ratio of the LAE number density in groups with respect to the mean of the overall population, we first compute the evolution of the mean LAE density with redshift. We partition the 2837 sky free redshift bins into 5 redshift bins with a similar numbers of LAEs and compute their mean LAE densities. The mean density decreases approximately linearly with redshift from 0.0201 \invmpc\ in the $z$=[2.86$-$3.45] bin to 0.0063 \invmpc\ in the $z$=[5.75$-$6.65] last bin.

For each group we can then estimate its overdensity ($\mathrm{\delta}$) by comparing the number of group members ($\mathrm{N_{lae}}$) to the number of expected LAEs using their mean density ($\mathrm{\overline{\rho}_{lae}}$) and the group volume ($\mathrm{\Delta V = 8 \, cMpc \times S}$), with S the area of the \mosaic\ field\footnote{$\rm S = 5.7^2 - 7.3^2 \, cMpc^2$ at z=3 and 6, respectively}: $ \mathrm{\delta = {N_{lae}} / \left( \overline{\rho}_{lae} \, \Delta V \right) }$.

\begin{figure*}[htbp]
\begin{center}
\includegraphics[width=0.9\paperwidth]{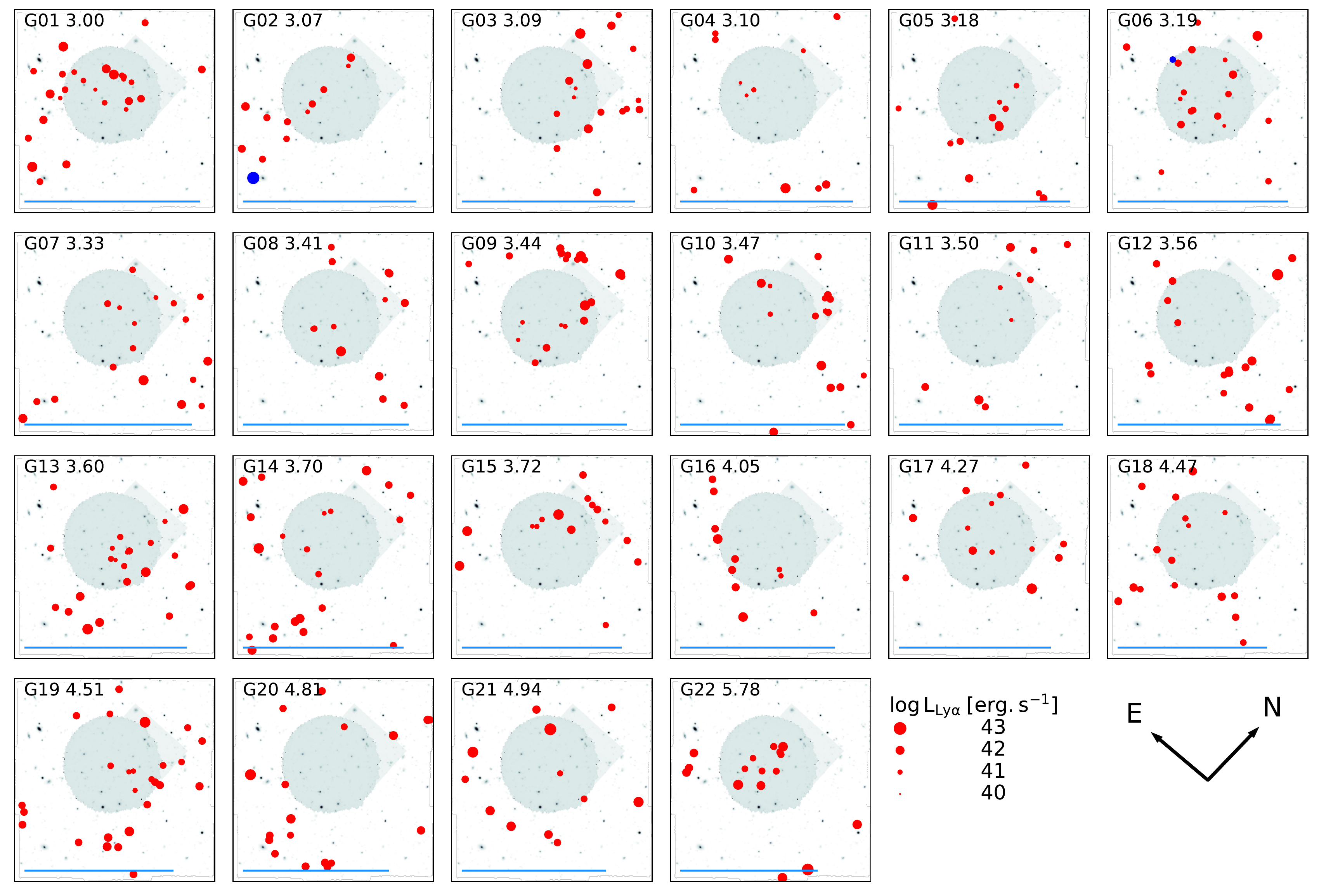}
\caption{Spatial distribution of LAEs in overdensities. LAEs are shown as red circles sized according to the log of their \lya\ luminosity. The blue circles in groups 2 and 6 identify the AGN (also sized according the log of their \lya\ luminosity). The location of each group member is overlayed on the HST F775W grey scale image. The field locations are shown as a grey circle (\mxdf) superimposed on a square (\udft). Group ID and redshift are labeled. Fields have the \mosaic\ size and orientation, i.e. 3 arcmin and 42\degree\ PA.
The blue line at the bottom of each image indicates the 5 cMpc scale.
}
\label{fig:clustering}
\end{center}
\end{figure*}

The group overdensities are shown in Fig.~\ref{fig:overdensity} and the corresponding spatial locations of LAEs in the groups in Fig.~\ref{fig:clustering}. 
Note that the group numbering follows the redshift.
There are 370 LAEs in groups, which is 29\% of the total LAE sample. The mean overdensity is 3.2 and the densest group is at $z$=4.5 and has $\mathrm{\delta = 5.0}$. 

\subsection{Properties of overdensities}
\label{subsec:prop}

Fig.~\ref{fig:overdensity} shows a clear trend with redshift: 2/3 of the groups are found at $z$<3.8, in 1/3 of the free-sky comoving total accessible volume. 

 
We examine the correlation of overdensities with the AGN population. The 7Ms Chandra Deep Field South (CDFS) catalog \citep{Luo2017} has 65 identified AGN at $z>$2.9 in the full area (484 arcmin$^2$). We found 10 overdensities with AGN at similar redshift ($\mathrm{\Delta z < 0.01}$) within the CDFS area. As might be expected, a significant fraction (45\%) of overdensities harbor an AGN. However, the UDF area is a tiny fraction (2\%) of the total CDFS area and there are only two overdensities with an AGN located within the UDF: group 2 at $z$=3.07 and group 6 at $z$=3.19.

\begin{figure*}[htbp]
\begin{center}
\includegraphics[width=0.85\paperwidth]{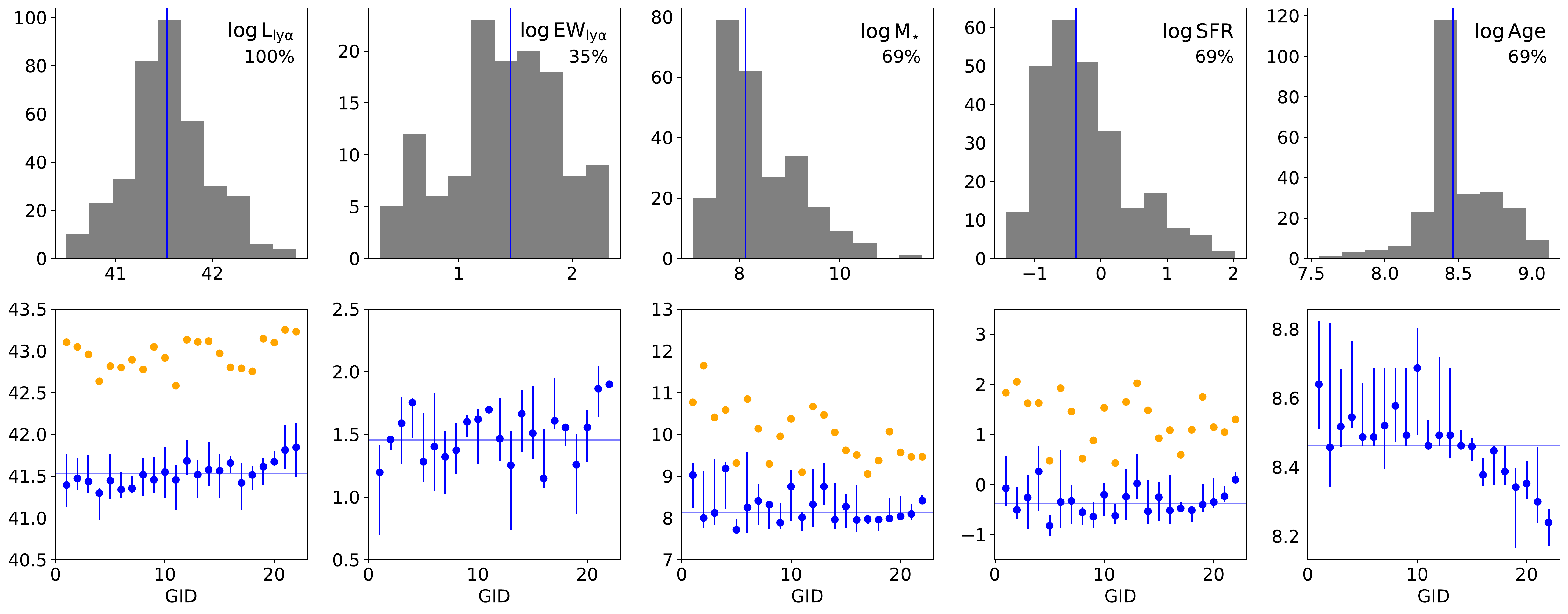}
\caption{Properties of LAEs in overdensities. From left to right the columns show: the $\rm \log$ \lya\ luminosity in \ergslum, the rest-frame $\rm \log$ \lya\ equivalent width in \AA, the  $\rm \log$ stellar mass in $\mathrm{M_\odot}$, the $\rm \log$ star formation rate in $\mathrm{M_\odot yr^{-1}}$ and the $\rm \log$ age in years. Except for the \lya\ luminosity, which used all LAEs, only subsets are used for the \lya\ equivalent width (35\%) and for Mass, SFR and Age (69\%). See Sect.~\ref{subsec:prop} for the subsample definition.
In the top row we show the histograms of these properties.
The bottom row displays the corresponding global properties for each group (GID): the blue circle symbols show the group medians and the 25\% and 75\% percentiles, the orange symbols are the summed property for all galaxies in the group. The median value for each property for the full sample is shown as a blue line in each panel. Note that the GID increases with redshift.}
\label{fig:properties}
\end{center}
\end{figure*}

As shown in the first panel of Fig.~\ref{fig:properties}, LAEs in overdensities are low luminosity \laes: $\mathrm{\log ( L_{Ly\alpha} \, \ergslum ) = 41.5 \pm 0.4}$.
We also display in panel 2 the \lya\ rest-frame equivalent width (EW) for the subset of \laes\ that are bright enough in their continuum to derive a meaningful equivalent width (i.e. with an EW SNR $>$ 3). Note that for the majority of \laes\ (65\%), their continuum is too faint, and thus we can only derive lower limits for the equivalent width. As already pointed out, many of these galaxies are so faint in the continuum that they are not even detected in the deepest HST images \citep{Maseda2018}.

An important fraction (69\%) of group members have an entry in the HST \cite{Rafelski2015} catalog. We use the exquisite ancillary information provided by the HST photometry to derive additional physical parameters for the groups.

The stellar mass, SFR and specific SFR (sSFR) are inferred via SED fitting using the high-$z$ extension of the code MAGPHYS \citep{daCunha2008, daCunha2015} and enabling for a minimum stellar mass of $\mathrm{10^6 \, M_{\sun}}$ (see Sec. 3.2 of \citealt{Maseda2017}). We use HST photometry from the UVUDF catalogue of \cite{Rafelski2015} which comprises WFC3/UVIS F225W, F275W and F336W; ACS/WFC F435W, F606W, F775W, and F850LP and WFC/IR F105W, F125W, F140W and F160W\footnote{Note, however, that many of these galaxies are very small and faint, and therefore do not have any reliable longer wavelength photometry. In those cases, the parameters derived from MAGPHYS are more uncertain.}

On average, the LAEs in overdensities are low mass ($\mathrm{1.4 \times 10^{8} \, M_\odot}$), young (0.3 Gyr) galaxies with high specific star formation rates ($\rm SFR > \mathrm{0.4 \,M_\odot yr^{-1}}$; $\rm sSFR > \mathrm{10^{-8.5} \,yr^{-1}}$), higher than the typical values at these redshifts \citep{Schreiber2015, Salmon2015}.

Regarding the remaining population (31\%) of LAEs not detected in the HST images, we cannot derive quantities like SFR or stellar mass for those galaxies without robust continuum detections from HST.
However, as demonstrated by \cite{Maseda2018,Maseda2020}, this selection yields faint ($\mathrm{M_{UV} \approx -15}$), star-forming ($\mathrm{\beta \approx -2.5}$) galaxies on average.  While we cannot directly determine the stellar masses from the UV continuum alone, an extrapolation to the \cite{Duncan2014} $\mathrm{M_{UV}-M_{\odot}}$ relation to $\mathrm{M_{UV}}$ = $-$15 implies that these galaxies should have stellar masses below $\mathrm{10^7 M_{\odot}}$ at these redshifts. 
Using this average mass, we estimate the contribution of the HST-undetected LAEs to the total stellar mass of the overdensities to be on average $\approx$1\%, with a maximum of 7\% for group 20. 
Fig.~\ref{fig:properties} displays the  main properties derived from the SED fitting, in addition to the \lya\ luminosity and equivalent width.

\begin{figure}[htbp]
\begin{center}
\includegraphics[width=0.95\columnwidth]{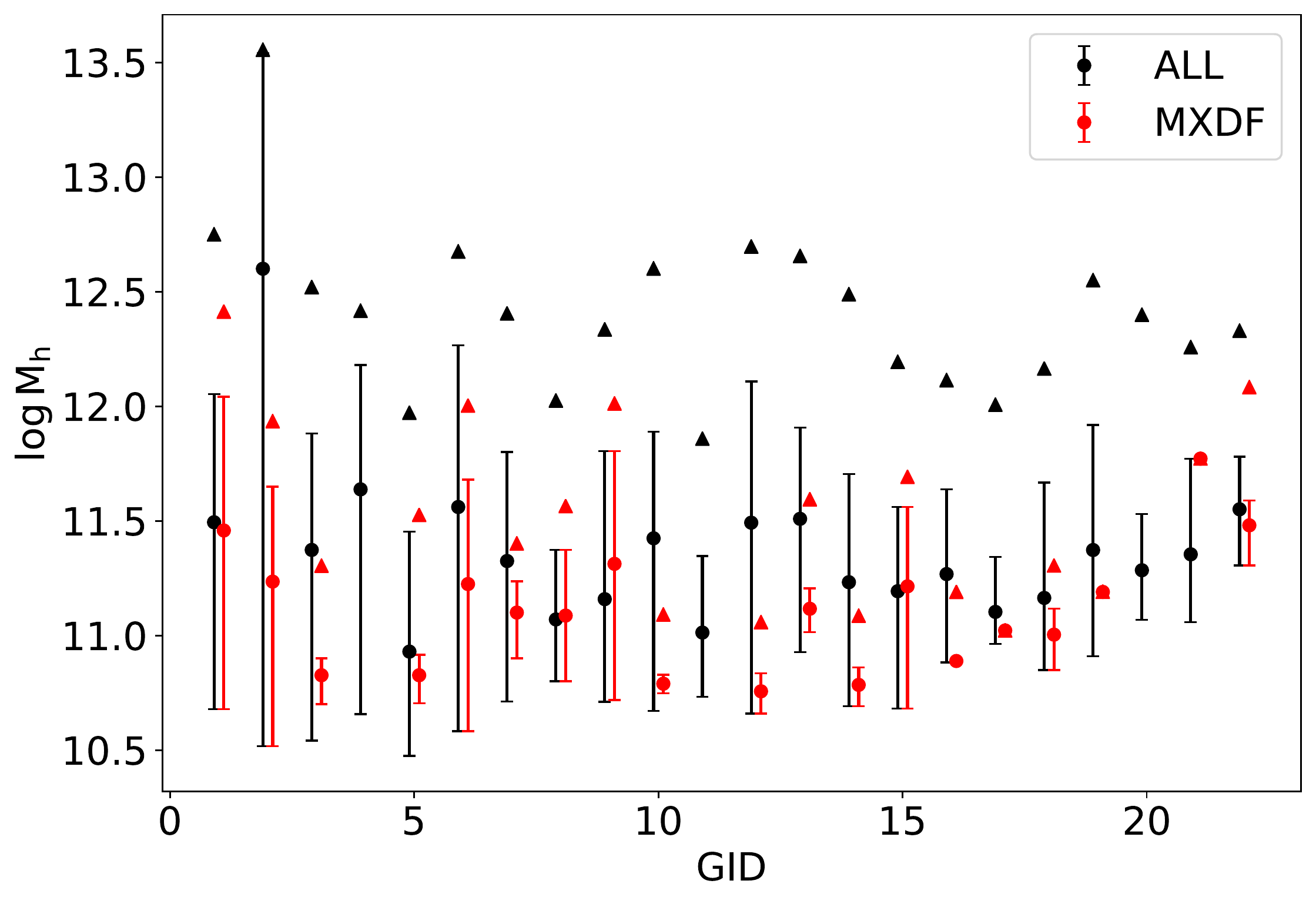}
\caption{Estimated dark matter halo masses in the overdensities (see section~\ref{subsec:prop}). The total (upper triangle), average (filled circle) and min-max (bars) values of dark matter halo masses in overdensities are displayed for all group members with an HST counterpart either within the \mosaic\ field of view (in black), or restricted to the \mxdf\ area (in red). Units are $\log \mathrm{M_\sun}$.
}
\label{fig:halodm}
\end{center}
\end{figure}

We estimate the individual dark matter halo masses associated with each group member with an HST counterpart using the galaxy stellar masses derived above and the stellar-to-halo-mass relation (SHMR) derived by \cite{Girelli2020}. The overdensity mass is then the sum of the halo masses of each galaxy.
We note that the observational data used by \cite{Girelli2020} to constrain their SHMR model only extend out to $\mathrm{z \approx 4}$ and consist of galaxies generally more massive than our sample ($\mathrm{M_{\rm star} > 10^{10}M_\odot}$ at $\mathrm{z=4}$). However, \cite{Girelli2020} show that their SHMR at $z=3$ is between the estimates of \citet{Behroozi2019} and \citet{Moster2018} who use deeper data. We thus chose the fits from \cite{Girelli2020} as a middle guess, and note that the errors on the SHMR are probably of order a factor 2, which is small with respect to uncertainties on our estimates of the stellar masses. Our results are shown in Fig. \ref{fig:halodm} where we see an average halo mass $\mathrm{\sim 10^{11.3} M_\odot}$, and almost no haloes more massive than $\mathrm{\sim 10^{12} M_\odot}$. This is typical of the LAE population we survey (e.g. \citealt{Garel2015a}), and our groups are likely progenitors of galaxies like our own \citep{Garel2015b}. We note the exception of group 2, which contains one very massive halo of $\mathrm{10^{13.5}M_\odot}$, {\it outside} the MXDF region. These massive environments, typical of proto-clusters, are known to host AGNs.

In the next section we use the discovered overdensities to search for extended \lya\ emission.


\section{Extended \lya\ emission}
\label{sect:diffuse}
 
Numerical simulations predict the \lya\ emission to occur along filaments that are 50-100 kpc wide and a few Mpc long (see e.g. Figure 7 of \citealt{Rosdahl2012}). The surface brightness of the filaments is predicted to be in the range \erglinesurf{1-10}{-20}, depending on the halo mass (e.g. \citealt{Gould1996}). Such a low surface brightness will not be detectable in the 10 hour deep \mosaic\ field, so we restrict the search to the \mxdf\ region.

In each overdense region, the \mxdf\ volume is a cylinder of  2.5 cMpc diameter in the transverse direction and 8 cMpc along the line-of-sight. Our goal is to detect faint \lya\ emission in between the galaxies.

To perform this detection we have developed a two-step method. In the first step, we start by performing a segmentation of the narrow-band signal-to-noise (SNR) image to isolate the regions of extended emission. In the second step we compute the \lya\ flux and its error in these segmented areas using the continuum-subtracted narrow band image. 
Extended \lya\ emission is marked as detected if the computed SNR and the area of the extended region are large enough. Here that means the estimated  probability that noise can explain such a SNR value by chance is sufficiently low.

Note that the two steps are somewhat independent. Alternative methods with different signal transformation to isolate the extended emission could be used to identify diffuse emission segments. Whatever the method, the flux and error computation are always performed in the original continuum subtracted narrow band image. The details of the methodology is described in the following sections.

\subsection{Multiscale analysis of narrow band images}
\label{sect:wavelet}

Detection and estimation of diffuse emission in the presence of sources which are locally two orders of magnitude brighter is a difficult signal processing problem. The additional complexity is that \lya\ emitters are themselves extended and thus cannot be considered as point sources. 

The geometry of the problem calls for a multiscale approach. An interesting tool in this respect is the wavelet transform and, given the isotropy of astronomical sources, in particular the Isotropic Undecimated Wavelet Transform (IUWT,  \citealt{Bijaoui1994, Starck1998}).
Compared to classical Gaussian filtering, this wavelet transform allows for a separation of the sources according to its spatial scale. In the classical version of the IUWT filterbank, the original image can be directly resynthesized by the simple sum of the multiscale approximation coefficients. This feature is very interesting for detection or denoising approaches operating scale by scale \citep{Starck1998}. Such wavelet signal decompositions have been successfully used in astronomy for various applications: e.g. X-ray source detection \citep{Finoguenov2020} , diffuse light study in compact group of galaxies \citep{DaRocha2005} and the estimation of faint and diffuse radio components  \citep{Dabbech2015, Ammanouil2019}.

For each overdensity, we start by building a narrow band SNR image by summing the continuum subtracted SNR cube over a window centered on the mean group velocity with a width corresponding to 8 comoving Mpc (see section \ref{sect:overdensities}). This corresponds to 9 and 18 wavelength channels at $z$=3 and $z$=6, respectively.

As the continuum subtracted datacube we use the PCA subtracted SNR cube provided by \origin. The \origin\ process \citep{Mary2020} is efficient for removing the continuum of bright galaxies without leaving many artifacts which can be problematic for low surface brightness detection.

\begin{figure*}[htbp]
\begin{center}
\includegraphics[width=0.7\paperwidth]{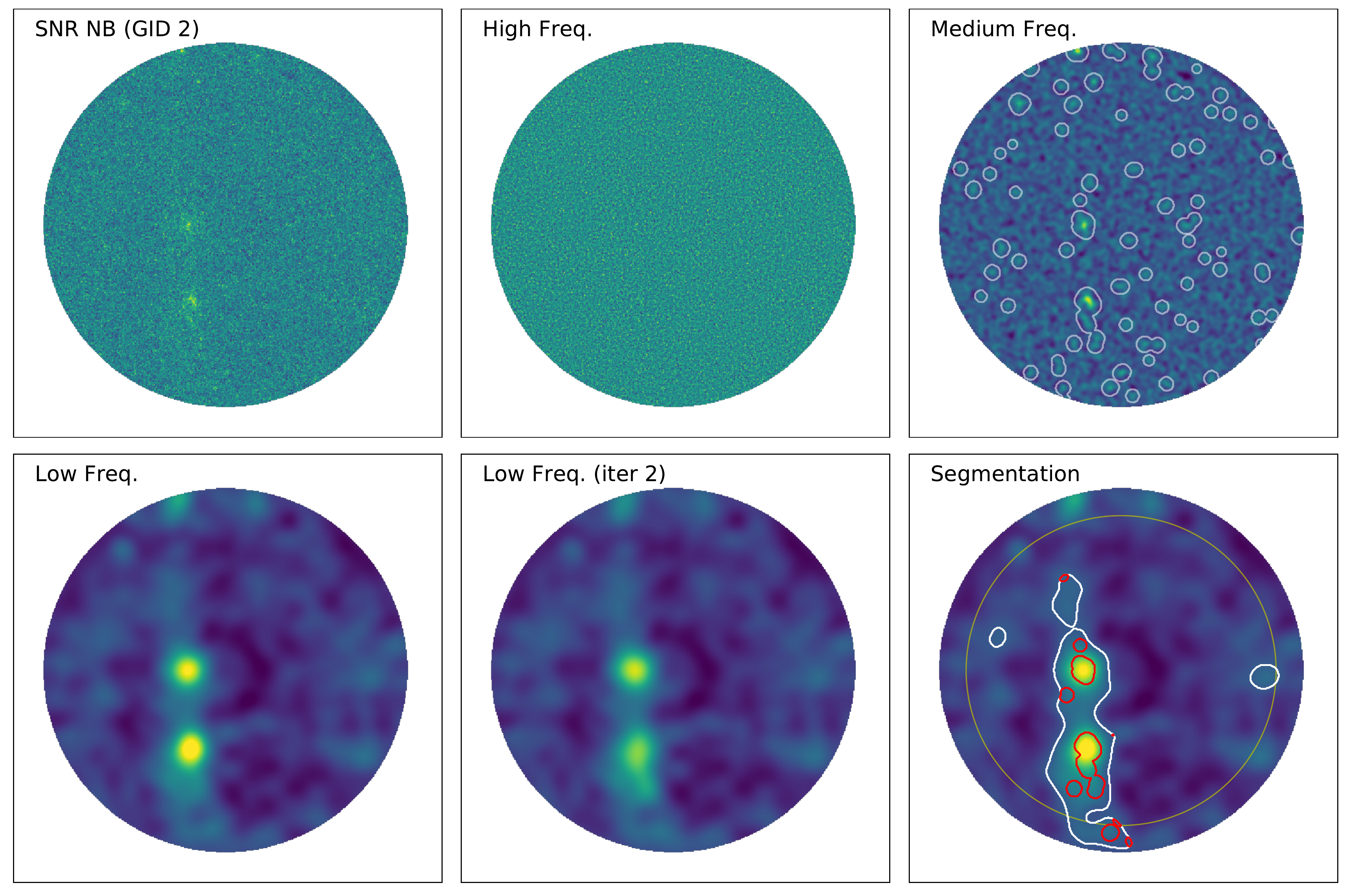}
\caption{Example of IUWT wavelet multiscale decomposition of the group 2 ($z$=3.07) SNR narrow-band image (upper left panel). The signal is split into three channels: high, medium and low spatial frequency. The central-lower panel (Low Freq. iter 2) shows the low frequency channel after masking the medium frequency peaks in the SNR image (see Sect.~\ref{subsec:shape}). 
The final segmentation is shown in the lower right panel. In white the "filaments", in red the area of compact sources included in the filaments. The diffuse area is defined as the area included in the filament (white contours) after exclusion of the compact sources (red contours).
The yellow ring shows the low SNR region with less than 60 hours exposure time.
The figure has the \mxdf\ orientation with North up.
}
\label{fig:wavelet}
\end{center}
\end{figure*}

We then decompose this narrow band image with the classical perfect reconstruction filterbank of the IUWT \citep{Bijaoui1994, Starck1998,Starck2007}. In the considered setting, the IUWT decomposes the image into eight component images\footnote{In IUWT the spatial extension of the filter grows as a power of 2 from one scale to another. Considering scales larger than 8 would lead to filters larger than the MXDF field.}, from high to low frequencies. The high frequency noise mostly lives in the first two scales,  bright and compact sources are well captured by the next two scales and the diffuse, large scale components of the image mostly live in the four remaining scales. 
The angular size of the spatial structures captured in each band is determined by the impulse responses of the IUWT analysis filters, whose FWHM are respectively in the range [0.2-0.6] arcsec (high frequency band), [0.6-2.2] arcsec (medium frequency band) and [2.2-37] arcsec (low frequency band).   
As shown in Fig.~\ref{fig:wavelet} for group 2 at $z$=3.07, this process filters out the noise from the signal (high freq. panel) and cleanly separates the LAE signal (medium freq. panel) from the large scale diffuse structure (low freq. panel).

\subsection{Shape and flux of $Ly\alpha$ diffuse emission}
\label{subsec:shape}

We now use the IUWT SNR images obtained by the process described above to separate the signal of the extended \lya\ emission from the \lae\ signal.

We start to identify compact sources by segmenting the medium frequency images using a threshold of 3.5$\sigma$ (right upper panel of Fig.~\ref{fig:wavelet}). We use this segmentation map to mask all outliers and compact sources in the original SNR image, replacing the mask values by an average SNR value obtained from a first estimate of the diffuse components. This first crude estimate is the average value of coefficients above a threshold of 1.5$\sigma$ in the low frequency SNR image. We then again apply the IUWT decomposition to this SNR masked and filled image.
This iteration removes most of the low frequency signal due to the compact sources (see lower left and central panels of Fig.~\ref{fig:wavelet}).

We then perform a segmentation of the resulting low frequency image, using a threshold of 1.5$\sigma$ (lower central panel of Fig.~\ref{fig:wavelet}). We discard all segments that have a size smaller than 200 pixels (i.e. 8 arcsec$^2$). This last step is motivated by the desire to keep only contiguous surfaces which are extended with respect to the PSF size ($\mathrm{\approx 5 \times FWHM}$).

At the edge of the \mxdf\ field, the exposure time drops rapidly from 140 to 10 hours. Besides having much lower SNR, this outer zone is also subject to larger systematics since it results from a smaller number of individual exposures. In this region, the probability of false detection is thus much larger.  Therefore, we have discarded all segments which have a mean depth (averaged over the full segment area) less than 60 hours (see the corresponding location in the lower right panel of Fig.~\ref{fig:wavelet}).

To avoid potential pollution by low-$z$ emission line interlopers, we identify all galaxies in the full source catalog with emission lines other than \lya\ having SNR$>$5 and inside the group window. Any interloper which falls inside the filaments is masked out for the filament's flux estimation.

The final segmentation (lower right panel of Fig.~\ref{fig:wavelet}) is composed of a series of extended structures, called "filaments" in the rest of the text. 
We also identify the area covered by the compact sources detected in the first step within each filament. The remaining area within the filaments is called the diffuse extended emission in contrast to the compact source area. We will use this terminology in the rest of the document, referring to filament, compact and diffuse area.

These segmentations obtained in the wavelet space are now used to compute, in the data space, the total flux of each area. 
The computation is performed on the narrow band original images, obtained from the \origin\ PCA continuum-subtracted datacube.

\subsection{Noise and SNR estimation}
\label{subsec:noise}

Although the datacube noise properties have been carefully validated, including the impact of noise covariance, one cannot rely on formal noise propagation when working at such faint surface brightness especially after continuum subtraction processes, which may modify the noise distribution. Any remaining systematics like low background fluctuations can have a significant effect when summing the signal over a large area.

In order to estimate the standard deviation of our flux measurement, we have performed the following computation:
we mask all spaxels contained within the filament segments or within the compact source segments identified in the first step (see upper and lower right panels of Fig.~\ref{fig:wavelet}). Depending on the group this masks 10 to 30\% of the total area.
We start by computing the average offset and its standard deviation of the main unmasked area (1\arcmin\ diameter or 120 hours depth). This mean offset is subtracted from the flux image. The computed offsets are always small: \erglinesurf{-0.9 \pm 1.2}{-20} on average for all groups.

We then perform the following bootstrap experiment: 
we sum the flux of $N$ spaxels selected by random permutation of the list of unmasked spaxels. $N$ is the number of spaxels of the segmented area. At each iteration, we add a single offset flux value randomly drawn from a Normal distribution with zero mean and standard deviation equal to the previously computed offset error.  
This allows us to take into account the uncertainty of the subtracted offset value. We repeat this 1000 times and compute the standard deviation of the sum values. This gives the empirical noise estimate for the sum of N spaxels. We perform this experiment for 3 different values of $N$, corresponding to the number of spaxels of the segmented area covered by the entire segments, the compact sources and the diffuse area, respectively. 

The standard deviation derived from the bootstrap experiment is then scaled to take into account the correlation of the noise introduced during the data reduction process (see section 4.6 of \citealt{Weilbacher2020}). This scale factor, estimated during the data reduction, is slowly evolving with wavelength, with values in the range 1.9$-$2.1. This results in our empirical standard deviation of the flux values for the filament, compact and extended segmented area.

This empirical noise estimation is performed for each overdensity.
We compare the computed values with the ones derived from noise propagation, using the values given in the datacube (already corrected for noise correlation). We note that, even after taking the noise correlation into account, the propagated values underestimate the noise by a factor 2 ($2 \pm 0.3$). This is likely due to small systematics left by the continuum subtraction. 
In the rest of the document the SNR is defined as the flux divided by the standard deviation of the noise estimated empirically as described above.

\subsection{Process validation}
\label{subsec:conf}

While the empirical error computed in the previous section gives an estimate of the expected error for the measured flux within a given diffuse emission segment identified by the algorithm, it says nothing about the probability of being fooled by noise, i.e. of estimating a diffuse emission segment with the same or higher SNR when there is  no diffuse emission. For each narrow band, the SNR value of the diffuse emission computed by the estimation algorithm can be used as a test statistic to discriminate between a null hypothesis (there is no diffuse emission segment) and the presence of diffuse emission. Owing to the nature of the data and to the various preprocessing leading to the computed SNR values, the distribution of these test statistics under the null hypothesis is unknown and in particular if this is not a Gaussian  as we will see shortly. One must consequently resort to Monte Carlo simulations to evaluate this distribution in order to quantify the significance of our findings.

\begin{figure}[htbp]
\begin{center}
\includegraphics[width=\columnwidth]{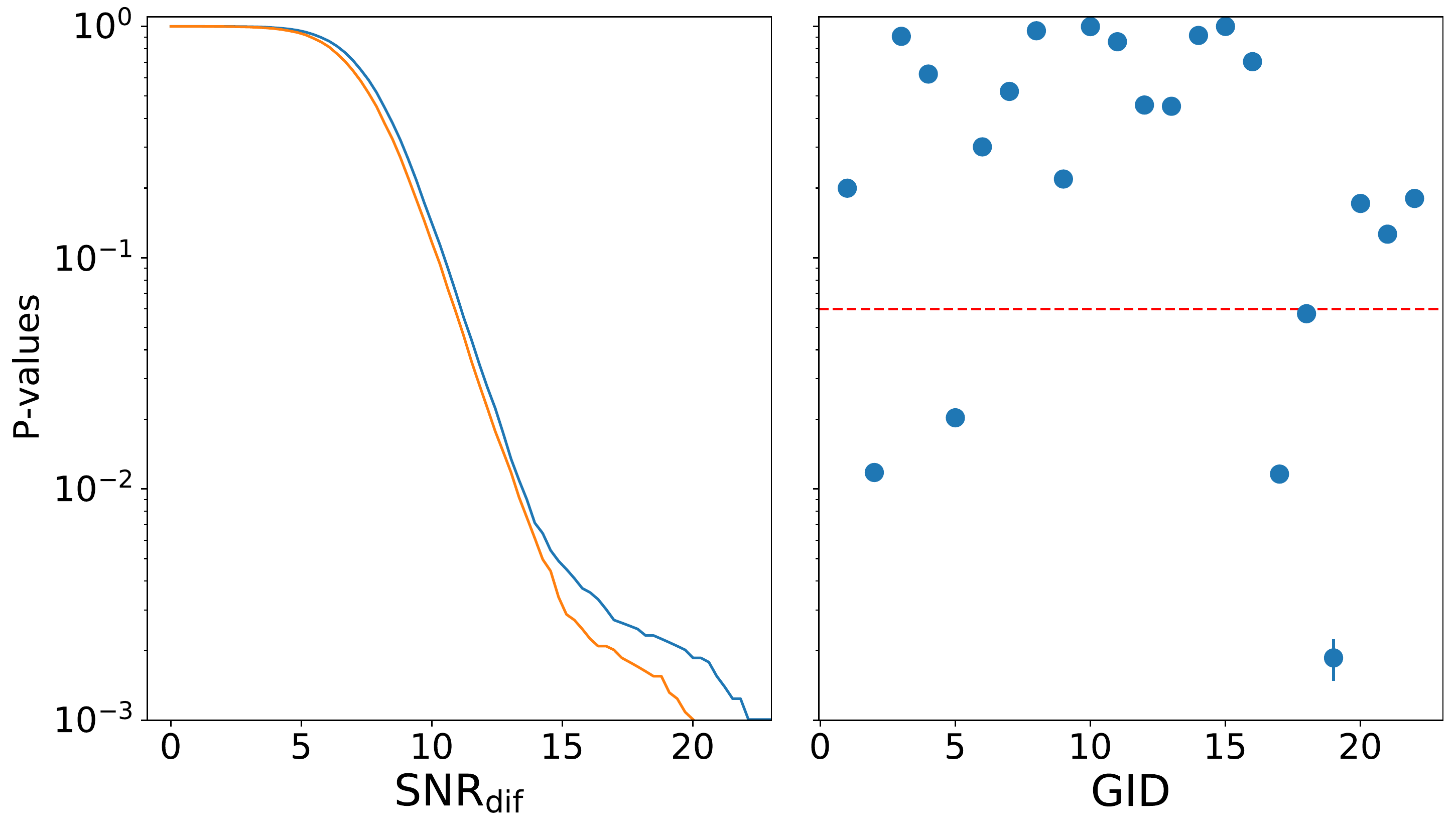}
\caption{Estimated P-values for the diffuse component. Left plot: P-values as a function of the diffuse component's SNR for the control samples (see Sect.~\ref{subsec:conf}). The samples with 10 and 20 wavelengths are shown in, respectively, blue and orange colors. Right plot: estimated P-values for all overdensities. The dotted horizontal red line show the 0.06 P-value (see Sect.~\ref{subsec:diffuse}). 
}
\label{fig:pfa}
\end{center}
\end{figure}

For that purpose we perform the following experiment. We select all wavelength slices free from sky pollution and outside the 22 groups' wavelength regions. This effectively leaves 2582 wavelengths among the 3721 cube wavelengths. We then perform a random permutation in wavelength of this sample and split it into 258 groups of 10 slices. The random permutation breaks the spectral continuity ensuring that no genuine extended structure at other redshifts than that of the overdensities can add coherently in the groups of shuffled slices. In contrast, we expect that the contribution of some diffuse emission in contiguous slices might, even if they are very faint in each slice, combine to something detectable in some of the 22 groups under examination. No spatial permutation is attempted because this would break the structure of numerous emission line sources that are present in all spectral regions of the datacube. These sources are also present in the groups under examination and even if the algorithm attempts to remove them in order to estimate diffuse emission, they do affect the algorithm's results.  The estimation algorithm (i.e. wavelet transformation, segmentation and empirical noise estimation) is then applied to each group and the resulting SNR of the diffuse emission segment are saved. We repeat this process 50 times in order to increase the size of the control sample to 12,900 groups. From the resulting statistical distribution of the SNR, one can estimate the P-value for a given overdensity ($\rm P_{k}$). The P-value $\rm P_{k}$, associated to a SNR value $\rm SNR_k$, is the probability of obtaining a SNR value equal or larger to $\rm SNR_{k}$ when there is no diffuse emission. We estimate the P-values by counting the empirical fraction of control groups with $\rm SNR > SNR_{k}$, where $\rm SNR_{k}$  is the SNR of the overdensity under examination  ($k$=1, ... ,22). In practice, to take into account the increase of the wavelength range with redshift (from 9 to 18 slices), we repeat the experiment with a similar sample of 12,900 groups of 20 wavelength slices. The resulting P-values as a function of SNR are given in the left panel of Fig.~\ref{fig:pfa}.


Further validation of the method using independent datasets and simulations are presented in Sections ~\ref{sect:group2} and \ref{subsec:galics}, respectively.

\subsection{Extended and diffuse $Ly\alpha$ emission in overdensities}
\label{subsec:diffuse}

\begin{figure*}[htbp]
\begin{center}
\includegraphics[width=0.77\paperwidth]{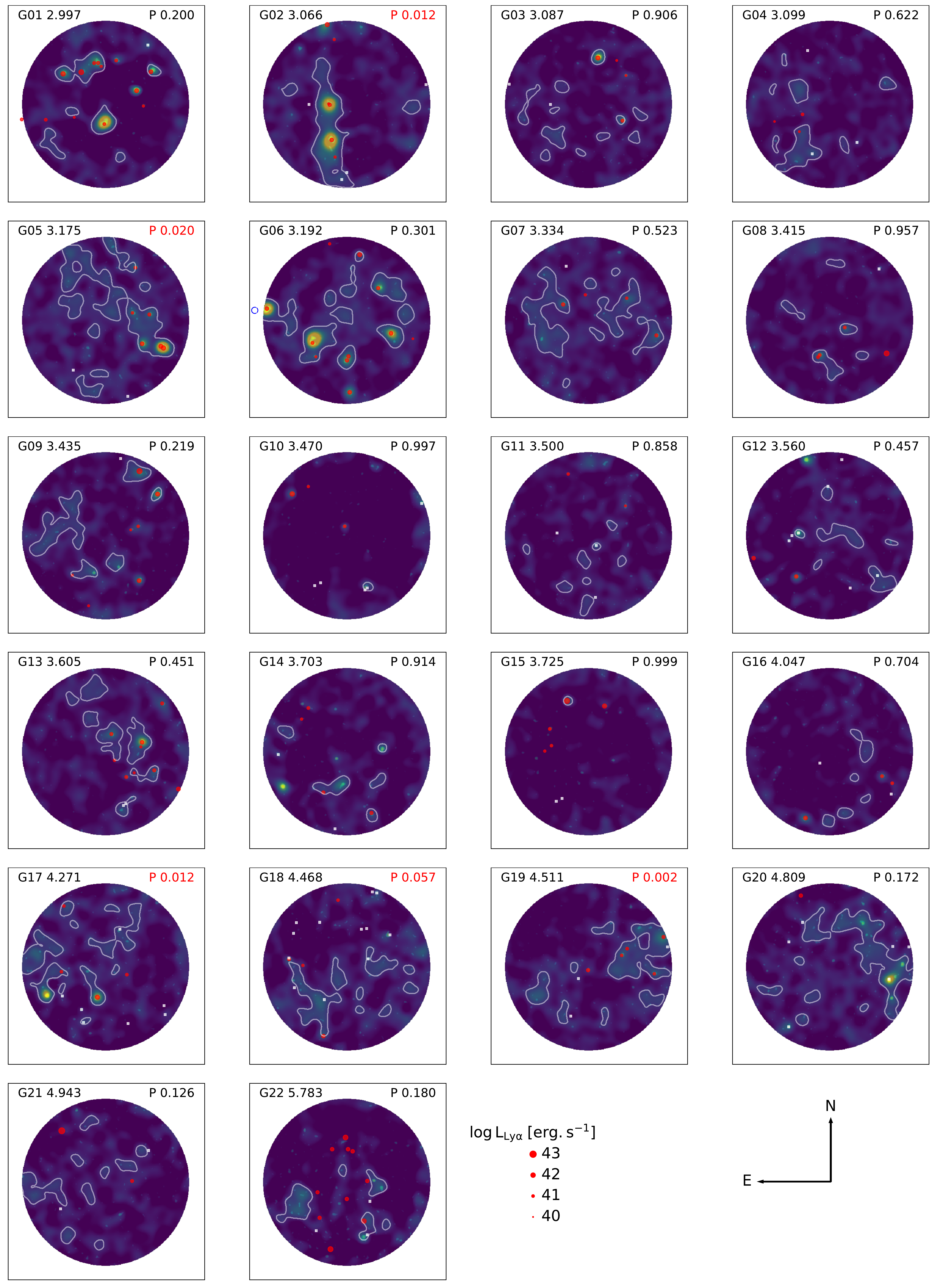}
\caption{Searching for extended \lya\ emission in groups within the \mxdf\ field. 
The images are composites of low and mid-frequencies IUWT wavelets SNR components. The solid contours display the segmentation corresponding to the identified structures (named "filament" in the text). LAE group members and potential low-$z$ interlopers are shown, respectively, as red and white symbols. 
The group number, its corresponding redshift and the diffuse emission P-value (P) (see section~\ref{subsec:conf}) are labeled.The P-value is written in red for groups with confidence 1. The blue circle symbol in group~6 indicates the location of the AGN ID 788 in the Chandra catalog \citep{Luo2017}. The \mxdf\ field shown here corresponds to a depth of 16+ hours. It has a diameter of 80\arcsec\ or 2.5 cMpc at z=3.
}
\label{fig:diffuselya}
\end{center}
\end{figure*}

\begin{figure*}[htbp]
\begin{center}
\includegraphics[width=0.85\paperwidth]{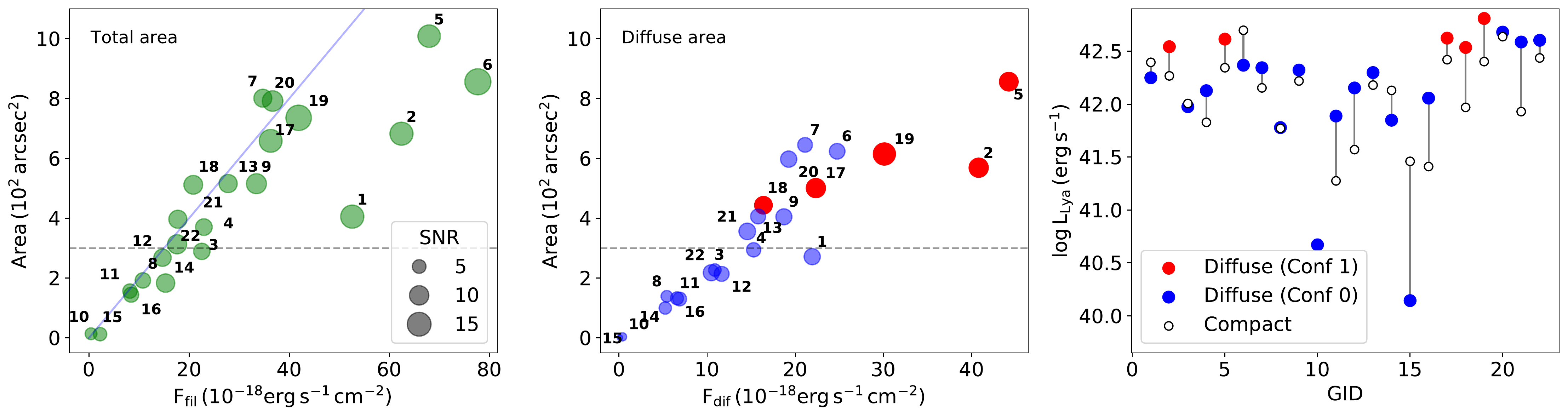}
\caption{Filament properties. Left panel: angular area covered by the entire segments (filaments) as a function of the corresponding \lya\ flux. The group IDs are labeled and displayed as green filled circles with radius proportional to the total flux SNR. The dotted horizontal line and the blue diagonal line show respectively the 300 arcsec$^2$ area limit and a constant surface brightness of \erglinesurf{5}{-20}.
Central panel: same as left panel for the diffuse \lya\ emission (diffuse area). The filled circles are colored according to the detection confidence in blue and red for low and high confidence, respectively (see Sect.~\ref{subsec:diffuse}). 
The right panel shows the total luminosity of diffuse \lya\ in filaments (filled circle symbols in red for high confidence, in blue for low confidence) compared with the luminosity of the compact \lya\ emission (black open circles) within the same filaments. Note that group 10 has no compact source detected within the filament area.
}
\label{fig:filprop}
\end{center}
\end{figure*}

The detection process described in Section~\ref{subsec:shape} has been  applied to the 22 groups. The resulting images are displayed in Fig.~\ref{fig:diffuselya}. As shown by this figure and the left panel of Fig.~\ref{fig:filprop}, extended emission\footnote{The extended emission refers to the total flux within the filaments including the compact sources area, whereas the diffuse emission refer to the filament area after exclusion of compact sources.}
is found in most of the overdensities. If we arbitrarly set a lower limit of 300 arcsec$^2$ (or $\rm 1.9 \times 10^{4}$  $\rm pkpc^2$ at $z$=3) as the minimum area, we have 14 groups (63\%) with "extended" emission.  The most extended group is group 5 with an area of 1008 arcsec$^2$ or $\rm 6 \times 10^{4} \, pkpc^2$, that is 20\% of the total available \mxdf\ area.
Note that groups 10 and 15 show almost no extended emission. As we will see later in Sect.\ref{subsec:galics}, this is likely due to the size difference between the \mosaic\ and \mxdf\ field-of-views.

We now review the diffuse area identified within each filament (central panel of Fig.~\ref{fig:filprop}). To evaluate their statistical significance, we translate their SNR to a P-value by using the relation between P-value and SNR derived from the control sample (Fig.~\ref{fig:pfa} left panel). 
For each overdensity we linearly interpolate the P-values at the corresponding number of wavelength slice of the group. The computed P-values are given in the right panel of Fig.~\ref{fig:pfa}. We split the groups in two subgroups: the high confidence overdensities (confidence 1) with diffuse emission signal detected with a P-value smaller than 0.06, and the rest of low confidence overdensities without statistically significant detection (P-value $>$ 0.06, confidence 0). 
Among the 14 groups with extended \lya\ emission, we have 5 groups with diffuse emission detected at high confidence: two at $\rm z \approx 3$ (groups 2 and 5) and three at higher redshifts (groups 17,18,19 at $\rm z \approx 4.5$). It is worth mentioning that the diffuse \lya\ emission in the high redshift groups is detected around 7000\AA, at the peak of MUSE sensitivity where the noise is minimal (column SB of Table~\ref{tab:over}).

It should also be pointed out that group 6, which is globally the brightest in \lya\ flux and one of the most extended (left panel of Fig.~\ref{fig:filprop}), is classified as low confidence ($\rm P = 0.3$) for diffuse \lya\ emission. The reason is that the flux and area attributed to compact LAEs prevail over its diffuse emission. The SNR for the diffuse area flux (central panel of Fig.~\ref{fig:filprop}) is then too low to pass the P-value threshold ($\rm P < 0.06$). For similar reasons, groups 1,7 and 20 were ranked as low confidence despite their total flux brightness and area.

The average surface brightness of the diffuse emission in the high confidence groups is \erglinesurf{5.1 \pm 1.2}{-20}. 
A statistical overview of the parameters of the high confidence filaments is given in Table~\ref{tab:sumfil} while the details for each overdensity are given in Table~\ref{tab:ext}.

Note that, while the algorithm results in the detection of extended and diffuse emission with a given confidence, it does not tell us if this extended emission is indeed \lya\ emission.
There are other possible sources of diffuse emission in the datacube due to other emission lines such as MgII (\citealt{Burchett.2020sgo}, Wizotski in prep, Leclercq in prep), [OIII] \citep{Johnson2018} or even galactic diffuse emission (e.g. $\rm H\alpha$, [SII]). However, the former are systematically associated with bright continuum galaxies which are not present in the identified overdensities and the latter are at known wavelengths and can thus be easily excluded. 
In addition, the fact that the identified diffuse emission overlaps spatially with the known \laes\ found independently during the construction of the catalog (Sect.~\ref{subsec:catalog}), suggest that this diffuse emission originates from \lya\ emission.

\begin{table}
\begin{center}
\caption{Measured physical parameters for the 5 high confidence filaments (average values)}
\begin{tabular}{rrrrr}
\label{tab:sumfil}
Name & Mean & Std & Min & Max \\
\hline
$\mathrm{F_{fil}}$ & 45.9 & 17.3 & 20.8 & 67.9 \\
$\mathrm{F_{comp}}$ & 15.1 & 7.0 & 4.5 & 23.7 \\
$\mathrm{F_{dif}}$ & 30.8 & 10.6 & 16.4 & 44.2 \\
\hline
$\mathrm{SNR_{fil}}$ & 13.5 & 2.5 & 9.4 & 17.1 \\
$\mathrm{SNR_{comp}}$ & 10.9 & 2.5 & 5.9 & 12.5 \\
$\mathrm{SNR_{dif}}$ & 9.9 & 1.6 & 8.1 & 12.8 \\
\hline
$\mathrm{log\, L_{fil}}$ & 42.8 & 0.1 & 42.6 & 43.0 \\
$\mathrm{log\, L_{comp}}$ & 42.3 & 0.2 & 42.0 & 42.4 \\
$\mathrm{log\, L_{dif}}$ & 42.6 & 0.1 & 42.5 & 42.8 \\
\hline
$\mathrm{S_{fil}}$ & 719.6 & 162.6 & 511.9 & 1008.8 \\
$\mathrm{S_{comp}}$ & 122.6 & 32.1 & 68.4 & 157.9 \\
$\mathrm{S_{dif}}$ & 597.0 & 142.3 & 443.5 & 856.5 \\
\hline
$\mathrm{SB_{dif}}$ & 5.1 & 1.2 & 3.7 & 7.2 \\
$\mathrm{F_{dif}/F_{fil}}$ & 68.5 & 6.1 & 61.5 & 78.6 \\

\end{tabular}
\tablefoot{ Mean, standard deviation, min and max values are given for the following 3 components (c):
fil = full filament, comp = compact area within the filament, dif = diffuse area within the filament.
$\mathrm{F_{c}}$: \lya\ flux in $\mathrm{10^{-18}}$\ergsline\ for the component c, 
$\mathrm{SNR_{c}}$: Flux SNR for the component c,
$\mathrm{\log\, L_{c}}$: Log of \lya\ luminosity in \ergslum\ for the component c.
$\mathrm{S_{c}}$: Surface in $\mathrm{arcsec^2}$ for the component c.
$\mathrm{SB_{dif}}$: Average surface brightness of the diffuse \lya\ emission in \erglsurf{-20}.
$\mathrm{F_{dif}/F_{fil}}$: Fraction of \lya\ flux in the diffuse area (in \%).
For individual values by overdensities see Table~\ref{tab:ext} in appendix~\ref{app:tables}.
}
\end{center}
\end{table}

\begin{figure}[htbp]
\begin{center}
\includegraphics[width=0.4\columnwidth, angle=90 ]{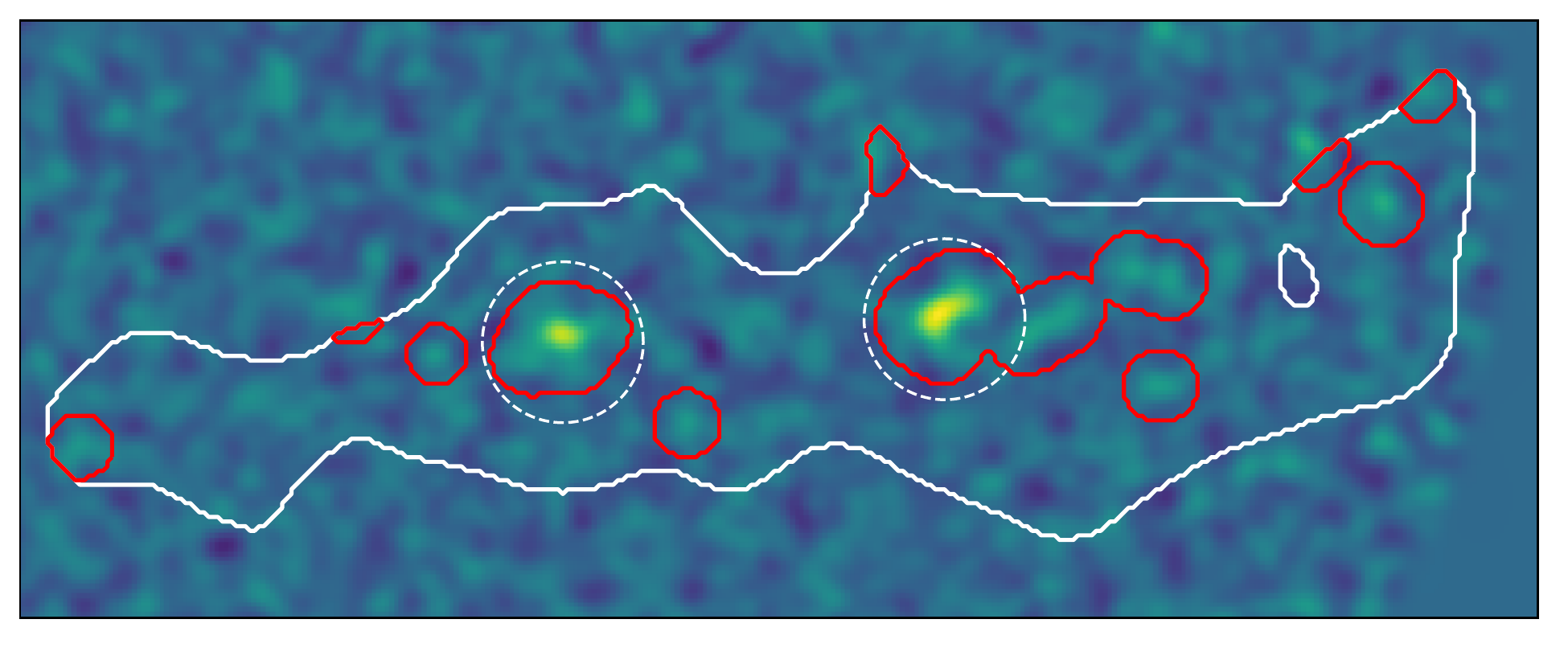}
\caption{Medium frequency NB image of the main group 2 filament (rotated by 90\degree\ counterclockwise). The compact source segmentation contours are shown in red. The two white dashed circles display the 3.5\arcsec\ radius area (27 pkpc) where the diffuse emission from the CGM of the two LAEs (bright yellow clumps) identified in the catalog becomes insignificant.
}
\label{fig:filgroup2}
\end{center}
\end{figure}

The boundary between compact and extended sources is based on the considered multiscale analysis. An important question is to what extent the segmentation performed on compact sources is representative of the galaxy itself only, or of the galaxy plus its surrounding CGM. In other words, what fraction of the filament flux can be explained by the CGM of the identified galaxies? In a few cases, e.g. group 5, the question is not relevant given that a high fraction of the filaments have no detected LAEs. But in some cases, such as group 2, there are two bright LAEs in the main filament which might bias the filament flux measurement. In Fig.~\ref{fig:filgroup2} we show that the area of group 2 covered by compact source segments extends up to 27 pkpc from the galaxy center. At such a distance, a very high fraction of the galaxy CGM is already included (see Figure 15 of \citealt{Leclercq2017}) and the average surface brightness of the \lya\ halos falls below \erglsurf{-20} (see Figure 2 of \citealt{Wisotzki2018}). We then expect the CGM of these \laes\ to have only a small contribution to the measured \lya\ emission flux in the diffuse area. 

In addition, as shown in Fig.~\ref{fig:filgroup2}, the compact source segmentation map (red contours) is not restricted to the detected LAEs, but covers all medium frequency peaks identified in the NB image. 
Note also that the flux of these compact sources is on average three times fainter than the \lya\ flux of the detected LAEs (Fig.~\ref{fig:limflux}).
We conclude that our extended diffuse emission flux is a conservative lower limit on the \lya\ emission shining outside the CGM of identified LAEs.

For each group we compute the total \lya\ luminosity in the diffuse area and compare it to the flux measured in the compact structure area (right panel of Fig.~\ref{fig:filprop}, Table \ref{tab:ext}).
For most of the groups, the measured \lya\ luminosity in the diffuse area is significantly larger than in the compact source area. This is true in particular for the five high confidence groups.
As seen above, if we assume that the compact source flux is a good proxy of the total \lya\ emission of identified galaxies, including their CGM emission, we conclude that a high fraction of the total \lya\ emission measured in the filaments is coming from either undetected faint \laes\ and their CGM and/or intrinsic diffuse \lya\ emission. 
For the high confidence groups the average log \lya\ luminosity (\ergslum) in the diffuse area amounts to $\mathrm{42.6 \pm 0.1}$ and represents between 61\% and 79\% (average 68\%) of the total filament \lya\ luminosity.

\subsection{The case of the z=3.07 overdensity}
\label{sect:group2}

\begin{figure}[htbp]
\begin{center}
\includegraphics[width=0.8\columnwidth]{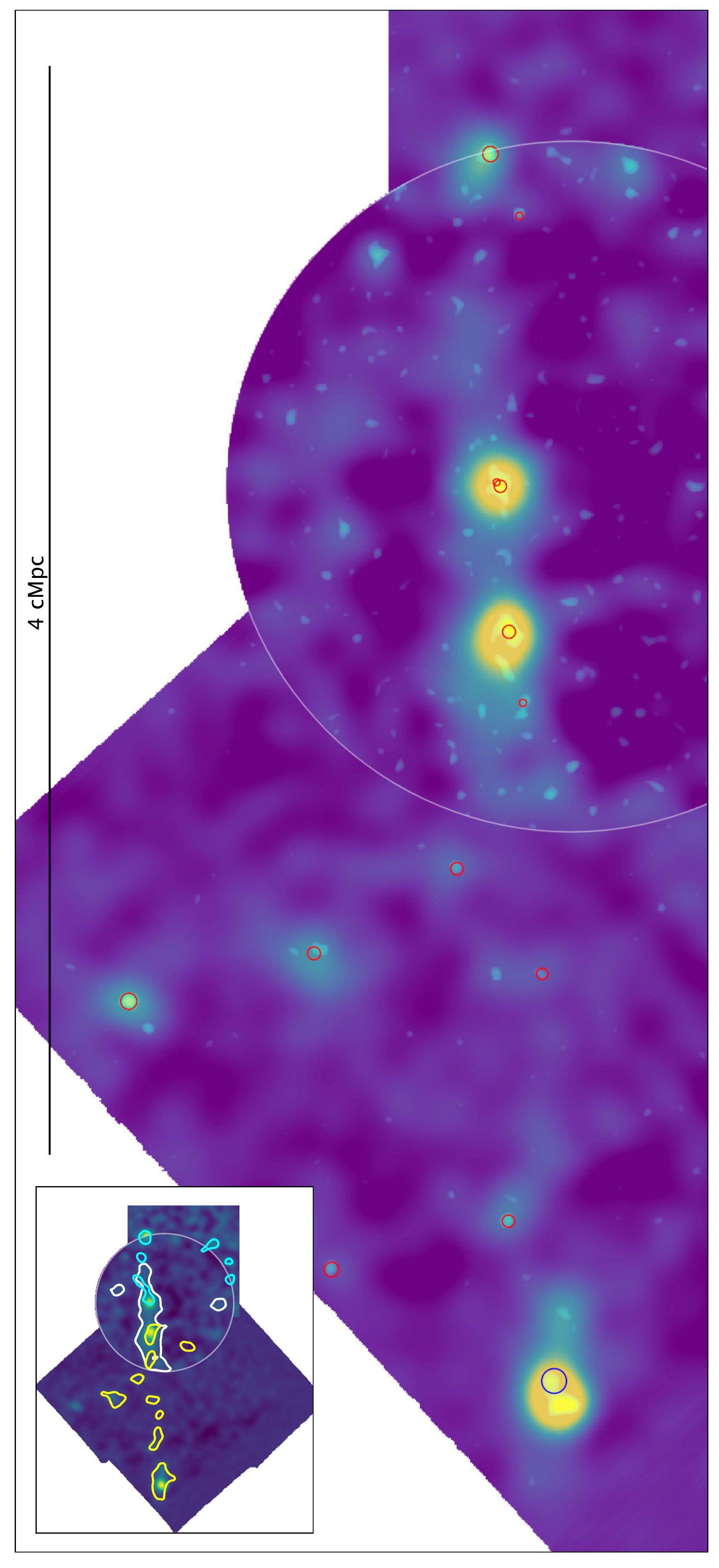}
\caption{Extended diffuse \lya\ emission in the z=3.07 overdensity (GID 2). Composite IUWT-SNR image formed with part of the \mosaic\ 10h depth exposure (bottom rotated square field), the UDF10 30h depth exposure (small upper square) and the \mxdf\ 140h depth exposure (circle). The locations of the identified \laes\ are displayed as red circles with size proportional to the log of their luminosity. The blue circle at the bottom shows the location of the AGN and \lae\ blob. The 4 comoving Mpc (1 pMpc) scale is shown. 
The inset displays the contours of the filaments in yellow, white and cyan for, respectively, the \mosaic\ (only a fraction of the full field), \mxdf\ and \udft\ exposures. 
}
\label{fig:group2}
\end{center}
\end{figure}

Although we did not expect to detect diffuse \lya\ emission other than the \lya\ halos in the \mosaic\ and \udft\ exposures with 10 and 30 hours depth, respectively,  we nevertheless check for possible exceptions in the high confidence groups.

The estimation algorithm shows no convincing emission in all groups 
with the notable exception of group 2 which {displays a clear signal extending the main \mxdf\ filament to the south. Fig.\ref{fig:group2} displays the complete structure which now extends to a total length of 4.6 cMpc (1.1 pMpc) with a width of 191 ckpc (47 pkpc).

As shown in the inset of Fig.~\ref{fig:group2}, the filament is detected in each of the three independent exposures. This observational evidence constitutes an additional process validation  and reinforces our confidence that the structure is real. 
The \mosaic\ and \udft\ exposures show a second filament crossing the main one at 45\degree.

A bright patch of \lya\ emission is seen at the south end of the filament (bottom of Fig.~ \ref{fig:group2}). The source of \lya\ emission coincides with a luminous, QSO-level type~II AGN identified as ID~746 in the Chandra 7 Ms catalog \citep{Luo2017}, hereafter referred to as CID~746. The extended \lya\ emission surrounding the AGN was already reported by \cite{Brok2019} in their study of extended \lya\ emission around type II AGN. The authors trace extended \lya\ out to 80~pkpc from the AGN (see their Figure 1). Note that their analysis is derived from the same dataset (i.e. the \mosaic\ MUSE datacube), but with a different post-processing. Our analysis confirms this extension and adds the finding of fainter diffuse \lya\ emission at much larger distance. We consider the possible importance of this AGN in Sect.~\ref{subsec:agn} below.

Fig.~\ref{fig:group2} is also illustrative of the density increase of compact sources seen at the \mxdf\ depth with respect to the \mosaic\ field. Of course we are not able to prove that all these compact sources are indeed faint {\lae}s, and some of them are possibly noise peaks, but the fact that a low fraction of them were independently identified as \laes\ during the construction of the catalogue suggests that the filaments are populated by a high number of faint \laes. This property is not specific to  group 2, but can be seen in the other groups as well (Fig.~\ref{fig:diffuselya}).

\section{Analysis and discussion}
\label{sect:analysis}

Part of the \lya\ emission that we detect overall undoubtedly comes from galaxies and their CGM. Emission from the CGM may be powered by a variety of processes, probably combined: dissipation of gravitational energy through cooling radiation, fluorescence from the UV background or from ionising radiation emitted by local sources, scattering of galactic \lya\ through neutral gas in the CGM, extended star formation, or undetected satellite galaxies. Regardless the origin of this emission, we have shown that our analysis separates this signal from the diffuse emission (see e.g. Fig.~\ref{fig:filgroup2}). It is the origin of this diffuse component, which represents the largest fraction (70\%) of the flux, that we try to explain in the present section. Possible sources for this emission are: hydrogen gas heated and ionized by the external UV background, undetected faint \laes, or gravitational compression. We explore these options in the following sections.

\subsection{\lya\ fluorescence}

\subsubsection{\lya\ fluorescence by the cosmic UV background}
\label{subsec:uvb}

Optically thick clouds illuminated by a diffuse intergalactic UV radiation field will emit \lya\ emission by the fluorescent conversion of H-ionizing photons into \lya\ \citep{Hogan1987, Gould1996}. A ubiquitous source of such photons is the cosmic UV background (UVB), i.e. the integrated UV emission from AGN and star-forming galaxies (e.g. \citealt{Haardt.1996, Haardt2012}). 



Here we follow the recipes from \cite{Cantalupo2005} to estimate the expected \lya\ surface brightness due to UVB fluorescence, in a similar way as in \cite{Gallego2018}. We adopt the photoionization rate of neutral hydrogen $\Gamma_{\ion{H}{i}}^{\mathrm{HM}}$ and its redshift evolution predicted by the synthesis model of \cite{Haardt2012} as a baseline value. To assess the uncertainties we also consider the more recent model calculations by \cite{Faucher2020} ($\Gamma_{\ion{H}{i}}^{\mathrm{FG}}$), as well as the empirical estimates by \cite{Becker.2013} ($\Gamma_{\ion{H}{i}}^{\mathrm{BB}}$). We emphasize that the \cite{Cantalupo2005} recipes consider \lya\ photons produced in Lyman limit systems, i.e. in optically thick clouds with \ion{H}{i} column densities above $\rm 10^{17.2} cm^{-2}$ with a covering fraction of 1.

As shown in Table~\ref{tab:uvb}, the fluorescent \lya\ emission powered by the UVB cannot explain the totality of the observed surface brightness, but its contribution is not negligible at $z\approx$3, with $\approx$30\% of the total surface brightness. At higher $z$, the contribution falls to $\approx$10\% (or 20\% if we use the \cite{Becker.2013}'s photoionization rate). It is important to stress that these values are upper limits as they assume a covering fraction of 1 for the Lyman-limit systems.

Note, however, that the photoionization rates given in Table~\ref{tab:uvb} are average values. Within the overdensities, given the higher number of \laes, one might expect the local ionising radiation field to be enhanced relative to the background value. However, because the mean free path for ionizing photons is much greater than the size of the overdense region, the typical local enhancement of the ionizing flux is expected to be much smaller than the local overdensity of sources of ionizing radiation (e.g.\citealt{Schaye2006}). The cosmological radiative transfer simulations performed by \cite{Rahmati2013} confirm that local sources only dominate over the UVB on very small scales. Their Figure~4 (left panel) displays the various contributions to the photoionization rate as a function of distance from  the centers of $\rm 10^{10.5}$-$\rm 10^{11} M_{\sun}$ dark matter haloes at $z=3$. One can observe that the contribution of local stellar radiation ($\Gamma_{\ion{H}{i}}^{\mathrm{LSR}}$) is already a factor 5 below $\Gamma_{\ion{H}{i}}^{\mathrm{UVB}}$ at the virial radius and then drops to negligible values at larger distances.
Given that the measured diffuse \lya\ emission {\em excludes} by construction most of the local source's CGM (Section~\ref{subsec:diffuse}), one can then safely ignore the contribution of local ionizing radiation to the \lya\ fluorescence in the diffuse emission area.

\begin{table*}
\caption{Estimation of maximum fraction of \lya\ surface brightness due to the Cosmic UV background in the five groups with high confidence diffuse emission.}   
\label{tab:uvb}
\centering
\begin{tabular}{ccrrrrrrr}
ID & z & $\mathrm{SB_{dif}}$ & $\mathrm{\Gamma_{HI}^{HM}}$ & $\mathrm{\Gamma_{HI}^{FG}}$  & $\mathrm{\Gamma_{HI}^{BB}}$ & $\mathrm{SB^{BB}_{uv}}$ & $\mathrm{SB^{HM}_{uv}}$ & R \\
\hline
2 & 3.07 & 7.17 & 8.08 & 8.96 & $8.12 \pm 0.68$ & $2.03 \pm 0.17$ & 2.02 & 28 \\
5 & 3.18 & 5.16 & 7.75 & 8.64 & $7.93 \pm 0.66$ & $1.78 \pm 0.15$ & 1.74 & 34 \\
17 & 4.27 & 4.46 & 5.19 & 6.17 & $9.22 \pm 1.05$ & $0.82 \pm 0.09$ & 0.46 & 10 \\
18 & 4.47 & 3.69 & 4.89 & 5.80 & $9.53 \pm 1.22$ & $0.73 \pm 0.09$ & 0.37 & 10 \\
19 & 4.51 & 4.90 & 4.84 & 5.73 & $9.50 \pm 1.26$ & $0.70 \pm 0.09$ & 0.36 & 7 \\

\end{tabular}
\tablefoot{
ID: Group ID.
z: redshift. 
$\mathrm{SB_{dif}}$: Average surface brightness of the diffuse \lya\ emission in units of \erglsurf{-20}.
$\mathrm{\Gamma_{HI}^{src}}$: Photoionization rate of neutral hydrogen from {\it src}, in units of $\mathrm{10^{-13}s^{-1}}$;
{\it src} takes the following values: HM \citep{Haardt2012}, FG \citep{Faucher2020}, BB \citep{Becker.2013}
$\mathrm{SB_{uv}^{src}}$: Expected \lya\ surface brightness from the UV background in units of \erglsurf{-20}.
R: relative fraction $\mathrm{SB^{HM}_{uv} / SB_{dif}}$ in \%.
}
\end{table*}

\subsubsection{\lya\ fluorescence from Active Galactic Nuclei}
\label{subsec:agn}

In the vicinity of a luminous AGN, the intense UV radiation can boost the \lya\ emission by large factors \citep{Haiman2001, Cantalupo2005, Kollmeier2010}. Two high-redshift AGN are known within the \mosaic\ field of view, both of which are outside the \mxdf\ but match the redshifts of two of our groups. The spectacular case of group~2 at $z=3.07$ was already mentioned in Sect.~\ref{sect:group2}; the other AGN is associated with group~6 at $z=3.19$, immediately adjacent to the \mxdf\ footprint. We now consider the importance of AGN for explaining the observed diffuse \lya\ emission in the \mxdf, beginning with group~2.

While an AGN as luminous as CID~746 is intrinsically a copious producer of UV radiation, the object is classified as a highly obscured type~2 AGN. In order to produce any fluorescent emission in the \mxdf\ filament the obscuration must be negligible in the transverse direction towards the \mxdf\ filament, which (in line with \citealt{Brok2019}) we assume in the following. We consider two simplified scenarios:

If the neutral hydrogen column density is high enough to make the \mxdf\ filament fully self-shielded and if we at the same time assume negligible absorption between the QSO and the filament, its \lya\ surface brightness will scale directly with the geometrically diluted intensity of the incident UV radiation. We can then estimate a boost factor, defined by \cite{Cantalupo2005} as the enhancement of UV illumination due to the AGN and quantified in their equation 14. The northern extension of the \lya\ nebula at 80 pkpc from the AGN has a surface brightness of \erglinesurf{1}{-18}, which translates into a boost factor of 50 if we adopt a baseline surface brightness of \erglinesurf{2}{-20} for fluorescence at an optically thick \ion{H}{i} cloud due to the UVB alone (see Table~\ref{tab:uvb} in Sect.~\ref{subsec:uvb}). This boost factor decreases with the square of the distance to the AGN, and at the \mxdf\ location in 700 pkpc its value becomes $\mathrm{1 + 50\times \left( 80 / 700 \right)^2 \approx 1.6}$. Note that this is an upper limit given that the 3D distance is likely to be larger than the projected separation. In this scenario, the extra diffuse \lya\ emission within the \mxdf\ region caused by the AGN would fall short of reproducing the measured value by a factor $\ga$4 (see Table.~\ref{tab:ext}).

If on the other hand the hydrogen within the Mpc-scale surroundings of the AGN including the \mxdf\ region is optically thin to ionizing radiation, it is less straightforward to estimate the emergent \lya\ surface brightness. Judging by the substantial X-ray luminosity of the AGN, the near-zone around the AGN should be much larger than the $\sim$1~pMpc distance relevant here (Bolton \& Haehnelt 2007, Eilers et al. 2017), implying that the \lya\ surface brightness enhancement is dominated by radiative recombinations and scales with the square of the local density. In this case it would indeed be possible that the AGN provides most or all of the additional photons needed to match the observed \lya\ emission. Since we have no way to distinguish between these scenarios, we must leave the final judgement open for this particular group.

In case of group~6, the AGN is located so close ($\approx 5$ arcsec) to the region of diffuse emission
that, although the AGN is much fainter, the emission can be explained with both scenarios. Note, however, that this group was ranked as low confidence for diffuse emission (see section \ref{subsec:diffuse}). In addition, one can observe in Figure~\ref{fig:diffuselya} that the diffuse emission is not specifically strong at the immediate vicinity of the AGN.

In the other overdensities with extended emission we did not find any evidence for any nearby AGN, neither based on characteristic features in the MUSE spectra nor from the Chandra catalog \citep{Luo2017}. Given the extraordinary depth of this catalog, this implies that we can confidently exclude any further strong AGN in this region. Low luminosity AGN below the Chandra flux limit or very highly obscured objects -- possibly even Compton-thick ones -- are still possible, but these would not be significant contributors of escaping UV photons capable of powering the observed diffuse \lya\ emission. Overall we conclude that AGN could boost extended \lya\ emission in two of our overdensities, but that they are unlikely to be important in the majority of cases.

Finally we note that AGN are expected to flicker on time scales of $\sim 10^{5}$ yr (e.g. \citealt{Schawinski2015}), in which case the presence or absence of an AGN may not be correlated with AGN-induced \lya\ emission at distance $\rm >100  \, pkpc$.

\subsection{Contribution of undetected \laes\  to the diffuse \lya\ emission}
\label{subsec:undetlae}

In spite of the depth of the MXDF observations, we expect a significant fraction of LAEs to fall beyond our detection limit. It is then indisputable that at least a fraction of the observed diffuse Ly$\alpha$ flux is coming from these undetected LAEs. But can these galaxies be responsible for the totality of the measured flux?  

In Sec. \ref{subsec:schechter}, we first address this question with simple modeling based on the luminosity function of LAEs. In Sec. \ref{subsec:galics}, we then use a more sophisticated approach based on the semi-analytic model \galics\ \citep{Garel2012, Garel2015a}. 

\subsubsection{ Luminosity function toy model} \label{subsec:schechter}

The luminosity function (LF) of LAEs is not observed to evolve strongly between $z=3$ and $6$ (e.g. \citealt{Ouchi2008, Cassata2011, Herenz2019}), and it is well described by a \cite{Schechter1976} function $\phi(L)dL = \phi_\star (L/L_\star)^\alpha \exp(-L/L_\star) dL/L_\star$. The faint end behavior of the LF is given by the parameter $\alpha$. Thanks to MUSE observations which have provided a large sample of faint \laes\ (e.g. \citealt{Bacon2017, Urrutia2019, Richard.2020}), we  now have better constraints on $\alpha$ down to $\log (L_{\rm Ly\alpha} / \ergslum) = 41.5$ in deep fields and $40.5$ in lensing cluster fields. \cite{Herenz2019} found $\alpha = -1.84^{+0.42}_{-0.41}$ with no evolution with redshift,  \cite{Drake2017} measured $\alpha = -2.03^{+1.42}_{-0.07}$ at $z$=3.5 and $\alpha = -2.86^{+0.78}_{-\infty}$ at z=5.5 and \cite{Vieuville2019}, using MUSE lensing cluster observations, found $\alpha = -1.63^{+0.13}_{-0.12}$ at $z$=3.5 and a similar value at $z$=4.5.
In the following we have selected the \cite{Herenz2019} intermediate values
with $\rm log (\phi_\star / cMpc^{-3}) \approx -2.71$, $\alpha \approx -1.84$, and ${\rm log}(L_\star / \ergslum) \approx 42.6$ as our fiducial model, but we will also explore later different values of $\alpha$.
Assuming to first order that the amplitude of the LF changes with environment but not its shape, we may write the mean surface brightness (hereafter SB) contributed by undetected LAEs as : 

\begin{equation}\label{eq:SB_LF}
{\rm SB_{LF}}(<L_{\det}) = \Delta l \times \zeta \times \int_0^{L_{\det}} \delta \times L \times \phi(L) dL,
\end{equation}
where $\phi$ is the field LF, $\delta$ is the over-density measured in each group (Table \ref{tab:over}), and the integral runs over luminosities below a luminosity $L_{\rm det}$ corresponding to a flux limit of  $10^{-18.5}$\ergsline{} at the redshift of each group. 
This flux limit is derived as the peak of the \lya\ flux distribution (Fig.~\ref{fig:limflux}) of compact sources identified by the detection algorithm (Sect.~\ref{subsec:shape}).
$\Delta l = 8 \, {\rm cMpc}$ is the depth of the narrow-band slice, and $\zeta \geq 1$ is a factor which accounts for angular clustering. We define $\zeta$ as the inverse of the filling factor of LAEs in a field: $\mathrm{\zeta = S_{\rm MXDF} / S_{\rm LAE}}$ where $S_{\rm MXDF} \simeq 3632\ {\rm arcsec}^2$ is the projected area of the field and $S_{\rm LAE}$ is the area occupied by diffuse emission from unresolved LAEs. 
A value $\zeta=1$ corresponds to sources uniformly distributed, and a higher value means a stronger angular clustering of sources. At the scales involved in the present study ($< 1$ arcmin), the angular clustering of LAEs is not constrained. Moreover, clustering constraints are statistical by nature and would not allow us to predict a value of $\zeta$ group by group. We thus estimate a value of $\zeta$ for each group as $\mathrm{\zeta = S_{\rm MXDF}/S_{\rm Fil}}$, where $S_{\rm Fil}$ is taken from Table 1. This assumes that all undetected galaxies are distributed within the observed diffuse emission mask. This is probably an upper estimate of $\zeta$ as one expects fainter sources to cluster less strongly \citep[e.g.][]{Mo1998} and hence to occupy a larger area. The values of $\zeta$ we obtain in this way range from 4 to 8 depending on the group, with a mean of 6.4.

The results of this simple model, with the parameters set to the values discussed above, are shown in Fig. \ref{fig:SBLF} with the blue plusses. This Figure shows the ratio of the expected \lya\ surface brightness due to undetected galaxies to the one measured in the diffuse component ($\mathrm{SB_{dif}}$ from Table~\ref{tab:ext}) for each high confidence group. We find that undetected galaxies account for about 40\% of the signal in groups 2 and 5, about 55\% for group 17, and 80-90\% for groups 18 and 19. Thus, for most groups studied here, the contribution of undetected LAEs is likely dominant. If we add the contribution from fluorescence of the UVB (Table~\ref{tab:uvb}), we can explain  about 70\% of the signal for groups 2,5 and 17 and 90-100\% for groups 18,19.

\begin{figure}[]
\begin{center}
\includegraphics[width=0.9\columnwidth]{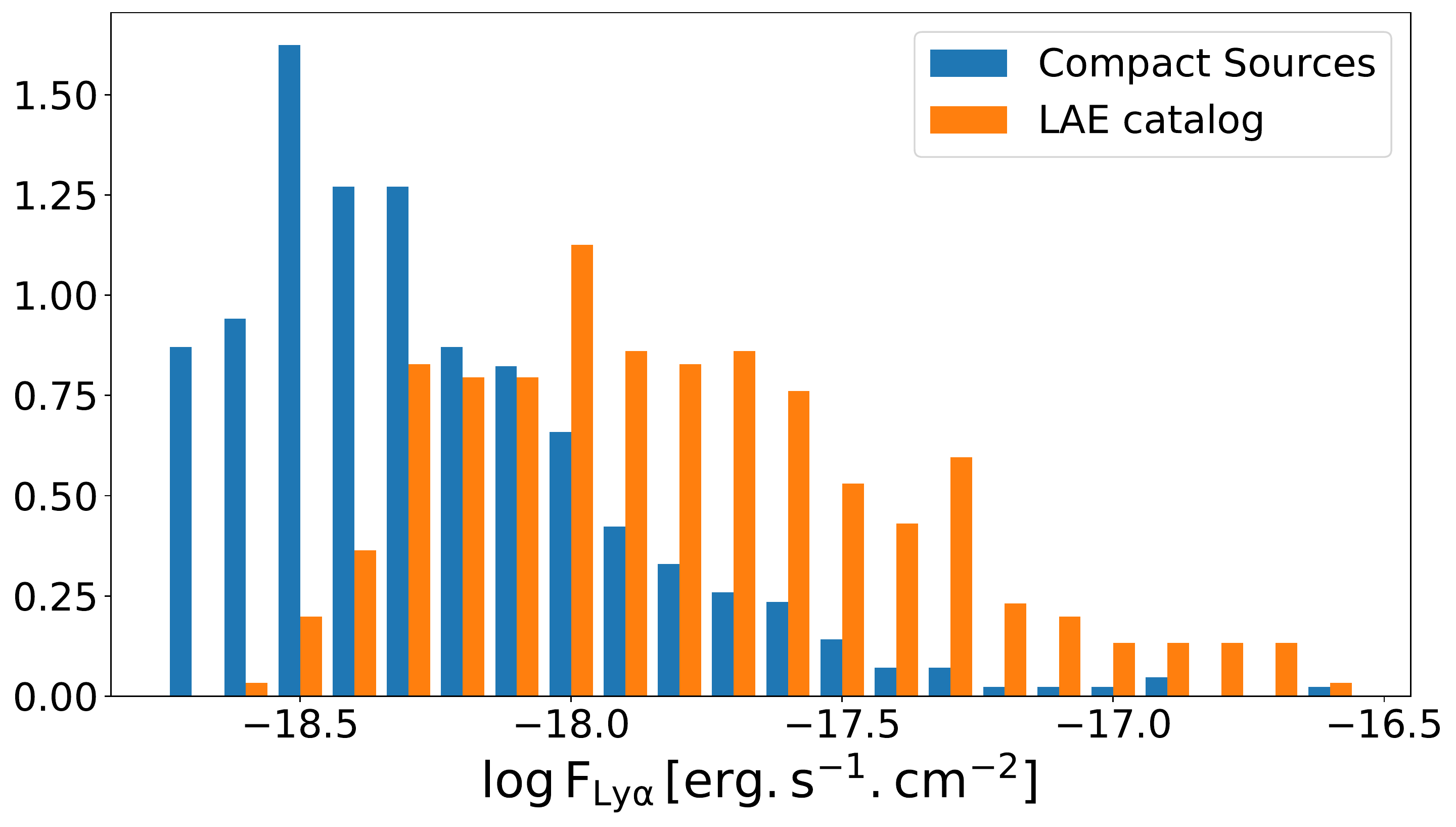}
\caption{Comparison of the \lya\ flux probability density for the compact sources detected by the algorithm (Sect.~\ref{subsec:shape}) and the \mxdf\ LAE catalogue.}
\label{fig:limflux}
\end{center}
\end{figure}

\begin{figure}[]
\begin{center}
  \includegraphics[width=\columnwidth]{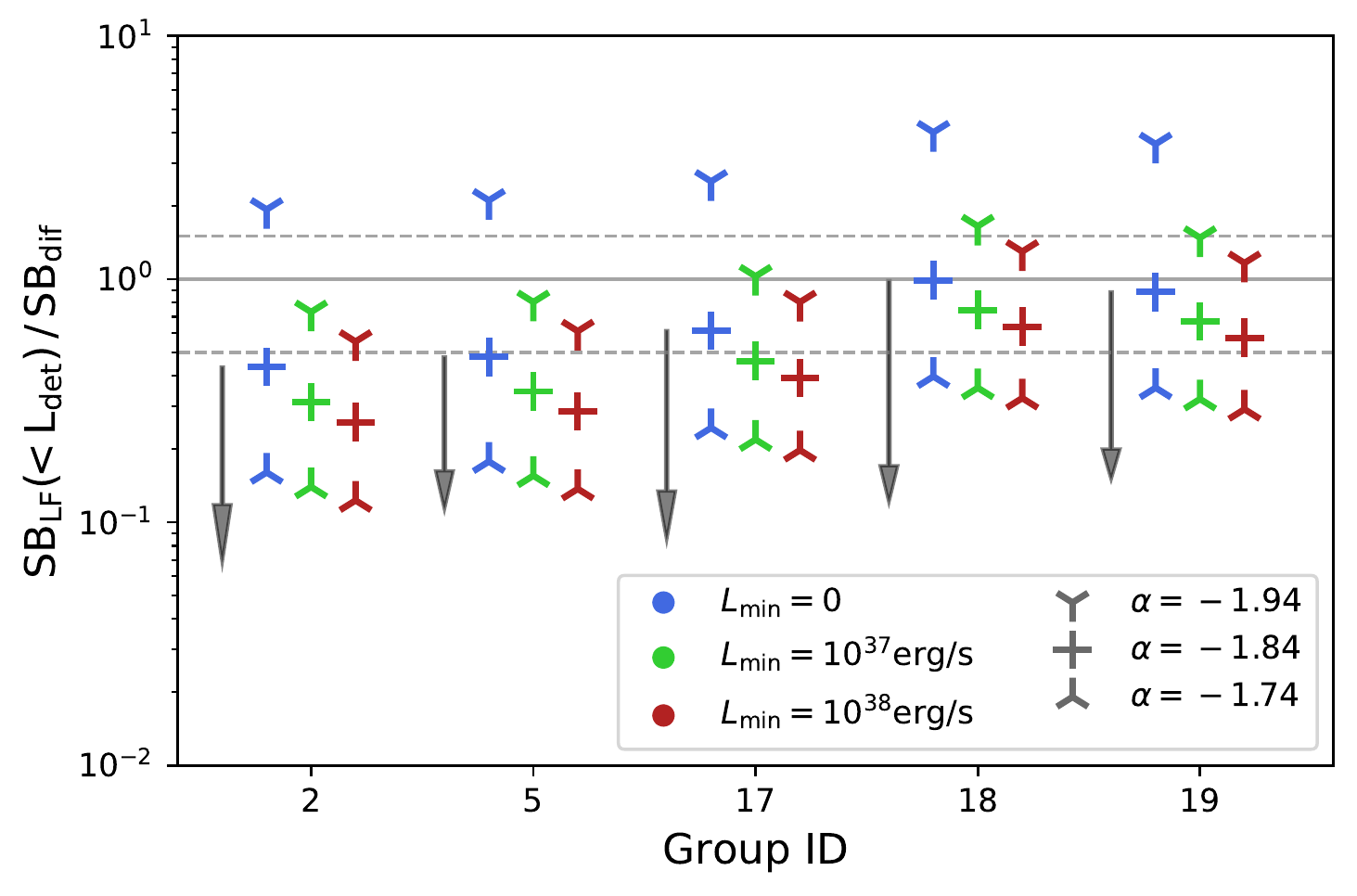}
\caption{Ratio of the expected SB due to undetected LAEs to the mean SB of detected diffuse emission, as a function of group ID for the high confidence groups. The blue plusses show the prediction for a fiducial LF with a faint-end slope of $\alpha = -1.84$ \citep{Herenz2019} extrapolated to zero luminosity. The other symbols show the predictions for steeper ($\alpha = -1.94$) or shallower ($\alpha = -1.74$) slopes as indicated on the plot. The red and green symbols show the result of the model when using a low luminosity cut ($L_{min}$) of respectively ${\rm 10^{38}}$ and ${\rm 10^{37}}$ \ergslum. The arrows show the amplitude of the multiplicative $\zeta$ correction that was included (see Section~\ref{subsec:schechter}). The horizontal dotted lines display the 0.5 and 1.5 values.
}
\label{fig:SBLF}
\end{center}
\end{figure}

Of course these estimates are crude at best and very uncertain due to many poorly constrained parameters. The steepness of the faint end of the \lya\ LF, in particular, is the parameter which has the largest impact\footnote{For simplicity, in the analysis presented here, we only vary $\alpha$ and keep the other two parameters of the LF fixed. We have verified that our conclusions are unchanged when changing $\phi_\star$ and $L_\star$ consistently with $\alpha$ to take into account the correlation between these three parameters shown by \cite{Herenz2019} (their Fig. 12). For the steep limit, our results are barely distinguishable, and for the shallow limit, changing  $\phi_\star$ and $L_\star$ would increase our model's SB by about 25\%.}. Assuming a steep, yet reasonable, slope $\alpha=-1.94$, we find that the expected SB due to undetected LAEs actually overshoots our observations by factors $\approx 2-3$ (top blue symbols in Fig. \ref{fig:SBLF}). From Eq. \ref{eq:SB_LF}, it is clear that this tension may easily be resolved by reducing the product $\zeta\times\delta$ by a factor 3. As mentioned above, our fiducial value of $\zeta$ is an upper limit and the true value may be anywhere between 1 and our estimate, either due to small-scale clustering variance or to projection effects. The amplitude of the multiplicative $\zeta$ correction is shown as an arrow in Fig. \ref{fig:SBLF} for each group. The value of $\delta$, in turn, is evaluated on the full extent of the MOSAIC field, and it is not obvious that the same over-density would be measured within the smaller MXDF area. It is interesting to note that a shallower slope of the LF, say $\alpha = -1.74$, produces at best 30\% of the observed diffuse emission (lower blue symbols on Fig. \ref{fig:SBLF}). This would imply that the diffuse emission we observe is for the most part not due to stellar irradiation. 
The precise value of the faint-end slope of the \lya\ LF is thus crucial to understand our observations. Its uncertainties are still large, although a steep slope appears more likely \citep[][]{Herenz2019}. 

Given the potential role of faint LAEs in producing the diffuse emission we detect, it is interesting to understand down to which luminosities the contribution of these LAEs matters. This depends on the slope of the LF: the steeper the LF the more important the contribution of very faint objects. As an example, for the slopes $\alpha = -1.74$, $-1.84$, and $-1.94$, 50\% of the luminosity below $L_{\rm det}$ is contributed by sources fainter than $0.07\times L_{\rm det}$, $0.01\times L_{\rm det}$, and $10^{-5}\times L_{\rm det}$, respectively. At the redshifts of our groups, $L_{\rm det}$ is in the range $\approx 2-7\times 10^{40} \ {\rm erg\, s^{-1}}$, and so half of the light from undetected LAEs is contributed by sources brighter than $\approx 1-5\times10^{39}$, $2-7\times 10^{38}$, and $2-7\times 10^{35}$ \ergslum\ for the three slopes, respectively. There are no observational constraints on the \lya\ LF at such faint luminosities. 

The deepest UV LFs at $z \approx 3-6$ are consistent with a powerlaw extending to magnitudes as faint as $\rm M_{AB} \approx -13$ \citep{Alavi.2016, Bouwens2017, Livermore2017, Atek2018}.
Making the crude approximation that \lya\ and UV emission are linearly related to the star formation rate and neglecting the effect of dust\footnote{Here, we make the reasonable assumption that $\mathrm{L_{\lya} [erg s^{-1}] = 10^{42} (SFR / \msunyr)}$ 
and that very faint galaxies are likely to be metal-poor, such that dust attenuation is expected to be much less significant than for brighter objects (e.g. \citealt{Garel2015a, Maseda2020}).},
this value corresponds to a \lya\ luminosity slightly fainter than $10^{40}$ erg s$^{-1}$ (e.g. \citealt{Garel2015a}).
With the same assumptions, a \lya\ luminosity of $\mathrm{L_{\lya} \approx 10^{38}}$ erg s$^{-1}$ corresponds to a UV magnitude as faint as $\rm M_{AB} \approx -8$, or equivalently to a star formation rate (SFR) of approximately $10^{-4}$ \msunyr. Note that systems with such a low SFR are observed in the local Universe and that the UV LF at $z=0$ keeps rising (at least) dwon to this level \citep{Bothwell.2011}. Extrapolating these local constraints at high redshift is difficult but it is worth pointing out that \cite{Weisz.2014} have been able to reconstruct the very faint-end of the UV LF at $z=3-5$ down to $\rm M_{AB} \approx -5$ using the star formation histories of Local Group dwarf galaxies, showing no evidence for a break of the LF. 

Predictions from semi-analytic models typically do not go much fainter than $\rm L_{\lya} \approx 10^{40}$ erg s$^{-1}$ or $\rm M_{AB} \approx -13$, often because of the mass resolution of the parent N-body simulation (e.g. \citealt{Garel2015a, Gurung2019}). Alternative models, based on the Extended Press-Schechter formalism, predict that the UV LF at $z\approx4$ keeps rising until UV magnitudes of $\approx -8$, corresponding to galaxies hosted by haloes at the atomic cooling limit \citep{Yung.2019}. Beyond the atomic cooling limit, it is unclear whether the processes that regulate galaxy formation are strong enough to break the slope of the LF, especially at redshifts 3$-$4 when the fully ionized intergalactic medium easily resists the gravitational pull of low-mass dark matter haloes (e.g. \citealt{Okamoto2008}).

To understand how our model is affected by the uncertainty on how faint the power law behavior of the Ly$\alpha$ LF extends, we show in Fig.~\ref{fig:SBLF} the results of our model when we ignore the contribution of galaxies fainter than $L_{\rm Ly\alpha} = 10^{38}$ \ergslum\ (red symbols) and $\rm 10^{37}$ \ergslum\ (green symbols). With these cuts and a slope $\alpha=-1.84$, the expected SB due to undetected LAEs is in the range 30-70\% of the detected emission and thus remains an important source of luminosity. Assuming a steeper slope ($\alpha=-1.94$), we see that even when the LF is truncated at $L_{\rm Ly\alpha} = 10^{38}$ \ergslum, the signal from undetected LAEs may fully explain our observations for most groups. 
Alternatively, we clearly see from Fig.~\ref{fig:SBLF} that a slope shallower than $\alpha = -1.74$ would only reproduce 10 to 30\% of the diffuse \lya\ flux, regardless of the value of the low luminosity cut. In this case, the contribution of undetected LAEs is unable to account for the diffuse emission, unless a very strong clustering is assumed (i.e. a very high $\zeta$).
Our model with a sharp cut of the LF below a given luminosity is of course unrealistic as one can expect a smooth transition between the power law behavior and the LF decline. Unfortunately we have no observational clue for the \lya\ LF shape at such low luminosity, but one can point out that using a more realistic LF shape will simply increase the required steepness of the LF.

\subsubsection{\galics\ semi-analytical model} 
\label{subsec:galics}
The arguments developed in the previous section demonstrate that the faint, undetected LAEs can be sufficiently numerous to contribute most or even all of the observed diffuse \lya\ emission.
We now use \galics, a semi-analytic model of galaxy formation coupled to numerical \lya\ radiation transfer models \citep{Hatton2003, Garel2012, Garel2015a}, to test this conclusion with a more elaborated model.
In the following we give a brief description of the method used to generate mock \lya\ narrow band images. A full description of the model is given in Appendix~\ref{app:galics}.

\galics\ relies on a cosmological N-body simulation to follow the hierarchical growth of dark matter structures and on semi-analytic prescriptions to describe the physics of the baryonic component.
We generate 100 mock lightcones that mimic the geometry and redshift range of the MUSE HUDF survey. We then reproduce the detection method described in Section~\ref{sect:over_method} to identify LAE overdensities in the mock fields. In total, this yields 13253 mock groups, of which 2475 have $\delta \ge 2$ and $N_{\rm LAE} \ge 7$.

In the previous section we have shown that a steep faint end slope ($\alpha \lessapprox -1.84$) of the \lya\ LF is required to explain the observed diffuse \lya\ emission. The model requires integrating the LF down to very low halo luminosity: $10^{37}$-$10^{38}$ \ergslum. However, the current dark matter mass resolution of the cosmological simulation used in \galics\ is $2 \times 10^9$ \msun. This mass cut corresponds approximately to a \lya\ luminosity of $10^{40}$-$10^{41}$  \ergslum, depending on the redshift. To overcome this limitation, we produce a large number of low luminosity ad-hoc LAEs by extrapolating each group LF down to $10^{37}$ \ergslum. The Schechter LF model already presented in \ref{subsec:schechter} with $\alpha = -1.84$ is used.
Note that from the recipes of star formation in \galics\ presented in \ref{subsec:galics_slope}, we predict that the faint-end slope of the \lya\ LF should be about -1.8.
These galaxies are then spatially placed to produce a given level of angular clustering (i.e. the factor $\zeta$ in Eq.~\ref{eq:SB_LF}). A Gaussian kernel with a width of 0.1 arcmin is used to randomly place the ad-hoc sources around \galics\ LAE in the mock fields (see Appendix~\ref{app:galics} for a detailed description).

By construction, the spatial extent of \lya\ sources is not modeled in \galics. In order to build a simulated \lya\ narrow band image we proceed as follows: 

An empty image with the \mxdf\ field of view is created.
For each source in the selected \galics\ group catalog, we derive a spatial profile using the galaxy-halo decomposition performed by \cite{Leclercq2017}. Surface brightness is modeled as a sum of two circular, 2D exponential profiles. Statistical information and/or correlation with some measured properties is also available from the \galics\ output, and are used to constrain the four model parameters. Details of the model can be found in Appendix~\ref{app:halomodel}. 

The simulated \lya\ image is then convolved with the \mxdf\ Moffat PSF model and  
Gaussian noise is added using the \mxdf\ datacube variance values summed over the corresponding spectral window. To take into account the increased noise with respect to the propagated values, as measured  in Section~\ref{subsec:noise}, a factor 2 is applied to the noise standard deviation.
In addition, a Gaussian spatial filter with a FWHM of one spaxel is performed on the noisy image to simulate the noise correlation present in the datacube.
This image is finally processed using the same detection parameters as the one used in Section \ref{sect:wavelet}. 

We present two examples of simulated groups in Fig.~\ref{fig:simufil}. These examples have been selected to be roughly representative of our observations, with overdensity values of 3 and 6, redshifts of 3 and 5 and filament \lya\ extended flux of ${\rm \approx 2 \times 10^{-17}}$ \ergsline.

Figure~\ref{fig:simufil} shows that the overall distribution of faint sources is well captured by the detection algorithm. One can see the power of the method by comparing the noisy \lya\ narrow band image in the third column with the wavelet decomposition in the right panel.  While it is hard to see more than the bright sources in the narrow band, the extended emission is clearly apparent in the low frequency IUWT images. 

We note that the results obtained for these two show-case structures are illustrative of the success but also of the limitation of the detection method.  In both cases, the algorithm appears sufficiently powerful to identify a diffuse emission structure when there is indeed one, even if it is hardly or not at all visible by-eye in the data. However, the second emission is so faint that the estimated SNR (see Table \ref{tab:simufil}) may only be marginally higher than the SNR that the algorithm would attribute to structures identified by mistake in the noise. In effect, if similar structures and SNR were obtained for some groups of the MXDF datacube, their associated P-values would be 0.3\% and 40\% respectively (see Fig.~\ref{fig:pfa}). Hence, in a noise-only situation, the SNR  would be found to be larger than the SNR of the second show-case 40\% of the time, showing that such a source is close to the detection limit. Note, however, that this reasoning assumes that the relation between P-values and SNR shown in Fig.~\ref{fig:pfa} is also valid for the \galics\ simulated narrow bands, which is probably pessimistic given the difference in noise properties between the real and the simulated data\footnote{This difference between the simulation and our observations could be explained by the idealized nature of the noise properties in the simulation. Although care was taken to properly scale the noise to reach the variance level observed in the data (see Section~\ref{subsec:noise}), the Gaussian noise injected in these simulations is different from the actual distribution of the noise. In particular, the imperfect continuum subtraction may leave behind observed narrow band \lya\ image systematics, which are not present in the simulated narrow bands.}.
The existence of such a limit is indeed inherent to any detection method. These remarks help to interpret the results of Fig.\ref{fig:diffuselya}, where many groups show diffuse emission that (if real) resembles the examples of the two show-cases of Fig.~\ref{fig:simufil}.

In Figure~\ref{fig:simustat} we show the derived filament parameters obtained when running the detection process on the 2475 \galics\ groups with $\delta \ge 2$ and $N_{\rm LAE} \ge 7$.
The simulations present a large scatter in the filaments properties. A large part of it is due to field-of-view effects. For statistical reasons we have used the large \mosaic\ field to search for overdensities, but depending on the relative location of the filaments with respect to the smaller \mxdf\ field, one might miss part of the structure. This is obvious in Fig.~\ref{fig:simufil} where one can see that moving the \mxdf\ location will eventually lead to no filament detection.

Compared to the \galics\ simulation, the observations (black labels in Fig.~\ref{fig:simustat}) are located in the upper tail of the distribution of the simulated filament properties. This is particularly true for the diffuse \lya\ flux and surface brightness (second and right panels in Fig.~\ref{fig:simustat}). As we can see from the figure, in most cases, the simulation is able to reproduce the observed \lya\ surface brightness in diffuse areas ($\rm SB_{dif}$) only in very overdense regions ($\rm \delta > 10$).

Fig.~\ref{fig:simufil} indicates that the Ly$\alpha$ haloes surrounding the
numerous faint LAEs in each overdensity should significantly overlap, leading to
a projected covering factor of unity over the entire area. We recall that in our
previous study of the incidence rates of Ly$\alpha$ emission
\citep{Wisotzki2018} we found that without spatial clustering, a surface
brightness level of $5\times 10^{-20}$~cgs corresponds to an incidence rate of
the order of 0.5 per unit redshift, comparable to that of high column density
\ion{H}{i} absorbers. In the overdense regions considered here this incidence
rate presumably increases further. It is therefore plausible (but unfortunately
untestable without bright background sources) that each line of sight through
these overdensities would also reveal strong \ion{H}{i} absorption.

The median surface brightness estimated in the \galics\ simulations is \erglinesurf{1.7}{-20} at $\rm \delta < 5$, a value significantly smaller than the minimum surface brightness in the high confidence groups (\erglinesurf{3.7}{-20}).
Only 8\% of the simulated groups have a surface brightness brighter than this value. This is a factor of three below the 23\% (5/22) success rate of the observations which implies that the \galics\ model is producing a lower average \lya\ surface brightness. 
The ad-hoc Gaussian model we use to extrapolate \galics\ results and to distribute faint sources is likely responsible for part of this disagreement. 
Indeed, (brighter) galaxies are typically observed to have a power-law correlation function, steepening at small scale due to the 1-halo term (e.g \citealt{Hildebrandt.2009, Harikane.2017}). 
The Gaussian kernel we use smooths out the 1-halo term at scales below $\approx$0.1 arcmin which results in spreading the luminosity of faint galaxies too widely.

Note also that our \galics\ mock catalogs are based on many other approximations, including an extrapolation of the \lya\ LF by 3 orders of magnitude and an empirical model of galaxy/halo decomposition. 
It is beyond the scope of the present paper to present a self-consistent physical model compatible with our observations. 
Nevertheless, despite these approximations, the analysis of the mock data confirms that faint LAEs cannot be ignored in the photon budget and that they can produce a significant fraction or even most of the observed diffuse emission.


\begin{table*}
\caption{Example of filament detection in simulated \galics\ overdensities.}             
\label{tab:simufil}
{
\begin{tabular}{rrrrrrrrrr}
 ID & z & $\mathrm{\delta}$ & $\mathrm{F_{cat}}$ & $\mathrm{F_{fil}}$ & $\mathrm{F_{dif}}$ & $\mathrm{SN_{dif}}$ & $\mathrm{S_{dif}}$ & $\mathrm{SB_{dif}}$ & $\mathrm{F_{dif} / F_{fil}}$ \\
 \hline
2563 & 3.10 & 2.8 & 133.3 & $113.2 \pm 0.7$ & $20.0 \pm 0.6$ & 15.5 & 670 & 3.0 & 17.7 \\
12195 & 4.96 & 6.5 & 80.2 & $33.2 \pm 1.4$ & $23.2 \pm 1.4$ & 8.8 & 611 & 3.8 & 69.9 \\

\end{tabular}
}
\tablefoot{
ID: \galics\ group ID.
z: redshift. 
$\mathrm{\delta}$: overdensity factor.
$\mathrm{F_{cat}}$: Total \lya\ flux from catalog.
$\mathrm{F_{fil}}$ and $\mathrm{F_{dif}}$: \lya\ flux measured in the filaments and the diffuse components.
$\mathrm{SN_{dif}}$: SNR of diffuse \lya\ emission.
$\mathrm{S_{dif}}$: Area of the diffuse component ($\mathrm{arcsec^2}$). 
$\mathrm{SB_{dif}}$: Average surface brightness of the diffuse \lya\ emission (\erglsurf{-20}). 
$\mathrm{F_{dif} / F_{fil}}$ : Fraction of flux in the diffuse component in \%.
Flux unit are $\mathrm{10^{-18}}$\ergsline.
}
\end{table*}

\begin{figure*}[]
\begin{center}
\includegraphics[width=0.80\paperwidth]{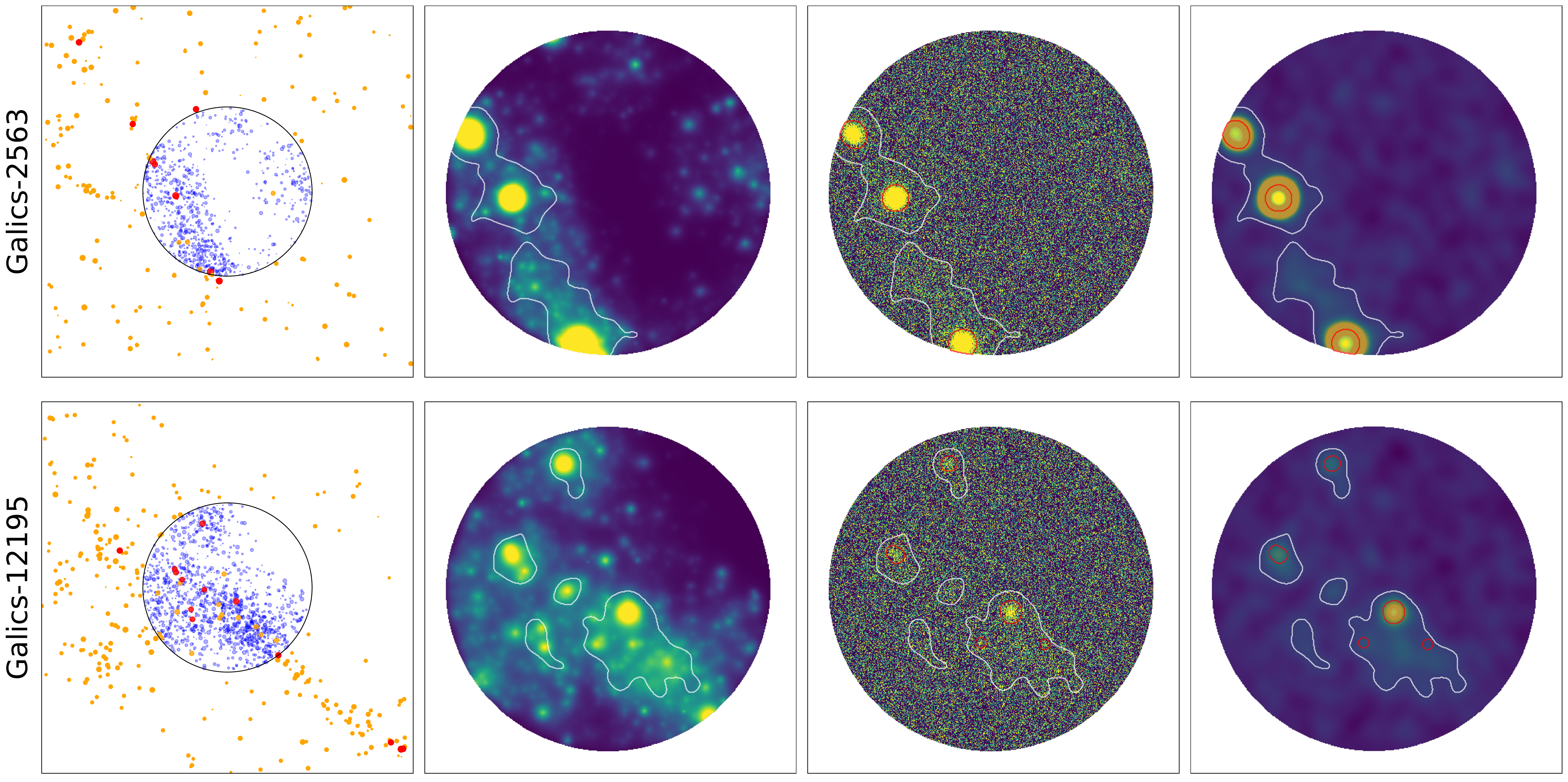}
\caption{Two examples of \galics\ simulations of diffuse \lya\ emission in overdensities. Top row: ID 2563  ${\rm \delta=2.8}$, z=3.10. Bottom row: ID 12195 ${\rm \delta=6.5}$, z=4.96.
First column: distribution of simulated galaxies within the \mosaic\ and \mxdf\ field of view. Bright \laes\ with $\mathrm{F_{\lya} > 10^{-18}}$ are marked with red symbols and those with $10^{-19} < \mathrm{F_{\lya} < 10^{-18}}$ with orange symbols. In the \mosaic, the limit for bright \laes\ is set to $10^{-17.5}$. Flux units are \ergsline. Low luminosity sources extrapolated from the LF (see text) are shown in blue.
Second column: Corresponding noiseless \lya\ narrow band image. Third column: Simulated \mxdf\ narrow band image with noise. Fourth column: Composite of low and mid frequencies IUWT wavelets SNR components. The contours of the identified filament and compact source area are shown respectively in white and red. The \mxdf\ field of view has an 82 arcsec diameter or 2.6 cMpc at z=3.}
\label{fig:simufil}
\end{center}
\end{figure*}

\begin{figure*}[]
\begin{center}
\includegraphics[width=0.9\paperwidth]{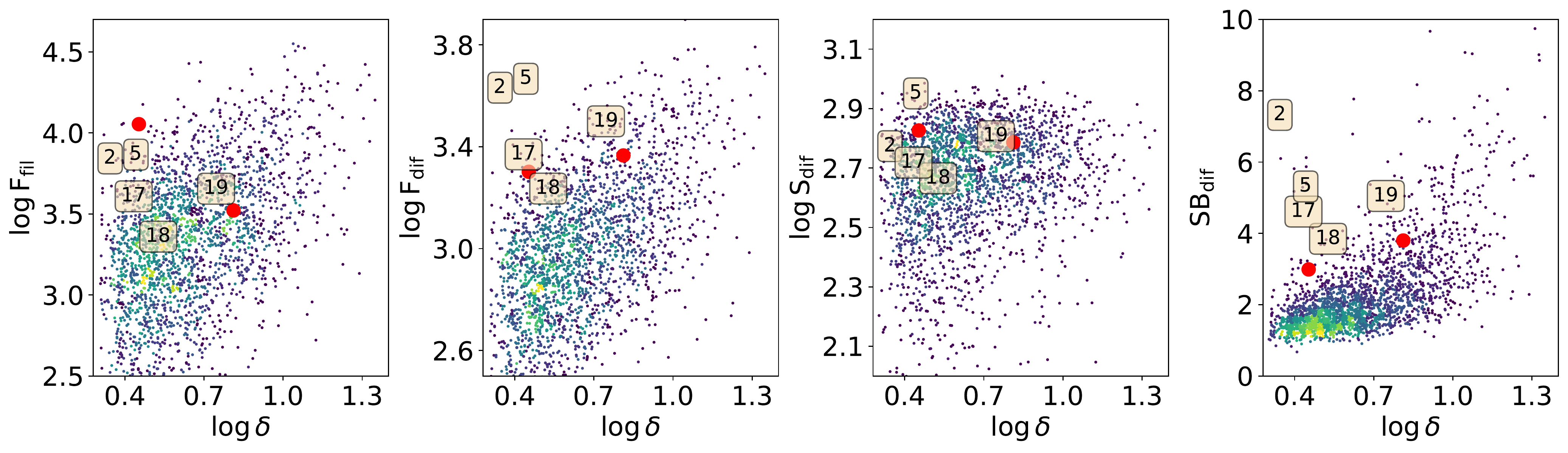}
\caption{Filament properties derived from \galics\ mocks. From left to right we show, as a function of the overdensity ($\delta$), the filament total and diffuse \lya\ flux in units of $10^{-20}$\ergsline, the area (arcsec$^2$) and surface brightness ($10^{-20}$\ergsurfb) of the diffuse component. Color coding correspond to the number of points in the 2D histogram bins with increasing values from blue to yellow. The group IDs for the corresponding observed filaments properties are shown as black labels. The red points display the 2 show cases presented in Fig.~\ref{fig:simufil} and Table~\ref{tab:simufil}.}
\label{fig:simustat}
\end{center}
\end{figure*}

\subsection{Gravitational heating}
\label{subsec:cold}

Gravitational compression of gas during structure formation is a net heating term which may be dissipated through the emission of \lya\ radiation. In the low density IGM, the thermal state of the gas is determined by a competition between cooling and heating both from gravitational compression and from the UVB \citep[e.g.][]{Hui1997}. At values of the density contrast\footnote{We refer here to the overdensity in the density field, not to the overdensity in the number of LAEs as used elsewhere in the paper.} typical of large-scale filaments ($\approx$10), the temperature of the gas is mostly the result of the competition between photo-heating from the UVB and radiative cooling (through \lya\ emission). From e.g. \citet[][their Fig. 9]{Rosdahl2012}, we see that the \lya\ emissivity of gas in this regime (i.e. at densities $<0.01$ cm$^{-3}$) is well described by a steep powerlaw $\propto n_H^{2.5}$. 
That the slope is steeper than $n_H^2$ is due to the fact that the ionized fraction rises towards low densities. Thus, at low densities, the joint contribution of the UVB and gravitational compression to \lya\ emission drops rapidly and is not likely to contribute significantly to the emission we observe.

Within dark matter halos, or in their immediate vicinity, part of the IGM has condensed into denser cold accretion streams \citep[e.g.][]{Keres2005}. These streams can produce \lya\ emission by dissipating their gravitational energy \citep{Fardal2001,Dijkstra2009,Faucher2010,Rosdahl2012}. In this regime, gravitational heating largely dominates over fluorescence from the UVB \citep[][their Fig. 6]{Rosdahl2012}. This mechanism has been suggested as a possible source of the bright \lya\ extended emission observed in \lya{} blobs (e.g. \citealt{Haiman2000,Fardal2001,Nilsson2006, Rosdahl2012}). For example, in their study of the R0-1001 \lya\ nebulae, \cite{Daddi2020} conclude that the gravitational energy associated with gas infall is the most likely source of power for the observed extended \lya\ emission ($\mathrm{1.3 \times 10^{44}}$ \ergslum). 

The same mechanism should also occur in less massive galaxies when cold gas falls into the potential wells of their dark matter halos. This may explain part of the extended \lya{} emission observed around most LAEs with MUSE \citep{Wisotzki2016,Leclercq2017}. Again, by construction, the diffuse \lya{} emission we measure excludes the CGM of all LAEs detected in the field, and we can thus rule out that cooling radiation is responsible for the diffuse emission we report.

\section{Source of diffuse \lya\ emission}
\label{sect:lyasource}

In the previous section, we discussed three possible sources of energy that could power the diffuse \lya\ radiation we observe: (1) fluorescence from the UVB (2) \lya\ emission from an abundant population of undetected small \laes, and (3) dissipation of gravitational heating.

We find that \lya\ fluorescence powered by the cosmic UV background can explain at most 28-34\% of the observed signal at $z\approx 3$, and less than 10\% at $z\approx 4.5$ (Sec. \ref{subsec:uvb}). Even if our estimate is relatively uncertain 
(e.g. we use the maximum value of 1 for the LLS covering fraction as in \citealt{Cantalupo2005} and
we neglect pumping from other lines which may enhance the signal by $\approx$20\%, \citealt{Furlanetto2005}), these results suggest that this process is not responsible for the bulk of the emission we observe. Perhaps more importantly, we note that \lya\ fluorescence from the UVB is mostly produced by relatively dense gas, with column densities typical of Lyman Limit Systems (LLS) or larger. This gas is mostly located in the CGM of galaxies (or within their dark matter haloes, e.g. \citealt{Vandevoort2012}) and its \lya\ emission may thus be associated with galaxies in \lya\ surveys, in the form of \lya{} haloes. Moreover, while increasing the intensity of the local UVB will increase the overall emissivity of this dense gas, it will also decrease the emissivity of lower-density gas (at $\rm n_H < 0.05 \, cm^{-1}$, see \citealt{Rosdahl2012}, their Fig. 16), thus pushing the emission even closer to the galaxies. It is thus not clear whether fluorescence from the UVB produces a signal which is distinguishable from that produced by galaxies and their CGM.

Similarly, we have argued that gravitational heating will lead to dissipation through \lya{} emission only in the densest parts of the intergalactic medium, in the CGM. This emission may contribute to the extended \lya{} haloes around LAEs, in addition to fluorescence from the UVB or local star formation, to scattered \lya{} photons from galaxies, or to extended star formation (e.g. \citealt{MasRibas2017}). For more diffuse gas, the heating source is a combination of UVB and gravitational heating, which cannot be disentangled, but which has been measured in simulations to produce extremely faint emission, decreasing rapidly with decreasing density \citep{Rosdahl2012}. Thus, it appears that fluorescence or cooling radiation, while they may play a significant role in lighting up the CGM of galaxies, are not able to explain the levels of extended emission that we detect beyond this CGM.

We therefore find that the most likely explanation for the diffuse emission we observe is an abundant population of ultra-faint undetected LAEs (section \ref{subsec:undetlae}). This result is in line with the recent findings of \citet{Mitchell2020} who use high-resolution radiation-hydrodynamical simulations to show that the very extended part of \lya\ haloes around galaxies is mostly due to undetected neighbouring galaxies. 
Our conclusion relies on a number of assumptions: the faint-end slope of the \lya{} LF must be steep enough ($\alpha \lessapprox -1.84$), this power-law behaviour must extend down to at least $\sim 10^{38} - 10^{37}$ \ergslum\ before turning over and the clustering of faint \laes\ must be favourable (filling factor $\sim 1/6$). While these assumptions seem reasonable, there are today no observational constraints on the \lya\ LF shape below $10^{40.5}$ \ergslum\ or on the angular clustering of extremely faint \laes.  Our observations may provide a new original and indirect constraint on this population of extremely faint LAEs.

To appreciate the nature of the extremely faint objects which may contribute to the observed signal, we remind the reader that a single massive star of 40 $\rm M_\odot$ will produce of order $10^{49}$ ionizing photons per second for $\approx$6 Myr (e.g. \citealt{Geen2018}). Assuming these photons are all processed to yield \lya\ through case B recombination, this single star would produce $\sim 10^{38}$\ergslum\ in the \lya\ line. The detailed simulations of molecular clouds by \cite{Kimm2019} show a more complete picture. These authors predict that a cloud of total mass $\rm 10^5\ M_\odot$ with $\rm 10^4\ M_\odot$ in stars has a \lya\ luminosity which evolves from $3\times 10^{39}$ \ergslum\ at the onset of star formation to $5\times 10^{36}$ \ergslum\ when the cloud is disrupted 6~Myr later. A cloud of the same mass, but which forms ten times less mass in stars (i.e. with stellar mass $\rm 10^3\ M_\odot$) is found to have a \lya\ luminosity evolving from $6\times 10^{38}$ to $3\times 10^{35}$ \ergslum\ over 20~Myr. In both cases, the \lya\ luminosity of the cloud drops faster than the production of ionizing photons by the stellar population because the disruption of the cloud allows ionizing photons to escape without being reprocessed into \lya. Systems with \lya\ luminosities as faint as $10^{37}-10^{38}$ \ergslum\ thus require very few massive stars to produce such luminosities. While some of these objects may be explained by stronger star formation events seen at later stages (when the ionizing radiation has dropped and neutral gas has been blown away), or by very low \lya\ escape fractions, it is likely that sources with $\rm L_{\lya}\sim 10^{37-38}$ \ergslum\ are indeed extremely small systems, with stellar masses as small as a few thousand solar masses, which are seen at the very first moments of their formation.  These could be nascent galaxies or even (compact) star clusters, which have been observed at similar implied continuum magnitudes at higher redshifts (e.g. \citealt{Boylan2017, Vanzella2018, Vanzella.2020}), although currently we cannot differentiate between the two possibilities.

While assessing the baryonic content of these ultra-faint LAEs is highly uncertain, we do not expect that these objects make a significant contribution to the global stellar mass budget. By extrapolating the SFR-M$^\star$ relation at $z \approx 3-4$ (e.g. \citealt{Salmon2015}) and assuming our fiducial LF slope $\alpha=-1.84$, we estimate that LAEs with $10^{38} < L_{\rm Ly\alpha} < 10^{42}$ \ergslum\ should only contribute $\lesssim 10-20$ \% to the cosmic stellar mass density.

The existence of a large number of faint galaxies has also some implications for synthesis models of the UVB. One of the key ingredients of these models is the total emissivity of galaxies, which scales with the integral of the UV LF down to faint luminosities : $0.01 L_{\star}$ \citep{Haardt2012} or $M_{AB} = -13$ \citep{Faucher2020}. Extending these integrals down to $M_{AB} \sim -8$ as suggested above would increase the global UV emissivities of these models by $\sim 50\%$ \citep{Haardt2012} and $\sim 20\%$ \citep{Faucher2020}. While these models can probably absorb this difference by re-adjusting their free parameters to reproduce the same observed values of $\Gamma_{HI}$, it is interesting to note that FG20’s model requires values of the escape fractions of ionising radiation as low as $1\%$ (at $z=3$), which is already in tension with observational results \citep{Steidel2018} and numerical works (e.g. \citealt{Rosdahl2018}). Increasing the total emissivity of galaxies by extending the LF to the faint-end makes this situation yet more uncomfortable. If confirmed, the very faint population of LAEs that we discussed above may thus require more important adaptations of these models.

Note that our analysis assumes that the Ly$\alpha$ emission from the CGM is counted as part of each galaxy's luminosity. Indeed, while previous measurements of the \lya\ LF were restricted to the \lya\ flux from galaxies, and were thus missing a large part of the \lya\ flux (70\% on average, \citealt{Leclercq2017}), the latest measurements with MUSE carefully include the halo flux (e.g. \citealt{Drake2017, Herenz2019}). Given the difficulty of measuring the extended \lya\ flux at large galactocentric distance, the \lya\ flux in the CGM may still be underestimated. This uncertainty affects the requirement stated above on the shape the LF should have to reproduce our observations.  

Including the CGM \lya\ emission in each galaxy's luminosity makes sense in a scenario where this emission is due to scattered light from the central galaxy. In that case, the total \lya\ luminosity better relates to the star formation rate. This is a strong assumption, however, which is still much debated in theoretical work \citep[e.g.][]{Mitchell2020,Byrohl2020}. It is interesting that the other sources of energy (UVB or gravitational compression) are also expected to produce most of their \lya\ emission in the CGM. In that respect, these different processes are not strictly additive. Indeed, it is likely that the signal we detect does include a contribution from these terms, and that this contribution is present mostly in the CGM of faint galaxies, which we account for in the LF regardless of its physical origin.

Finally, we note that our results do not confirm the conclusion of \cite{Elias2020} who have predicted that diffuse \lya\ emission from cosmic web filaments at z$\approx$3 and {\em restricted to the IGM}, should be easily detected by MUSE in a 30 hour deep exposure. Our measurements within the \mxdf, a 5 times deeper exposure than the one foreseen in their analysis, confirm that their prediction, as already pointed out by the authors, was overly optimistic. Moreover, we show that the majority of the diffuse flux is due to the \lya\ emission within the CGM of undetected galaxies and that only a small fraction can be coming from the IGM proper.

\section{Summary and conclusions}
\label{sect:conclusion}

We introduced and analyzed observations carried out with MUSE in the \mxdf, a single, 140 hour deep field located in the HUDF area, complemented by MUSE observations of the entire HUDF area at 10 hours depth.  This resulted in datacubes of exquisite sensitivity and a catalog of 1258 \laes\ covering redshifts 2.9 to 6.7. We analyzed this unique dataset to (i) detect and quantify \lae\ overdensities and (ii) search and characterize diffuse \lya\  emission within each detected overdensity. Our major findings are summarized below:

\laes\ are strongly clustered in redshift space, with 30\% of the 1258 \laes\ residing in 22 overdensities spanning redshifts 3.0 to 5.8. The overdensity, measured within volumes of  260 cMpc$^3$ ($\mathrm{z = 3.0}$) to 415 cMpc$^3$ ($\mathrm{z = 5.8}$),  is on average 3.2 and reaches 5.0 for the densest groups at $\mathrm{z = 4.5}$ and $\mathrm{z = 4.7}$.
LAEs in overdensities have a low average \lya\ luminosity ($10^{41.5}$\ergslum) and their mean number density is $\mathrm{0.055 \pm 0.017 \, cMpc^{-3}}$.

SED photometric analysis of the 67\% of the LAEs with an HST counterpart shows that the galaxy population in overdensities is mainly composed of low mass ($\mathrm{1.4 \times 10^{8} \, M_\odot}$), young (0.3 Gyr) galaxies with high specific star formation rates ($\mathrm{10^{-8.5} \,yr^{-1}}$).
The majority of the remaining LAEs without an entry in the \cite{Rafelski2015} photometric catalog display very high \lya\ equivalent widths and no visible HST counterpart. On average, these very faint ($\mathrm{M_{UV} \sim -15}$) star forming galaxies must have stellar mass below $\mathrm{10^{7} \, M_\odot}$.
Overdensities are expected to be populated by dark matter halos with masses around $\mathrm{10^{11.3} M_{\sun}}$, with the notable exception of the $z$=3.07 group 2 for which we estimate a halo mass of $\mathrm{10^{13.5} M_{\sun}}$.

A search for diffuse \lya\ emission within the \mxdf\ area with an original method based on multiscale undecimated isotropic wavelet transforms, resulted in the identification of 14 overdensities with extended \lya\ emission.
This extended \lya\ emission arises from filamentary structures of cMpc size and covers an area of $\mathrm{0.4 - 1.1 \, cMpc^2}$, corresponding to a significant fraction (10-20\%) of the total \mxdf\ area.
Group 2 at $z$=3.07 is the only group displaying a filament outside the \mxdf\ field, in an extended area covered with MUSE at 10 hours depth. This additional detection extends the \mxdf\ filament length to 4.6 cMpc, and reveals a second filament crossing the field. 

We use the mid and large spatial scale segmentation images resulting from the wavelet decomposition process to split filaments into areas corresponding to compact sources and diffuse emission. We show that the compact source area can account for the total \lya\ emission of identified galaxies, including the emission from their CGM. 
Using Monte Carlo simulations we show that 5 overdensities among the 14 with extended \lya\ emission display an extended diffuse \lya\ emission signal with a mean surface brightness of \erglinesurf{5.1 \pm 1.2}{-20} and a high confidence level (P-value $<$ 0.06). The \lya\ luminosity from this diffuse area represents a large fraction of the total filament flux: $\mathrm{68 \pm 6 \,\%}$.

We have investigated the potential impact of AGN and concluded that, except for groups 2 and 6 where we cannot rule out a boost in \lya\ surface brightness, AGN are unlikely to be a significant contributor to the observed extended \lya\ emission in the 14 groups. This is especially true for 4 of the 5 overdensities with high confidence diffuse \lya\ emission.

At $z\approx$3, a maximum of 28\% and 34\% of the observed surface brightness of the diffuse emission can be explained by \lya\ fluorescence powered by the cosmic UV background. At higher $z$, this fraction is reduced to below 10\%.

The measured diffuse \lya\ surface brightness can be reproduced by a population of undetected ultra low luminosity \laes, provided that the faint end of the \lya\ LF is steep enough ($\alpha \lessapprox -1.84$), that it extends to luminosities lower than $10^{38} - 10^{37}$ \ergslum\ and that the clustering of faint \laes\ is significant (filling factor $< 1/6$).

\bigskip

Narrow band \lae\ wide field surveys, followed-up by spectroscopic multi-object observations, have produced a wealth of information on LAE clustering and the large scale structure of galaxies at high $z$ (e.g. \citealt{Francis2004, Matsuda2005, Zheng2016, Ouchi2020}). They cover a large volume: e.g. $\mathrm{200 \times 200 \times 80 \, Mpc^3}$ for the SILVERRUSH survey \citep{Ouchi2017, Shibuya2017}, but at the expense of a bright limiting luminosity (typically $\sim 10^{43}$\ergslum)
and sparse sampling (e.g. 179 LAEs for the \citealt{Harikane2019} study at $z$=5.7 and 6.6, that is $\mathrm{5.6 \times 10^{-5} Mpc^{-3}}$).
With its two orders of magnitude higher average number density of LAEs (e.g. $\mathrm{6.3 \times 10^{-3} Mpc^{-3}}$ at z=6), our study for the first time probes the large scale structure at high $z$ with a much finer sampling. Thanks to this zoomed view of the cosmic web, we can resolve the filamentary structure in unprecedented detail. This is very complementary to the classical wide field approach.

Using both a simple analytical LF model and the \galics\ semi-analytical model, we show that our measurements imply that a large population of ultra low luminosity \laes\ ($<10^{40}$ \ergslum) powers the diffuse \lya\ emission within the filaments. The \lya\ luminosity of this ultra faint population corresponds to star formation rates smaller than $\mathrm{< 10^{-4} M_\odot yr^{-1}}$.
An important consequence of these results is that it is unclear whether intergalactic (as opposed to circumgalactic) gas  is even observable through its \lya\ emission, because the denser regions where this emission is most likely to be detected will be crowded with a high surface density of small \lya\ emitting galaxies. 
Even if the study of IGM proper in emission could be problematic, 
understanding how star formation proceeds in these dwarf galaxies, down to what \lya\ luminosity the LF extends, what their clustering properties are and what the relative importance is of the various processes that produce the \lya\ emission within the CGM and IGM, will be very valuable to understand galaxy formation.
Last - but not least - the population of faint \laes\ appears to be a good tracer of the cosmic web.

This first detection of the cosmic web structure in \lya\ emission in typical filamentary environments, i.e. outside massive structures typical of web nodes, is a milestone in the long search for the cosmic web signature at high $z$. This has been possible because of the unprecedented faint surface brightness of \erglinesurf{5}{-20} achieved by 140 hour MUSE observations on the VLT. However, because of the limited size of the \mxdf, we cannot trace the full length of the filaments, which extend at least to the Mpc (physical) scale as demonstrated in the case of the $z$=3.07 filament (Fig.~\ref{fig:filgroup2}).
Repeating this study on a larger field and different environments would be very valuable to increase the detection statistics and to provide better constraints on the filaments' physical parameters. This remains, nevertheless, a costly investment in telescope time.  In the longer term, the BlueMUSE project \citep{Richard2019} will allow us to probe the \lya\ redshift range 2-4 with a larger field of view. The lower \lya\ redshift of BlueMUSE will be very beneficial for such a study. With a wavelength range mostly free of bright sky lines, a reduced impact of redshift dimming, and access to the peak of the cosmic star formation history, 
BlueMUSE should be able to perform similar studies in less telescope time and over larger fields of view.


\begin{acknowledgements}
We warmly thank ESO Paranal staff for their great professional support during all \mxdf\ GTO observing runs. We thank the anonymous referee for a careful reading of the manuscript and helpful comments. We also thank Matthew Lehnert for fruitful discussions.
RB, AF, SC acknowledge support from the ERC advanced grant 339659-MUSICOS.
JB acnowledges support by Fundação para a Ciência e a Tecnologia (FCT) through the
research grants UID/FIS/04434/2019, UIDB/04434/2020, UIDP/04434/2020 and through the Investigador FCT Contract No. IF/01654/2014/CP1215/CT0003.
TG, AV acknowledges support from the European Research Council under grant agreement ERC-stg-757258 (TRIPLE).
DM acknowledges A. Dabbech for useful interactions about IUWT and support from the GDR ISIS through the Projets exploratoires program (project TASTY). 
AF acknowledges the support from grant PRIN MIUR2017-20173ML3WW\_001.
SLZ acknowledges support by The Netherlands Organisation for Scientific Research~(NWO) through a TOP Grant Module~1 under project number 614.001.652.
This research made use of the following open-source software and we are thankful to the developers of these:  GNU Octave \citep{Eaton.2018} and its statistics, signal and image packages, the Python packages Matplotlib \citep{Hunter.2007}, Numpy \citep{Walt.2010}, MPDAF \citep{Piqueras2017}, Astropy \citep{Astropy.2018}, PyWavelets \citep{Lee.2019}.
\end{acknowledgements}

\bibliographystyle{aa}
\bibliography{mxdf_biblio.bib}

\begin{appendix}


\section{GALICS}

\label{app:galics}

\subsection{Model description}

The \galics\ model \citep{Hatton2003} presented in \citet{Garel2015a} is specifically designed to study the formation and evolution of galaxies in the high redshift Universe. It relies on a cosmological N-body simulations to follow the hierarchical growth of dark matter structures and on semi-analytic prescriptions to describe the physics of the baryonic component. The simulation was run in a box of $100$ $h^{-1}$ cMpc on a side and it assumes a standard cosmology which is consistent with the WMAP-5 results. The simulation contains $1024^3$ dark matter particles with an individual mass of $\approx8.5\times10^7\, M_\odot$. Haloes are identified if they contain at least 20 dark matter particles which corresponds to a minimum halo mass of about $10^9\, M_\odot$. 

The intrinsic \lya\ emission from galaxies is estimated from the production rate of hydrogen-ionizing photons (estimated from the stellar spectral energy distributions) assuming case B recombination. To compute the observed \lya\ properties of galaxies, \galics\ is combined with the library of radiative transfer simulations of \citet{Schaerer2011} which predicts the escape fraction of \lya\ photons through dusty galactic outflows \citep[see][]{Verhamme2008,Garel2012}. The \galics\ \lya\ luminosities used in the analysis presented in Section 5.2.2 therefore correspond to the dust-attenuated luminosities that emerge from the CGM of galaxies. As shown in \citet{Garel2015a,Garel2015b}, the model was tuned to reproduce various fundamental observational constraints at $z \approx 3-7$, including the UV and \lya\ luminosity functions, the stellar mass functions and the SFR to stellar mass relation.
 
For the sake of the present study, we follow the procedure of \citet{Garel2015b} to generate 100 mock lightcones that mimic the geometry and redshift range of the MUSE HUDF survey (that is, a square field of $\approx 9$ arcmin$^2$ and $3 \leq z \leq 6.7$). We then reproduce the detection method described in Section \ref{sect:over_method} to identify LAE overdensities in the mock fields. In practice, we bin the redshift distribution in slices of 8 cMpc, count the number of detected LAEs using a LAE flux detection limit of $10^{-18}$ \ergsline\ inside the MXDF and $10^{-17.5}$ elsewhere, and compute the overdensity $\delta$ corresponding to each group. In total, this yields 13253 mock groups, among which 2475 have $\delta \ge 2$ and $N_{\rm LAE} \ge 7$.

\subsection{Faint-end extrapolation of GALICS mocks with ad hoc LAEs}

Due to the limited mass resolution of the cosmological simulation, dark matter haloes less massive than $2 \times 10^9$ \msun\ are not detected in \galics. As a consequence, the number density of LAEs can be under-predicted at the faint-end. Our LAE sample is estimated to be incomplete below a \lya\ flux limit of $F_{\rm lim} \approx 2 \times 10^{-19}$ erg s$^{-1}$ cm $^{-2}$ \citep[see][]{Garel2015b}. 

In an attempt to correct for this effect, we add ad-hoc LAEs in our mock fields (for simplicity, the \mxdf\ is assumed to lie exactly at the center of the HUDF field). We do so by extrapolating the \lya\ LF between $L_{\rm lim}$, the \lya\ luminosity limit corresponding to $F_{\rm lim}$ at the redshift of the group, and a minimum \lya\ luminosity, $L_{\rm min}$. In practice, the \lya\ luminosity limit varies from $1.6\times10^{40}$ erg s$^{-1}$ at $z = 3$ to $1.1\times10^{41}$ erg s$^{-1}$ at $z = 6.7$. The choice of $L_{\rm min}$ is somewhat arbitrary because the \lya\ LF is unconstrained in this luminosity range. These considerations are beyond the scope of the present study, therefore we assume a fixed value of $L_{\rm min}=10^{37}$ erg s$^{-1}$ which is more than three orders of magnitude smaller than the faintest observed LAEs. Based on the canonical relation between \lya, star formation rate and UV luminosity \citep[e.g.][]{Garel2015a}, this value would translate into $SFR \approx 10^{-5}$ \msun\ yr$^{-1}$ and $M_{\rm UV} \approx -6$.

\begin{figure}
\includegraphics[width=0.8\columnwidth]{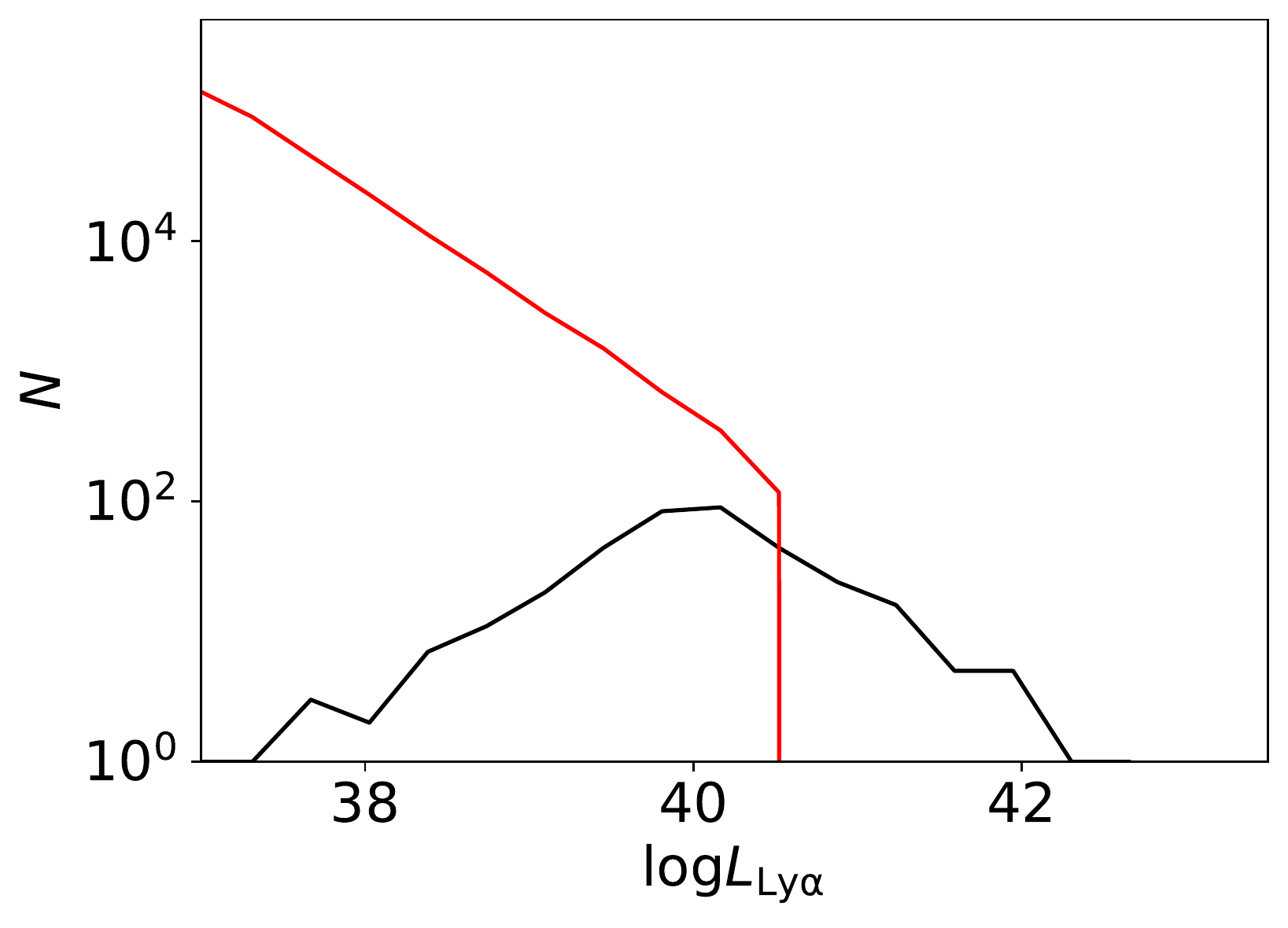}
\caption{Example of the distribution of \lya\ luminosities in a \galics\ group. The black and red histograms represent the number of true sources and ad hoc sources respectively. The luminosities of the ad hoc sources have been drawn from a power-law distribution with a slope $\alpha=-1.84$, in agreement with the \lya\ LF from \citet{Herenz2019}.}
\label{fig:galics_fig1}
\end{figure}

Using the best-fit parameters for the \lya\ LF at $3 < z < 6$ of \cite{Herenz2019}, log$\Phi_\star = -2.71$, log$L_\star = 42.66$, $\alpha=-1.84$, the number density of LAEs is given by the Schechter function, $\Phi(L)$, which can be expressed as $\Phi(L) \approx \frac{\Phi_\star}{L_\star} \left(\frac{L}{L_\star}\right)^\alpha$ for $L \ll L_\star$. Then, the extrapolated number density between $L_{\rm min}$ and $L_{\rm lim}$ is given by $n_{\rm ext} = \int_{L_{\rm min}}^{L_{\rm lim}}\Phi(L) dL = \frac{\Phi_\star}{(\alpha + 1)L_\star^{\alpha + 1}}(L_{\rm lim}^{\alpha+1}-L_{\rm min}^{\alpha+1})$. Then, we compute the total number of ad-hoc LAEs per group, $N_{\rm ext}$, as the product of $n_{\rm ext}$ by the comoving volume of the group, $V_{\rm gp}$, and the group overdensity, $\delta$. From there, we randomly draw $N_{\rm ext}$ luminosities between $L_{\rm min}$ and $L_{\rm lim}$ according to a power-law distribution with a slope $\alpha$. 

As an example, we present the result of this procedure in Figure \ref{fig:galics_fig1} which shows the luminosity distribution of LAEs in the mock GALICS group 12195 which is located at $\rm z=4.96$ and has an overdensity $\rm \delta = 6.5$. The black histogram corresponds to the number of \galics\ LAEs while the red curve shows the extrapolated distribution of ad hoc LAEs between $L_{\rm min}$ and $L_{\rm lim}$. In this case, there are 361 true sources in the initial \galics\ catalogue. As a result of the extrapolation between $L_{\rm min}$ and $L_{\rm lim}$, 328049 ad hoc sources are added to the catalogue. We note that \galics\ sources that are fainter than $L_{\rm lim}$ are discarded from the final catalogues such that we keep only true LAEs at $L > L_{\rm lim}$ and only ad hoc LAEs at $L \le L_{\rm lim}$.

\subsection{Expected faint-end slope index}
\label{subsec:galics_slope}

While \galics\ cannot reproduce the full population of very faint LAEs because of the mass resolution limit, galaxies with \lya\ luminosities $L_{\rm Ly\alpha} = 10^{37} - 10^{39}$ erg s$^{-1}$ do exist in the current simulation (see Fig.~\ref{fig:galics_fig1}). These are mostly located in young haloes at the mass resolution threshold that recently appeared in the simulation, while a smaller fraction ($\approx 25\%$) correspond to satellites sitting in more massive haloes. 
From scaling arguments, we expect that the faint-end slope index is of the order $-1.8$, which is consistent with the value assumed above.\\

The origin of this relation comes from the Kennicutt law for star formation used in \galics\ (see \citealt{Garel2012, Garel2015a}) which can be expressed as ${\rm SFR} \propto M_{\rm gas}^{1.4} / r_{\rm gal} ^{0.8}$ where $M_{\rm gas}$ is the gas mass reservoir in the galaxy and $r_{\rm gal}$ is the galaxy disc radius. Cooling operates on short timescales in low-mass haloes at high redshift, making the baryonic gas available for star formation before supernova feedback, and the $\lambda$ parameter that drives the conservation of angular momentum in DM haloes has a narrow distribution that peaks around 0.05 \citep{Mo1998}. Thus, assuming $M_{\rm gas}= f_{\rm b} M_{\rm vir}$ (where $f_{\rm b}$ is the universal baryonic fraction), $r_{\rm gal} = \lambda R_{\rm vir}$ (where $R_{\rm vir}$ is the halo virial radius) and $R_{\rm vir} \propto M_{\rm vir}^{1/3}$, the star formation rate is expected to scale with the halo virial mass as ${\rm SFR}  \propto  M_{\rm vir}^{1.13}$. \\

Linking the SFR with the \lya\ luminosity and neglecting the effect of dust in such faint galaxies (see Section~\ref{subsec:schechter}), we can write the \lya\ luminosity as $L_{\rm Ly\alpha}  \propto SFR \propto M_{\rm vir}^{1.13}$. With a low-mass function of haloes $\phi (M) dM \propto M^{-2} dM$, (for $\Lambda$CDM), this relation gives a low-luminosity LF : 

\begin{equation}
\phi (L_{\rm Ly\alpha}) dL_{\rm Ly\alpha}  \propto L_{\rm Ly\alpha} ^{-1.77}.
\end{equation}

So, the choice of connecting an analytical low-luminosity LF with a typical slope $-1.8$ to the \galics\ \lya\ LF at higher luminosity is consistent with the assumption on the SFR in the simulation and with expectations from first principles.

\subsection{Spatial distribution of ad-hoc LAEs}

Once the fake \lya\ luminosities have been computed, we then have to generate the spatial positions of our ad-hoc sources. On the one hand, the redshifts are simply drawn randomly between the minimum and maximum redshifts of the group. On the other hand, the celestial coordinates of ad-hoc sources are determined from the spatial clustering of \galics\ sources within the \mosaic. Here, we make the assumption that faint sources follow a clustering pattern similar to that of bright galaxies, rather than being purely randomly distributed. In practice, we start by making a projected map of \galics\ objects within the \mosaic\ and we apply a 2D gaussian smoothing to the image, assuming a standard deviation $\sigma$ of $0.1'$ . Although this value is arbitrary, it is justified by the fact that $0.1'$ corresponds roughly to the virial radius of the typical dark matter haloes in our \mosaic\ groups in the redshift range of our survey ($M_{\rm vir} \approx 10^{11}$ \msun; see Section 3.3). Then, the smoothed image is normalised such that it can be used as a 2D probability distribution function for the spatial sampling of ad-hoc LAEs. For each object, we draw random sky coordinates and implement a rejection method to sample our 2D distribution. We repeat this process until all ad-hoc sources in the field, $N_{\rm ext}$, have been successfully sampled. 

\begin{figure}
\includegraphics[width=0.8\columnwidth]{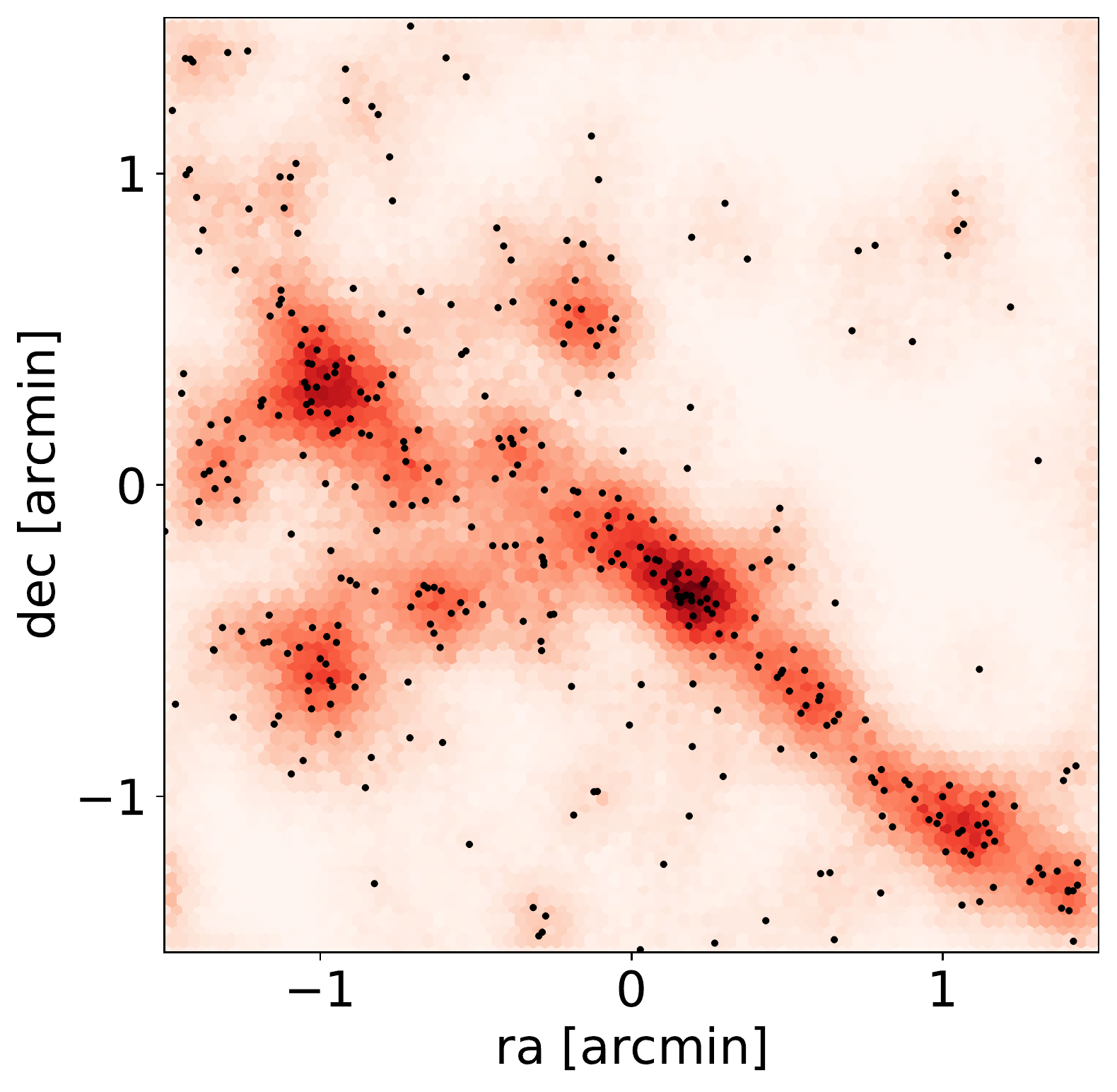}
\caption{Projected map of mock LAEs in the \mosaic. The black dots correspond to the true \galics\ sources. The red bins show the positions of the ad hoc sources that are used to extrapolate the number density of LAEs at the faint end. These ad hoc LAEs are distributed according to the spatial clustering of the true sources (see text).}
\label{fig:galics_fig2}
\end{figure}

\begin{figure}
\hspace{0.3cm}
\includegraphics[width=0.8\columnwidth]{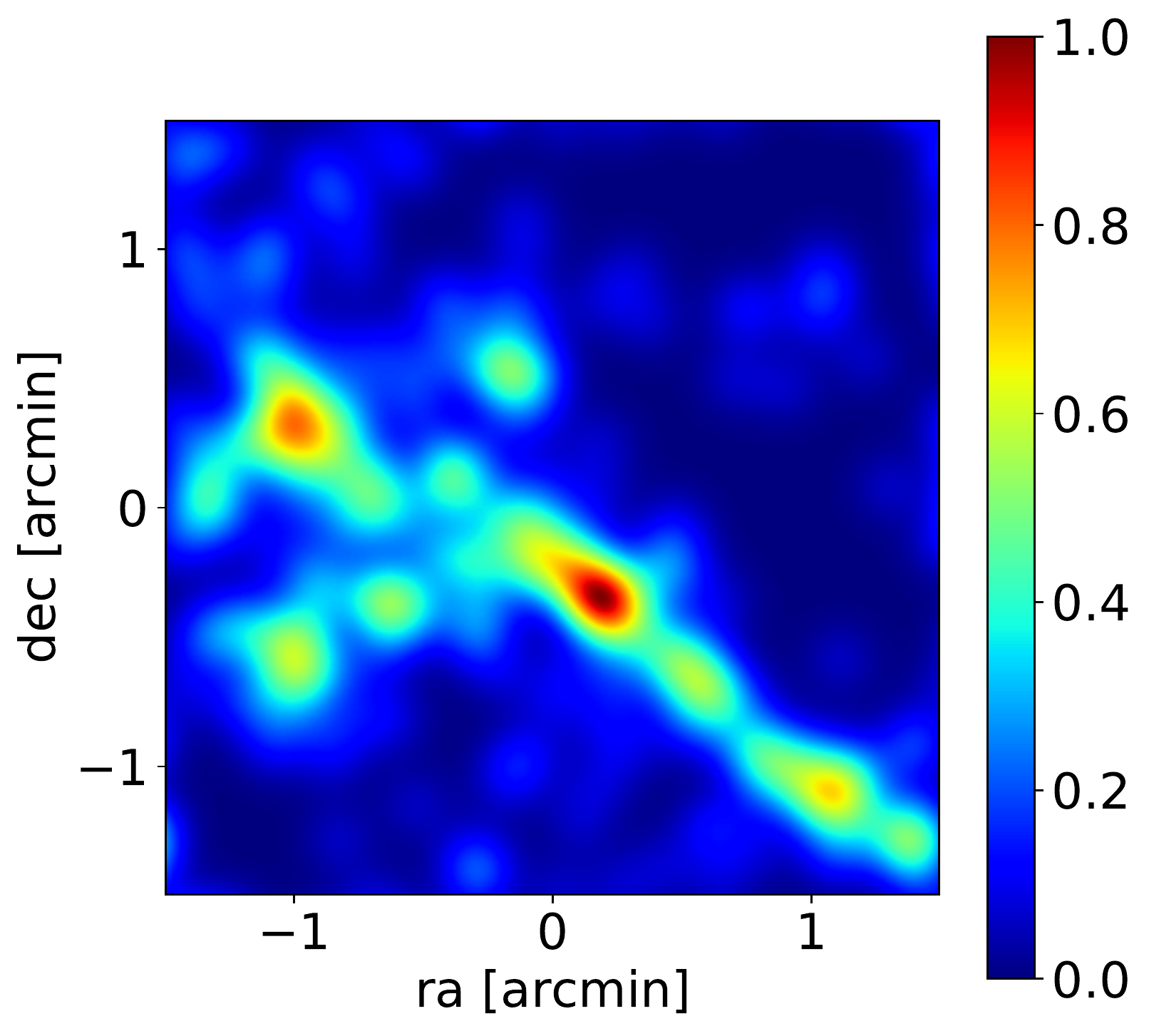}
\caption{2D image of the spatial distribution of \galics\ sources in the \mosaic\ convolved with a Gaussian ($\sigma = 0.1'$). This map is used as a 2D probability distribution function to draw the positions of ad hoc LAEs in our mock \galics\ groups. The color-code shown in the colorbar gives the probability per pixel.}
\label{fig:galics_fig3}
\end{figure}

Figure \ref{fig:galics_fig2} and Figure \ref{fig:galics_fig3} illustrate the successive steps of our spatial sampling procedure applied to the same mock group as in the previous section. The initial projected map of \galics\ LAEs within the \mosaic\ is represented by the black dots in Figure \ref{fig:galics_fig2}. Figure \ref{fig:galics_fig3} depicts the corresponding 2D probability distribution function obtained from the gaussian smoothing. Finally, the red dots in Figure \ref{fig:galics_fig2} show the resulting spatial distribution of ad-hoc LAEs.

\section{Galaxy Halo model}
\label{app:halomodel}
The surface brightness (SB) of the galaxy + halo is modeled as a sum of two circular, 2D exponential distributions, following \cite{Wisotzki2016} and \cite{Leclercq2017} prescription:
$$
\mathrm{
SB \left( r \right) = F_{\lya} \left[ \left( 1-X_h \right) e^{-r/r_g} + X_h e^{-r/r_h} \right],
}
$$
with ${\rm F_{\lya}}$ the total \lya\ flux given by \galics, ${\rm X_h}$ the fraction of flux in the halo, and the two scale lengths ${\rm r_g}$, ${\rm r_h}$ of, respectively, the galaxy and the halo.

${\rm X_h}$ is randomly drawn from a skewed normal distribution \citep{Ashour2010} fitted to the \cite{Leclercq2017} measurements of 145 \lya\ extended halos from the MUSE observations. The parameters of the distribution are loc=0.914, scale=0.298 and $\alpha$=-4.049, where loc and scale are the mean and standard deviation of the Normal distribution and $\alpha$ the skewness parameter.

The galaxy scale length ${\rm r_g}$  (in kpc) is derived from the ${\rm M_{uv}}$ \galics\ value, using the linear regression measured from the \cite{Leclercq2017} data (Figure 12 top). 
$$
\mathrm{r_g = -a_g M_{uv} + b_g},
$$
with $\mathrm{a_g = 0.080 \pm 0.017}$ and $\mathrm{b_g = -1.012 \pm 0.315}$.
The measured regression errors are used to produce ${\rm r_g}$ random values compatible with the measured dispersion.

Similarly the ${\rm r_h}$ halo scale lengths (in kpc) is derived using the computed linear regression  
 of Halo flux (in \ergsline) versus scale length (\citealt{Leclercq2017} Figure 8).
 $$
\mathrm{r_h = a_h \log F_{\lya} + b_h},
$$
with $\mathrm{a_h = 0.066 \pm 0.017}$ and $\mathrm{b_h = 3.477 \pm 0.264}$.
  
The surface brightness is truncated at the virial radius derived from the \galics\ halo mass.

Note that for the low luminosity sources below the mass cut of \galics\ and extrapolated from the \lya\ luminosity function (see section~\ref{subsec:galics}), we do not have UV magnitudes or halo masses. In those cases we use ${\rm M_{UV} = -15}$ and ${\rm M_{h} = 10^{8} M_{\odot}}$ to infer a value of the galaxy scale length and the virial radius. In practice these ultra faint galaxies are smaller than the \mxdf\ PSF and can be considered as point sources.

\section{Tables}
\label{app:tables}
\begin{sidewaystable*}
\caption{Overdensities of \lya\ emitters in the HUDF}         
\label{tab:over}
\centering
{
\doublespacing
\begin{tabular}{rrrrrrrrrrrrrrrrrr}
 ID & z & $\mathrm{\delta}$ & $\mathrm{\Delta}$ & $\mathrm{N_{s}}$ & $\mathrm{SB}$ & V &  $\mathrm{N_{l}}$ &  $\mathrm{N_{x}}$ &  $\mathrm{N_{r}}$ & $\mathrm{\log\,L_{Ly\alpha}}$ & $\mathrm{\log\,L_{Ly\alpha}^{T}}$ & $\mathrm{\log\,E_{Ly\alpha}}$ & $\mathrm{\log\,M_{\star}}$ & $\mathrm{\log\,M_{\star}^{T}}$ & $\mathrm{\log\,SFR}$ & $\mathrm{\log\,SFR^{T}}$ & $\mathrm{\log\,Age}$ \\
 \hline
1 & 2.9966 & 4.6 & 0.096 & 9 & 7.2 & 259.7 & 26 & 13 & 18 & $41.69 \pm 0.04$ & $43.10 \pm 0.01$ & $1.41 \pm 0.16$ & $9.51^{+0.18}_{-0.29}$ & $10.77^{+0.05}_{-0.05}$ & $0.58^{+0.15}_{-0.31}$ & $1.83^{+0.04}_{-0.06}$ & $8.71^{+0.18}_{-0.32}$ \\
2 & 3.0657 & 2.1 & 0.072 & 9 & 6.7 & 290.8 & 13 & 6 & 9 & $41.93 \pm 0.03$ & $43.05 \pm 0.01$ & $1.41 \pm 0.07$ & $10.69^{+0.31}_{-0.63}$ & $11.65^{+0.13}_{-0.13}$ & $1.10^{+0.48}_{-0.45}$ & $2.05^{+0.22}_{-0.11}$ & $8.69^{+0.21}_{-0.26}$ \\
3 & 3.0868 & 3.1 & 0.066 & 10 & 7.1 & 258.1 & 17 & 4 & 14 & $41.73 \pm 0.04$ & $42.96 \pm 0.01$ & $1.78 \pm 0.18$ & $9.26^{+0.27}_{-1.10}$ & $10.41^{+0.09}_{-0.12}$ & $0.48^{+0.35}_{-0.85}$ & $1.62^{+0.12}_{-0.11}$ & $8.69^{+0.17}_{-0.29}$ \\
4 & 3.0989 & 2.2 & 0.047 & 9 & 6.4 & 257.9 & 12 & 3 & 6 & $41.56 \pm 0.06$ & $42.64 \pm 0.02$ & $1.62 \pm 0.10$ & $9.81^{+0.15}_{-0.27}$ & $10.59^{+0.07}_{-0.09}$ & $0.85^{+0.25}_{-0.25}$ & $1.63^{+0.12}_{-0.08}$ & $8.71^{+0.17}_{-0.24}$ \\
5 & 3.1754 & 2.6 & 0.051 & 10 & 6.2 & 256.5 & 14 & 6 & 11 & $41.67 \pm 0.04$ & $42.82 \pm 0.01$ & $1.65 \pm 0.18$ & $8.27^{+0.29}_{-0.16}$ & $9.31^{+0.11}_{-0.04}$ & $-0.57^{+0.15}_{-0.14}$ & $0.47^{+0.05}_{-0.04}$ & $8.60^{+0.20}_{-0.31}$ \\
6 & 3.1924 & 3.2 & 0.066 & 9 & 5.9 & 288.2 & 19 & 11 & 13 & $41.52 \pm 0.06$ & $42.80 \pm 0.01$ & $1.59 \pm 0.10$ & $9.73^{+0.16}_{-0.11}$ & $10.84^{+0.05}_{-0.03}$ & $0.81^{+0.29}_{-0.45}$ & $1.92^{+0.10}_{-0.09}$ & $8.66^{+0.20}_{-0.35}$ \\
7 & 3.3344 & 3.2 & 0.063 & 10 & 5.5 & 285.1 & 18 & 4 & 12 & $41.64 \pm 0.05$ & $42.89 \pm 0.01$ & $1.48 \pm 0.14$ & $9.06^{+0.18}_{-0.78}$ & $10.14^{+0.06}_{-0.12}$ & $0.38^{+0.24}_{-\infty}$ & $1.46^{+0.08}_{-0.16}$ & $8.64^{+0.17}_{-0.27}$ \\
8 & 3.4149 & 2.1 & 0.045 & 10 & 5.7 & 314.6 & 13 & 4 & 9 & $41.66 \pm 0.05$ & $42.78 \pm 0.01$ & $1.54 \pm 0.08$ & $8.34^{+0.12}_{-0.25}$ & $9.29^{+0.04}_{-0.07}$ & $-0.44^{+0.12}_{-0.13}$ & $0.52^{+0.04}_{-0.04}$ & $8.63^{+0.18}_{-0.33}$ \\
9 & 3.4351 & 3.6 & 0.071 & 11 & 5.9 & 282.6 & 20 & 7 & 15 & $41.75 \pm 0.03$ & $43.05 \pm 0.01$ & $1.55 \pm 0.08$ & $8.78^{+0.04}_{-0.52}$ & $9.95^{+0.01}_{-0.09}$ & $-0.30^{+0.16}_{-0.11}$ & $0.88^{+0.05}_{-0.03}$ & $8.64^{+0.17}_{-0.26}$ \\
10 & 3.4703 & 2.8 & 0.054 & 10 & 5.6 & 313.1 & 17 & 3 & 15 & $41.69 \pm 0.04$ & $42.92 \pm 0.01$ & $1.63 \pm 0.11$ & $9.20^{+0.19}_{-0.64}$ & $10.37^{+0.06}_{-0.10}$ & $0.35^{+0.22}_{-0.32}$ & $1.53^{+0.07}_{-0.06}$ & $8.68^{+0.15}_{-0.26}$ \\
11 & 3.4998 & 1.9 & 0.036 & 11 & 5.9 & 281.1 & 10 & 2 & 7 & $41.58 \pm 0.06$ & $42.58 \pm 0.02$ & $1.70 \pm 0.11$ & $8.25^{+0.44}_{-0.43}$ & $9.10^{+0.22}_{-0.12}$ & $-0.42^{+0.14}_{-0.37}$ & $0.43^{+0.06}_{-0.11}$ & $8.54^{+0.20}_{-0.32}$ \\
12 & 3.5595 & 3.4 & 0.057 & 11 & 6.2 & 279.5 & 18 & 2 & 16 & $41.88 \pm 0.38$ & $43.13 \pm 0.09$ & $1.61 \pm 0.12$ & $9.46^{+0.16}_{-0.30}$ & $10.67^{+0.05}_{-0.06}$ & $0.44^{+0.30}_{-0.45}$ & $1.65^{+0.10}_{-0.08}$ & $8.62^{+0.18}_{-0.34}$ \\
13 & 3.6048 & 4.4 & 0.061 & 11 & 5.9 & 278.4 & 23 & 9 & 15 & $41.74 \pm 0.04$ & $43.11 \pm 0.01$ & $1.53 \pm 0.15$ & $9.32^{+0.44}_{-\infty}$ & $10.47^{+0.17}_{-0.17}$ & $0.87^{+0.24}_{-\infty}$ & $2.02^{+0.08}_{-0.22}$ & $8.57^{+0.26}_{-0.73}$ \\
14 & 3.7032 & 3.9 & 0.069 & 11 & 6.0 & 306.5 & 22 & 4 & 18 & $41.77 \pm 0.04$ & $43.12 \pm 0.01$ & $1.74 \pm 0.13$ & $8.79^{+0.33}_{-1.13}$ & $10.05^{+0.10}_{-0.11}$ & $0.23^{+0.49}_{-1.06}$ & $1.48^{+0.18}_{-0.11}$ & $8.53^{+0.21}_{-0.38}$ \\
15 & 3.7245 & 2.7 & 0.052 & 12 & 6.2 & 305.8 & 15 & 5 & 11 & $41.79 \pm 0.03$ & $42.97 \pm 0.01$ & $1.72 \pm 0.14$ & $8.62^{+0.35}_{-\infty}$ & $9.62^{+0.14}_{-0.19}$ & $-0.08^{+0.30}_{-0.30}$ & $0.92^{+0.12}_{-0.07}$ & $8.54^{+0.26}_{-0.87}$ \\
16 & 4.0469 & 2.0 & 0.034 & 13 & 5.7 & 325.7 & 11 & 3 & 8 & $41.76 \pm 0.05$ & $42.80 \pm 0.02$ & $1.43 \pm 0.08$ & $8.66^{+0.37}_{-0.86}$ & $9.51^{+0.18}_{-0.17}$ & $0.24^{+0.33}_{-0.29}$ & $1.09^{+0.16}_{-0.09}$ & $8.44^{+0.22}_{-0.39}$ \\
17 & 4.2714 & 2.4 & 0.035 & 13 & 5.0 & 347.0 & 13 & 4 & 8 & $41.68 \pm 0.05$ & $42.79 \pm 0.01$ & $1.84 \pm 0.12$ & $8.15^{+0.39}_{-0.37}$ & $9.05^{+0.18}_{-0.10}$ & $-0.31^{+0.22}_{-0.28}$ & $0.59^{+0.09}_{-0.08}$ & $8.41^{+0.22}_{-0.42}$ \\
18 & 4.4684 & 3.0 & 0.043 & 14 & 4.6 & 368.0 & 16 & 4 & 10 & $41.55 \pm 0.07$ & $42.75 \pm 0.02$ & $1.49 \pm 0.12$ & $8.37^{+0.39}_{-\infty}$ & $9.37^{+0.17}_{-0.20}$ & $0.09^{+0.38}_{-\infty}$ & $1.09^{+0.16}_{-0.17}$ & $8.38^{+0.23}_{-0.52}$ \\
19 & 4.5109 & 5.0 & 0.074 & 13 & 4.5 & 366.4 & 26 & 5 & 15 & $41.73 \pm 0.05$ & $43.15 \pm 0.01$ & $1.65 \pm 0.13$ & $8.89^{+0.45}_{-\infty}$ & $10.07^{+0.17}_{-0.18}$ & $0.57^{+0.46}_{-\infty}$ & $1.75^{+0.17}_{-0.13}$ & $8.34^{+0.29}_{-0.84}$ \\
20 & 4.8093 & 3.7 & 0.047 & 15 & 4.9 & 381.7 & 18 & 1 & 13 & $41.84 \pm 0.05$ & $43.10 \pm 0.01$ & $1.53 \pm 0.10$ & $8.46^{+0.38}_{-0.88}$ & $9.57^{+0.14}_{-0.12}$ & $0.03^{+0.34}_{-0.70}$ & $1.14^{+0.12}_{-0.11}$ & $8.38^{+0.22}_{-0.45}$ \\
21 & 4.9433 & 2.9 & 0.035 & 16 & 4.9 & 376.1 & 13 & 2 & 8 & $42.14 \pm 0.02$ & $43.25 \pm 0.01$ & $1.92 \pm 0.10$ & $8.56^{+0.19}_{-\infty}$ & $9.46^{+0.08}_{-0.23}$ & $0.14^{+0.13}_{-0.18}$ & $1.05^{+0.05}_{-0.06}$ & $8.30^{+0.27}_{-0.86}$ \\
22 & 5.7828 & 4.9 & 0.041 & 18 & 6.3 & 415.5 & 16 & 10 & 6 & $42.03 \pm 0.04$ & $43.23 \pm 0.01$ & $1.90 \pm 0.09$ & $8.69^{+0.33}_{-0.35}$ & $9.46^{+0.16}_{-0.11}$ & $0.52^{+0.19}_{-\infty}$ & $1.30^{+0.09}_{-0.23}$ & $8.22^{+0.27}_{-0.65}$ \\

\end{tabular}
}
\tablefoot{
ID: Group ID.
z: redshift. 
$\mathrm{\delta}$: overdensity factor. 
$\mathrm{\Delta}$: group density in $\mathrm{cMpc^{-3}}$.
$\mathrm{N_{s}}$ number of wavelength slices.
$\mathrm{SB}$ limiting surface brightness (1$\sigma$, 1 arcsec$^2$) in \erglsurf{-20} unit.
V: volume in $\mathrm{cMpc^{3}}$.
$\mathrm{N_{l}}$: number of LAE.
$\mathrm{N_{x}}$: number of LAE within \mxdf\ field.
$\mathrm{N_{r}}$: number of LAE with HST counterpart.
$\mathrm{L_{Ly\alpha}}$: \lya\ luminosity (group average value) in \ergslum\ unit.
$\mathrm{L_{Ly\alpha}^{T}}$: \lya\ luminosity (group total value).
$\mathrm{E_{Ly\alpha}}$: \lya\ Equivalent width in \AA.
$\mathrm{M_{\star}}$: stellar mass in solar mass (group average value) in $\rm M_{\sun}$ unit.
$\mathrm{M_{\star}^{T}}$:  stellar mass in solar mass (group total value).
$\mathrm{SFR}$: SFR (group average value) in $\rm M_{\sun}\, yr^{-1}$ unit.
$\mathrm{SFR^{T}}$: SFR (group total value).
$\mathrm{Age}$: age in year.  

Asymmetric error propagation was performed with the public code developed by \cite{Laursen2019}. 
}
\end{sidewaystable*}

\begin{sidewaystable*}
\caption{\lya\ diffuse emission in overdensities}            
\label{tab:ext}
\centering
\begin{tabular}{ccrrrrrrrrrrrrrr}
ID & Conf & $\mathrm{P_{dif}}$ & $\mathrm{F_{dif}}$ & $\mathrm{SN_{dif}}$ & $\mathrm{\log\, L_{dif}}$ & $\mathrm{SB_{dif}}$ & $\mathrm{S_{dif}}$ & $\mathrm{F_{comp}}$ & $\mathrm{SN_{comp}}$ & $\mathrm{\log\, L_{comp}}$ & $\mathrm{S_{comp}}$ & $\mathrm{F_{fil}}$ & $\mathrm{SN_{fil}}$ & $\mathrm{\log\, L_{fil}}$ & $\mathrm{S_{fil}}$ \\
\hline
1 & 0 & 0.200 & $21.9 \pm 3.1$ & 7.0 & $42.25 \pm 0.06$ & $8.1 \pm 1.3$ & 271.6 & $30.7 \pm 2.1$ & 14.7 & $42.39 \pm 0.03$ & 133.9 & $52.6 \pm 3.7$ & 14.2 & $42.63 \pm 0.03$ & 405.5 \\
2 & 1 & 0.012 & $40.8 \pm 4.2$ & 9.7 & $42.54 \pm 0.04$ & $7.2 \pm 0.8$ & 569.0 & $21.6 \pm 1.9$ & 12.4 & $42.27 \pm 0.04$ & 113.6 & $62.4 \pm 4.6$ & 14.1 & $42.73 \pm 0.03$ & 682.7 \\
3 & 0 & 0.906 & $10.9 \pm 2.6$ & 4.2 & $41.98 \pm 0.10$ & $4.8 \pm 1.2$ & 226.9 & $11.7 \pm 1.4$ & 8.3 & $42.01 \pm 0.05$ & 62.4 & $22.5 \pm 3.2$ & 7.1 & $42.29 \pm 0.06$ & 289.3 \\
4 & 0 & 0.622 & $15.3 \pm 2.8$ & 5.4 & $42.13 \pm 0.08$ & $5.2 \pm 1.0$ & 294.6 & $7.7 \pm 1.4$ & 6.2 & $41.83 \pm 0.08$ & 75.9 & $22.9 \pm 3.3$ & 7.4 & $42.30 \pm 0.06$ & 370.5 \\
5 & 1 & 0.020 & $44.2 \pm 4.9$ & 9.1 & $42.61 \pm 0.05$ & $5.2 \pm 0.6$ & 856.5 & $23.7 \pm 1.9$ & 12.5 & $42.34 \pm 0.03$ & 152.3 & $67.9 \pm 5.1$ & 13.3 & $42.80 \pm 0.03$ & 1008.8 \\
6 & 0 & 0.301 & $24.7 \pm 3.8$ & 6.5 & $42.37 \pm 0.07$ & $4.0 \pm 0.6$ & 624.0 & $52.9 \pm 2.3$ & 23.4 & $42.70 \pm 0.02$ & 232.3 & $77.7 \pm 4.3$ & 18.0 & $42.86 \pm 0.02$ & 856.2 \\
7 & 0 & 0.523 & $21.1 \pm 3.7$ & 5.8 & $42.34 \pm 0.08$ & $3.3 \pm 0.6$ & 645.3 & $13.6 \pm 1.7$ & 7.9 & $42.15 \pm 0.06$ & 156.0 & $34.7 \pm 4.0$ & 8.6 & $42.56 \pm 0.05$ & 801.4 \\
8 & 0 & 0.957 & $5.4 \pm 1.5$ & 3.6 & $41.78 \pm 0.12$ & $3.9 \pm 1.1$ & 139.2 & $5.3 \pm 0.9$ & 5.8 & $41.77 \pm 0.07$ & 53.1 & $10.8 \pm 1.7$ & 6.3 & $42.08 \pm 0.07$ & 192.3 \\
9 & 0 & 0.219 & $18.7 \pm 2.7$ & 6.8 & $42.32 \pm 0.06$ & $4.6 \pm 0.7$ & 404.9 & $14.7 \pm 1.4$ & 10.6 & $42.22 \pm 0.04$ & 110.7 & $33.4 \pm 3.1$ & 10.7 & $42.57 \pm 0.04$ & 515.6 \\
10 & 0 & 0.997 & $0.4 \pm 0.2$ & 1.8 & $40.67 \pm 0.24$ & $11.2 \pm 7.4$ & 3.6 & $-0.0 \pm 0.4$ & 3.2 & & 10.4 & $0.4 \pm 0.4$ & 3.7 & $40.67 \pm 0.46$ & 14.0 \\
11 & 0 & 0.858 & $6.6 \pm 1.5$ & 4.4 & $41.89 \pm 0.10$ & $5.0 \pm 1.2$ & 132.5 & $1.6 \pm 0.7$ & 3.7 & $41.27 \pm 0.18$ & 24.2 & $8.2 \pm 1.6$ & 5.5 & $41.98 \pm 0.09$ & 156.7 \\
12 & 0 & 0.457 & $11.6 \pm 2.0$ & 6.0 & $42.15 \pm 0.07$ & $5.4 \pm 1.0$ & 214.1 & $3.0 \pm 1.0$ & 6.8 & $41.57 \pm 0.14$ & 54.1 & $14.7 \pm 2.3$ & 8.0 & $42.25 \pm 0.07$ & 268.2 \\
13 & 0 & 0.451 & $15.8 \pm 2.6$ & 6.0 & $42.30 \pm 0.07$ & $3.9 \pm 0.7$ & 406.1 & $12.0 \pm 1.4$ & 8.4 & $42.18 \pm 0.05$ & 109.8 & $27.8 \pm 3.2$ & 8.8 & $42.54 \pm 0.05$ & 516.0 \\
14 & 0 & 0.914 & $5.3 \pm 1.3$ & 4.1 & $41.85 \pm 0.11$ & $5.2 \pm 1.4$ & 100.0 & $10.1 \pm 1.1$ & 8.8 & $42.13 \pm 0.05$ & 83.3 & $15.3 \pm 1.7$ & 9.0 & $42.31 \pm 0.05$ & 183.4 \\
15 & 0 & 0.999 & $0.1 \pm 0.1$ & 0.8 & $40.14 \pm 0.53$ & $10.7 \pm 15.1$ & 1.0 & $2.1 \pm 0.4$ & 5.0 & $41.46 \pm 0.09$ & 11.7 & $2.2 \pm 0.5$ & 4.8 & $41.48 \pm 0.09$ & 12.6 \\
16 & 0 & 0.704 & $6.9 \pm 1.4$ & 5.1 & $42.06 \pm 0.09$ & $5.3 \pm 1.1$ & 130.1 & $1.6 \pm 0.4$ & 3.6 & $41.41 \pm 0.12$ & 14.3 & $8.4 \pm 1.4$ & 6.0 & $42.14 \pm 0.07$ & 144.4 \\
17 & 1 & 0.012 & $22.3 \pm 2.3$ & 9.7 & $42.62 \pm 0.04$ & $4.5 \pm 0.5$ & 500.8 & $14.0 \pm 1.2$ & 12.2 & $42.42 \pm 0.04$ & 157.9 & $36.3 \pm 2.7$ & 13.8 & $42.83 \pm 0.03$ & 658.7 \\
18 & 1 & 0.057 & $16.4 \pm 2.0$ & 8.1 & $42.54 \pm 0.05$ & $3.7 \pm 0.5$ & 443.5 & $4.5 \pm 0.8$ & 5.9 & $41.97 \pm 0.08$ & 68.4 & $20.8 \pm 2.3$ & 9.4 & $42.64 \pm 0.05$ & 511.9 \\
19 & 1 & 0.002 & $30.1 \pm 2.3$ & 12.8 & $42.81 \pm 0.03$ & $4.9 \pm 0.4$ & 614.9 & $11.8 \pm 1.0$ & 11.7 & $42.40 \pm 0.04$ & 120.9 & $41.9 \pm 2.4$ & 17.1 & $42.95 \pm 0.03$ & 735.8 \\
20 & 0 & 0.172 & $19.2 \pm 2.7$ & 7.0 & $42.68 \pm 0.06$ & $3.2 \pm 0.5$ & 597.7 & $17.4 \pm 1.5$ & 11.7 & $42.64 \pm 0.04$ & 194.2 & $36.7 \pm 3.3$ & 11.2 & $42.96 \pm 0.04$ & 791.9 \\
21 & 0 & 0.126 & $14.6 \pm 2.0$ & 7.3 & $42.59 \pm 0.06$ & $4.1 \pm 0.6$ & 356.2 & $3.2 \pm 0.6$ & 5.0 & $41.93 \pm 0.09$ & 40.6 & $17.8 \pm 2.0$ & 8.7 & $42.67 \pm 0.05$ & 396.8 \\
22 & 0 & 0.180 & $10.5 \pm 1.5$ & 6.9 & $42.60 \pm 0.06$ & $4.8 \pm 0.8$ & 217.9 & $7.1 \pm 1.0$ & 7.6 & $42.44 \pm 0.06$ & 95.2 & $17.6 \pm 1.8$ & 9.8 & $42.83 \pm 0.04$ & 313.1 \\

\end{tabular}
\tablefoot{
ID: Group ID. Conf: confidence. $\mathrm{P_{dif}}$: P-value for the diffuse component, $\mathrm{F_{c}}$: \lya\ flux in $\mathrm{10^{-18}}$\ergsline\ for the component c, $\mathrm{SN_{c}}$: Flux SNR for the component i,
 $\mathrm{\log\, L_{c}}$: Log of \lya\ luminosity in \ergslum\ for the component i.
 $\mathrm{SB_{dif}}$: Average surface brightness of the diffuse \lya\ emission in \erglsurf{-20}. 
 $\mathrm{S_{c}}$: Surface in $\mathrm{arcsec^2}$ for the component c.
 Components (c), dif: diffuse emission, comp: compact source emission, fil: full filament. 
}
\end{sidewaystable*}

\end{appendix}

\end{document}